\documentclass{article}
\usepackage{emulateapj,timesfonts}
\usepackage{float,epsfig,psfig}
\usepackage{pstricks}
   \def\apj#1#2#3#4{\par #4 19#3, {ApJ,\/} {#1}, #2 }

   \def\apjs#1#2#3#4{\par #4 19#3, {ApJ (Supplement Series),\/} { #1}, #2 }
   \def\aj#1#2#3#4{\par #4 19#3, {AJ,\/} { #1}, #2}
   \def\pasj#1#2#3#4{\par #4 19#3, {PASJ,\/} { #1},#2 }

   \def\aa#1#2#3#4{\par #4 19#3, {A\&A,\/} { #1}, #2 }
   
   \def\aas#1#2#3#4{\par #4 19#3, {A\&A (Supplement Series),\/} { #1}, #2 }
   
   \def\mnras#1#2#3#4{\par #4 19#3, {MNRAS,\/} { #1}, #2 }

   \def\apjl#1#2#3#4{\par #4 19#3, {ApJ (Letters),\/} { #1}, #2 }

   \def\nature#1#2#3#4{\par #4 19#3, {Nature,\/} { #1}, #2 }
   
   \def\asr#1#2#3#4{\par #4 19#3, {\em Adv. Space Res.,\/} { #1}, #2 }

   \def\BIB {\par}

\def\hea4{{\it HEAO~A4}}
\def\heaoa2{{\it HEAO~A2}}
\def\heao1{{\it HEAO~1}}

\def\amin{$^\prime$}

\def\eg{{\it e.g.}~}

\def\h0{$H_{\rm o}=50$~km~s$^{-1}$~Mpc$^{-1}$}
\def\q0{$q_{\rm o}$}

\def\msun     {$M_{\odot}$}
\def\lsun     {$L_{\odot}$}

\def\etal    {{ et~al.}~}

\def\cms3  {~{cm$^{-3}$}}

\begin{document}

\submitted{Submitted to ApJ: August 11, 1999, Last revised \today}
\title{{An ASCA Study of the Heavy Element Distribution in Clusters of Galaxies}}
\author{A.~Finoguenov$^{1,4,5}$, L.P.~David$^2$ and T.J.~Ponman$^3$}
\affil{\vspace*{0.5cm} {$^1$ Astrophysikalisches Institut Potsdam, An der
Sternwarte 16, 14882 Potsdam, Germany}\\ 
{$^2$ Smithsonian Astrophysical Observatory, 60 Garden st., MS 2, Cambridge,
  MA 02138, USA}\\  
{$^3$ University of Birmingham, Edgbaston, Birmingham B15 2TT, UK}\\
{$^4$ Space Research Institute, Profsoyuznaya 84/32 117810 Moscow, Russia}\\
{$^5$ Alexander von Humboldt Fellow}}
%\authoremail{alexis@hea.iki.rssi.ru} 
\submitted{}

\begin{abstract}
  
  We perform a spatially resolved X-ray spectroscopic study of a set of 11
  relaxed clusters of galaxies observed by the ROSAT/PSPC and ASCA/SIS.
  Using a method which corrects for the energy dependent effects of the ASCA
  PSF based on ROSAT images, we constrain the spatial distribution of Ne,
  Si, S and Fe in each cluster.  Theoretical prescriptions for the chemical
  yields of Type Ia and II supernovae, then allow determination of the Fe
  enrichment from both types of supernovae as a function of radius within
  each cluster.  Using optical measurements from the literature, we also
  determine the iron mass-to-light ratio (IMLR) separately for Fe
  synthesized in both types of supernovae.  For clusters with the best
  photon statistics, we find that the total Fe abundance decreases
  significantly with radius, while the Si abundance is either flat or
  decreases less rapidly, resulting in an increasing Si/Fe ratio with
  radius.  This result indicates a greater predominance of Type II SNe
  enrichment at large radii in clusters.  On average, the IMLR synthesized
  within Type II SNe increases with radius within clusters, while the IMLR
  synthesized within Type Ia SNe decreases. At a fixed radius of 0.4
  $R_{virial}$ there is also a factor of 5 increase in the IMLR synthesized
  by Type II SNe between groups and clusters.  This suggests that groups
  expelled as much as 90\% of the Fe synthesized within type II SNe at early
  times. All of these results are consistent with a scenario in which the
  gas was initially heated and enriched by Type II SNe driven galactic
  winds. Due to the high entropy of the preheated gas, SN~II products are
  only weakly captured in groups. Gravitationally bound gas was then
  enriched with elements synthesized by Type Ia supernovae as gas-rich
  galaxies accreted onto clusters and were stripped during passage through
  the cluster core via density-dependent mechanisms (e.g. ram pressure
  ablation, galaxy harassment etc.). We suggest that the high Si/Fe ratios
  in the outskirts of rich clusters may arise from enrichment by Type II SNe
  released to ICM via galactic star burst driven winds. Low S/Fe ratios
  observed in clusters suggest metal-poor galaxies as a major source of SN
  II products.
  
\end{abstract}

\keywords{Abundances --- galaxies: clusters: general --- galaxies: evolution
  --- intergalactic medium  --- stars: supernovae --- X-rays: galaxies}

\section{Introduction}

%\doublespace

Understanding the formation and evolution of large structures like groups
and clusters of galaxies is one of the main goals of contemporary
astrophysics.  The presence of hot X-ray emitting gas in clusters is an
invaluable diagnostic of the dynamic, thermodynamic, and chemical state of
clusters.  The discovery of Fe-K line emission in cluster spectra by
Mitchell \etal (1976) and Serlemitsos \etal (1977) was the first indication
that some of the hot intracluster gas originated within galaxies.  Heavy
elements can be supplied to the cluster gas through galactic winds (De Young
1978; David, Forman, and Jones 1991; White 1991; Metzler \& Evrard 1994) and
ram pressure stripping (Gunn and Gott 1971). Estimates of the iron abundance
in clusters from HEAO-1, Exosat, Ginga, and ROSAT observations show that
most rich clusters have iron abundances that range between 25\%-50\% of the
solar value.  Arnaud \etal (1992) showed that there is a strong correlation
between the Fe mass in rich clusters and the luminous mass in E and S0
galaxies.  Such a correlation suggests that, at least in rich clusters, the
iron may predominantly originate from early type galaxies. Even though the
Fe abundances in clusters are typically sub-solar, there is a considerable
amount of Fe in the intracluster medium since the hot gas comprises
approximately 5 times the luminous mass in galaxies (David \etal 1990,
Arnaud \etal 1992). As noted by Renzini (1997), the Fe mass-to-light ratio
(IMLR) for the hot gas in clusters is typically $\sim$0.02 \msun/\lsun\, and
is several times larger than that locked up in stars.  Thus, not only does
the hot gas contain most of the baryons in clusters, it also contains most
of the heavy elements.

\begin{table*}
{
\begin{center}
\footnotesize
\tabcaption{\centerline{\footnotesize
X-ray and optical quantities of the sample}
\label{tab:opt}}

\begin{tabular}{lrrrrrcccrccc}
\hline
\hline
Name  &D$_{L}$ & 1\amin\  &$L_B$          & $R_{mes}$ &$L_{B, N_{0.5}}$ & $L_{B_{cD}}$   &
$R_c$ & $\beta_{out}$ & $r_{out}$ & $T_e$ & $\delta T_e$ & $R_{virial}$ \\
      &Mpc     & kpc      & $10^{11}$\lsun& Mpc       &$10^{11}$\lsun   & $10^{11}$\lsun & 
Mpc   &               & kpc       & keV   & keV          & Mpc \\
\hline
A496  & 199 &  53 &       &       & 10.9$^{\dag}$& 1.35 & 0.39 & 0.70 & 250 & 4.7 & 0.2 & 2.67 \\
A780  & 346 &  90 &       &       &  9.4$^{\dag}$&      & 0.24 & 0.66 & 120 & 4.3 & 0.4 & 2.56 \\
A1060 &  69 &  19 &       &       &  5.42        & 1.22 & 0.24 & 0.70 & 161 & 3.1 & 0.1 & 2.18 \\
A1651 & 520 & 128 &       &       & 15.9$^{\dag}$&      & 0.24 & 0.70 & 260 & 6.1 & 0.4 & 3.05 \\
A2029 & 470 & 118 &       &       & 22.53        & 2.53 & 0.22 & 0.68 & 280 & 9.1 & 1.1 & 3.72 \\
A2199 & 181 &  50 &       &       & 10.91        & 2.19 & 0.24 & 0.64 & 140 & 4.8 & 0.2 & 2.70 \\
A2597 & 520 & 129 &       &       &  9.7$^{\dag}$& 0.40 & 0.24 & 0.68 & 180 & 4.4 & 0.4 & 2.59 \\
A2670 & 465 & 117 & 26.18 &  3.10 &  9.1$^{\dag}$& 0.55 & 0.24 & 0.55 & 113 & 4.2 & 0.6 & 2.53 \\
A3112 & 461 & 116 &       &       & 12.9$^{\dag}$& 2.59 & 0.24 & 0.63 & 120 & 5.3 & 0.8 & 2.84 \\
A4059 & 291 &  77 &       &       &  9.7$^{\dag}$& 1.04 & 0.24 & 0.67 & 220 & 4.4 & 0.3 & 2.59 \\
MKW4  & 121 &  33 & 14.80 & 10.25 &              & 1.78 & 0.17 & 0.64 & 180 & 1.7 & 0.1 & 1.60 \\
AWM7  & 104 &  29 & 18.80 &  9.21 &  6.89        & 1.90 & 0.22 & 0.53 & 102 & 3.5 & 0.2 & 2.31 \\
HCG51 & 156 &  43 &  2.80 &  1.19 &              & ---  & 0.06 & 0.30 &  78 & 1.3 & 0.1 & 1.43 \\
HCG62 &  82 &  23 &  0.95 &  1.09 &              & ---  & 0.03 & 0.30 &  52 & 1.1 & 0.1 & 1.30 \\
N5044 &  54 &  15 &  2.30 &  0.5  &              & 0.68 & 0.18 & 0.51 &  11 & 1.2 & 0.1 & 1.37 \\
Virgo &  17 &   5 &  9.60 &  3.0  &  4.73        & 1.05 & 0.24 & 0.45 &  11 & 2.7 & 0.2 & 2.03 \\
\hline
\end{tabular}
\end{center}

%\begin{enumerate}
%\item[{$^{\dag}$}]{\footnotesize Interpolated value using Eq.\ref{nb-kt} }
\hspace*{2.6cm} {$^{\dag}$} \hspace*{0.3cm}{\footnotesize Estimated value using Eq.\ref{nb-kt} }
%\end{enumerate}

}
\end{table*}

To determine the exact origin of the heavy elements in clusters requires
information on the abundance of several heavy elements which is now possible
with ASCA.  Galactic winds primarily occur at early times and enrich the
intergalactic medium with type II SN ejecta which is abundant in
$\alpha$-process elements. Ram pressure stripping of galaxies occurs over a
longer period as clusters continuously accrete field galaxies and enrich the
intergalactic medium with type Ia SN ejecta which is abundant in Fe.
Loewenstein \& Mushotzky (1996) showed that the abundance ratios of
$\alpha$-process elements relative to Fe outside the central regions of four
rich clusters was similar to that of type II SN ejecta.  Fukazawa \etal
(1998) have shown that globally averaged Fe abundances outside cluster cores
are nearly independent of gas temperature, while Si abundances increase
significantly with gas temperature. This suggests that there is an
increasing predominance of type II SN in the enrichment process of richer
clusters. In general, the heavy elements that originate from type II SN are
a direct measure of the stellar initial mass function (IMF) and the gas
retention ability of clusters.

In Finoguenov \& Ponman (1999), we demonstrated that both SN~Ia and SN~II
make a significant contribution to the enrichment of the intergalactic
medium by comparing the metal content of groups with that of a cluster of
galaxies. We found that SN~II products are widely distributed within the
IGM, which probably indicates that the ejecta were released at early times
in the formation of clusters via energetic galactic winds. The observed iron
mass-to-light ratios indicate that rich clusters retained the SN~II products
more efficiently than groups at all radii from 0.05 to at least 0.4 of
$R_{virial}$. In contrast, the distribution of Fe attributed to SN~Ia is
more centrally peaked, with a core radius comparable to the optical radius
of the system.  This indicates that the primary gas enrichment mechanism is
strongly density-dependent, such as gas ablation, ram pressure stripping, or
galaxy-galaxy interactions.

In this {\it Paper}, we present an extensive study of the distribution of
heavy elements in groups and clusters and discuss the implications regarding
the chemical enrichment process of the intracluster medium.  The elemental
abundances are determined using 3-dimensional modeling of ASCA/SIS data.
This procedure takes advantage of the higher spatial resolution ROSAT
observations, which are used to model the effect of the broader, energy
dependent PSF of the ASCA XRT on the distribution of detected events in the
ASCA focal plane. We present new spectroscopic analysis for MKW4, A496,
A780, A1060, A1651, A2029, A2199, A2597, A2670, A3112, and A4059.  This
sample, contains almost all of the 4 keV clusters observed by ASCA, for
which we can determine the abundance of several heavy elements. Our sample
also contains MKW4 and A1060, which are slightly cooler, and A2029 and A1651
which are slightly hotter.  To complete our sample, we include previously
published results on M87 (Finoguenov \& Jones 2000); AWM7 (Finoguenov \&
Ponman 1999); and the groups NGC5044, HCG51, and HCG62 (Finoguenov \& Ponman
1999), which sample the low-temperature end of the cluster distribution.

This paper is organized as follows: in section {\it\ref{sec:data}} we
describe our analysis technique; we then present the radial abundance
profiles of Fe, Si, Ne and S in section {\it\ref{sec:radial}}, and
analyze the spatial distribution of abundance ratios in our cluster
sample in section {\it\ref{sec:ratios}}. In section
{\it\ref{sec:disc}} we discuss the relative importance in the overall
enrichment process of different types of supernovae.  Our main results
are summarized in section {\it\ref{sec:sum}} along with a discussion
about their implications for cluster evolution.  We assume {\h0} and
{\q0$=0.5$} throughout the paper.

\section{Data Reduction}\label{sec:data}

A detailed description of the ASCA observatory as well as the SIS detectors
can be found in Tanaka, Inoue \& Holt (1994) and Burke \etal (1991).  All
observations are screened using FTOOLS version 4.1 with standard screening
criteria. The effect of the broad ASCA PSF is treated as described in
Finoguenov \etal (1999), including the geometrical projection of the
three-dimensional distribution of X-ray emitting gas. Our minimization
routines are based on the $\chi^2$ criterion.  No energy binning is done,
but a special error calculation is introduced as in Churazov \etal (1996),
to properly account for small number statistics.

Model fits to ROSAT (Truemper 1983) surface brightness profiles are used as
input to the ASCA data modeling. The ROSAT results are used to define the
{\it shape} of the X-ray profile {\it within} each spatial bin used for the
ASCA analysis. Our analysis therefore assumes that the ASCA and ROSAT
surface brightness profiles have the same shape. This shape affects the
amount of coupling (due to projection and PSF blurring) between the ASCA
spectra, which is allowed for in the simultaneous fitting process. However,
the {\it normalization} of the flux in each ASCA bin is not determined by
the ROSAT profile, but is free to fit to the ASCA data. Hence any difference
in emission between the soft (ROSAT) and harder bands, due to possible
temperature gradients or the presence of multiphase gas, has only a second
order effect on our analysis, via changes in profile shape with
energy. Moreover, the effects of any such energy variations must reduce as
smaller spatial bins are used for the analysis, since as the radial bins
shrink towards zero width, these factors are known from the ASCA PSF
calibration alone, and does not depend on the ROSAT data at all. In checking
the effects of systematics, we have varied the spatial binning to quantify
the effect on our modeling of possible differences between profiles in the
ROSAT and ASCA bands, near the cluster center.

For regions with complex structure (\eg\ the central part of A2670 which has
a bimodal structure), we used the actual PSPC image (rather than just its
radial profile) as a basis for the ASCA modeling. For all ROSAT imaging
analysis, we use the software described in Snowden \etal (1994) and
references therein. The details of our minimization procedure for ASCA
spectral analysis are described in Finoguenov and Ponman (1999).

We adapted the XSPEC analysis package to perform the actual fitting and
error estimation.  Fluctuations in spectral properties are constrained using
a regularization technique (Press \etal 1992; Finoguenov and Ponman 1999),
but the solution returned is always required to lie within 90\% confidence
of the best fit. Since regularization effectively couples the values
returned for the spectral properties at neighboring points, we display the
error regions as continuous bands in the Figures below.

\begin{table}[H]
\begin{minipage}{8cm}
\footnotesize
\begin{center}
\tabcaption{\footnotesize
\centerline{ASCA SIS observational log.}
\label{tab:obslog}}

\begin{tabular}{clcc}
\hline
\hline
     &         & SIS0 & SIS1 \\
Name & seq. \# & exp. & exp. \\
     &         & ksec & ksec \\
\hline
A496  & 80003000 & 31 & 23 \\
      & 86069000 & 22 & 22 \\
      & 86069010 &  6 & 24 \\
A780  & 80015000 & 14 & 12 \\
A1060 & 80004000 & 30 & 22 \\
A1651 & 82036000 & 32 & 30 \\
A2029 & 81023000 & 29 & 28 \\
      & 83040010 &  6 & --- \\
      & 83040020 &  3 & --- \\
A2199 & 80023000 & 21 & 16 \\
      & 86068000 & 19 & 19 \\  
      & 86068010 & 22 & 22 \\
A2597 & 83062000 & 36 & 34 \\
A2670 & 82049000 & 17 & 13 \\
A3112 & 81003000 & 25 & 16 \\
A4059 & 82030000 & 28 & 27 \\
MKW4  & 52027000 & 47 & 50 \\
      & 82012000 & 12 & 6 \\
      & 82013000 & 13 & 6 \\
      & 82014000 & 11 & 4 \\
\hline
\end{tabular}
\end{center}
\end{minipage}
\end{table}

The MEKAL plasma code (Mewe \etal 1985, Mewe and Kaastra 1995, Liedahl \etal
1995) has been used for all of our spectral analysis.  Abundances are given
relative to the solar values in Anders \& Grevesse (1989).  The abundances
of He and C are fixed to solar values.  The remaining elements are combined
into five groups for fitting: Ne; Mg; Si; S and Ar; and Ca, Fe, and Ni. We
restrict our analysis to the energy range 0.8--7.0 keV to avoid the large
systematic uncertainties at low energies, which also prevent us from
determining the O abundance.  We do not report the Mg abundance due to the
proximity of the Mg K lines to the poorly understood 4-2 transition lines of
iron, which are strongest at temperatures of 2 to 4~keV (see Fabian \etal
1994 and Mushotzky \etal 1996).

The presence of a central cooling flow in some clusters further complicates
the spectral analysis. In such cases, we have applied three different models
for the cooling flow emission, within the central one or two spatial
bins: (a) a standard XSPEC cooling flow model ({\it mkcflow}) plus a thermal
component (MEKAL) of the same temperature as the maximum temperature in the
cooling flow model, (b) a realization of the code of Wise \& Sarazin (1993)
for cooling flow plus an isothermal component similar to that in (a), and (c) a
two-temperature model (Buote 1999). In each model, element abundances were
assumed to be the same in both components, except in the case of Virgo,
where they were fitted separately and shown to be the same (Finoguenov \&
Jones 2000). The two-temperature model was found to most strongly reduce the
luminosity of the primary temperature component. Using this model within the
cooling region, the gas distribution from the {\it hot} component was found
to be a good extrapolation of a beta-model fit to ROSAT data from the data
outside the cooling region.

The addition of a cooling flow component was found to have little effect on
element abundance determination in the hotter systems.  However, for systems
whose temperature and abundance determination relies on the Fe L-shell
emission, introduction of the two-temperature component strongly affects the
stability of the solution, due to the inadequate energy resolution of our
CCD measurements. Among the systems, studied in this paper, HCG51, HCG62,
NGC5044 and MKW4 fall into this category. We discuss the implication of this
effect for the credibility of the corresponding abundance determinations in
the Appendix. Following Finoguenov \& Jones (2000), we add 10\% systematic
error below 1.5 keV in spectral analysis, whenever we determine the Fe
abundance from the L-shell line complex.

No simple correlation was found between our cooling flow mass drop rate
estimations and the strength of abundance gradients. However, the suggestion
of Allen and Fabian (1998) that non-cooling flow clusters have lower central
iron abundance due to recent mergers, explains the abundance profile of
A2670. A2670 was included in our analysis first and in adding other clusters
to our sample we specially avoided indications of merger, so it is not
surprising that perhaps 'too many' our clusters show abundance gradients.

Following a suggestion of the referee, we explore the effects of ASCA PSF
systematics as well as the sensitivity of our results to changes in the
modeling of the ROSAT surface brightness distribution. The effects of PSF
variations were explored using simulations, as in Markevitch \etal
(1998). The calibration uncertainty in the ASCA PSF estimation was adopted
as 10\% at the 68\% confidence level. In our spectral analysis, we account
for the effects of the PSF using a scattering matrix (Finoguenov \etal
1999). In simulating the effect of systematic PSF errors, we create a number
of different scattering matrices. The resulting systematic uncertainty in
the interesting parameters was calculated as the RMS of their best-fit
values obtained using different scattering matrices.

\begin{figure*}

\includegraphics[width=1.8in]{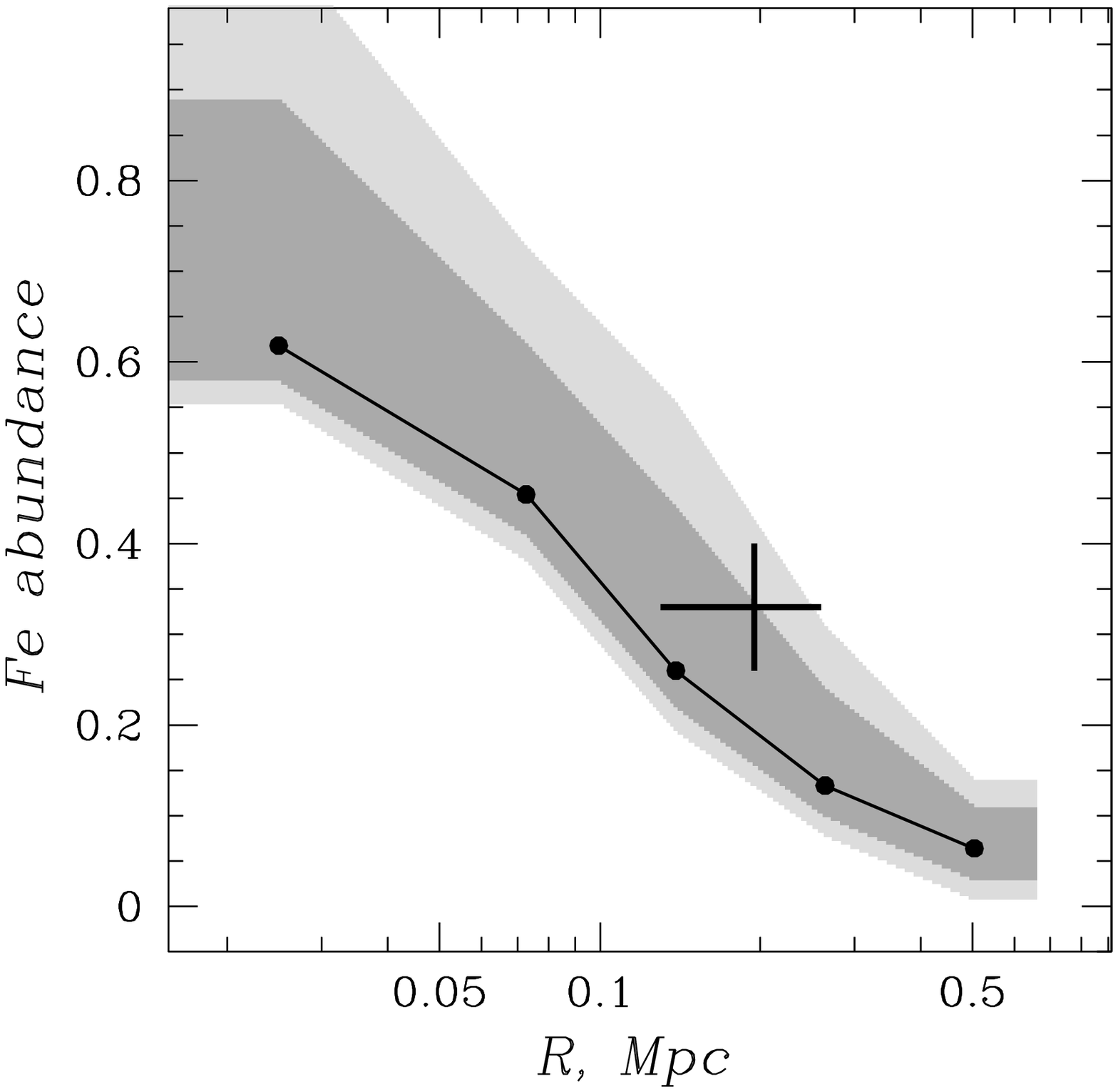} \hfill \includegraphics[width=1.8in]{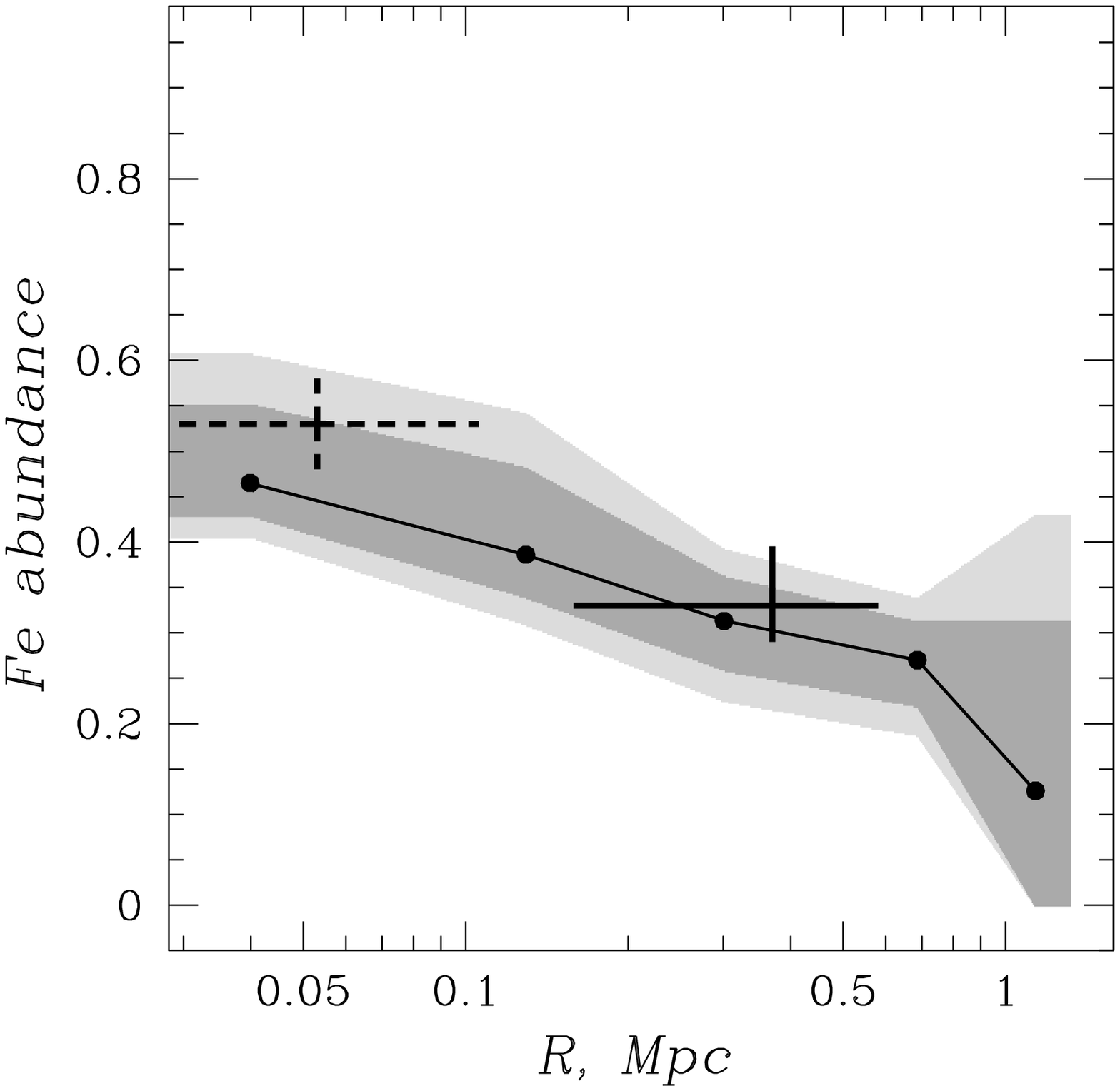} \hfill \includegraphics[width=1.8in]{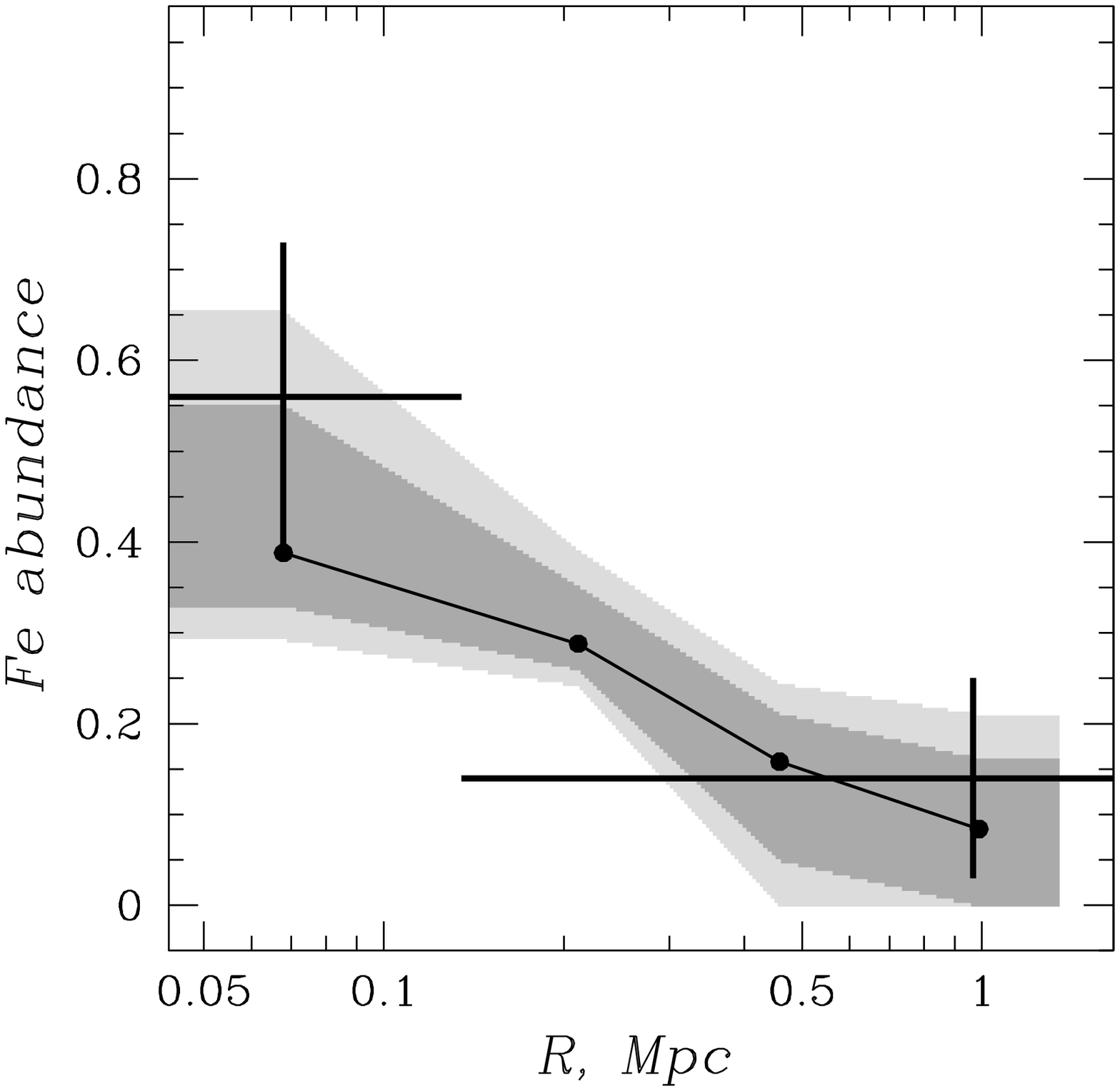} \hfill \includegraphics[width=1.8in]{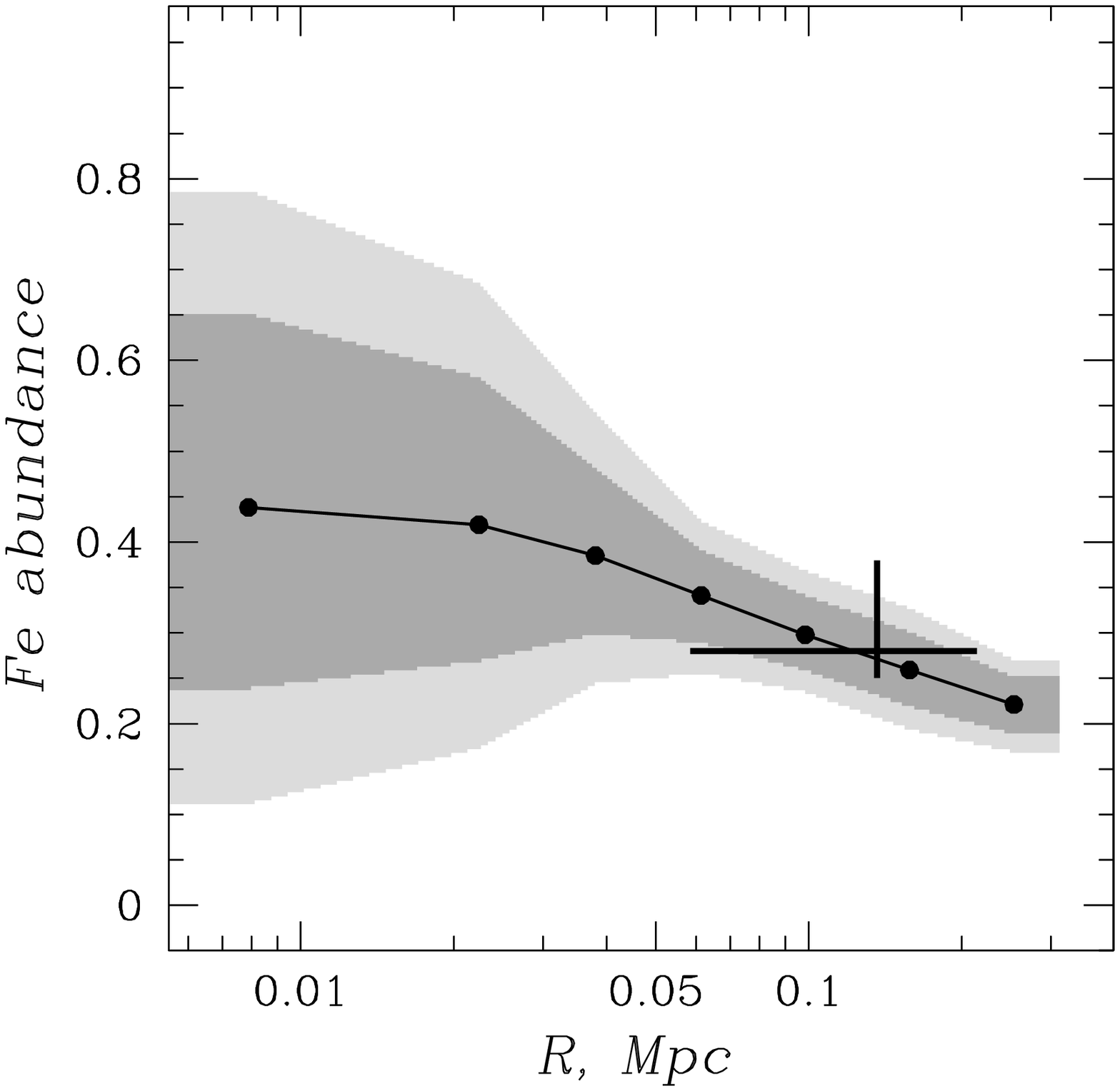}

\includegraphics[width=1.8in]{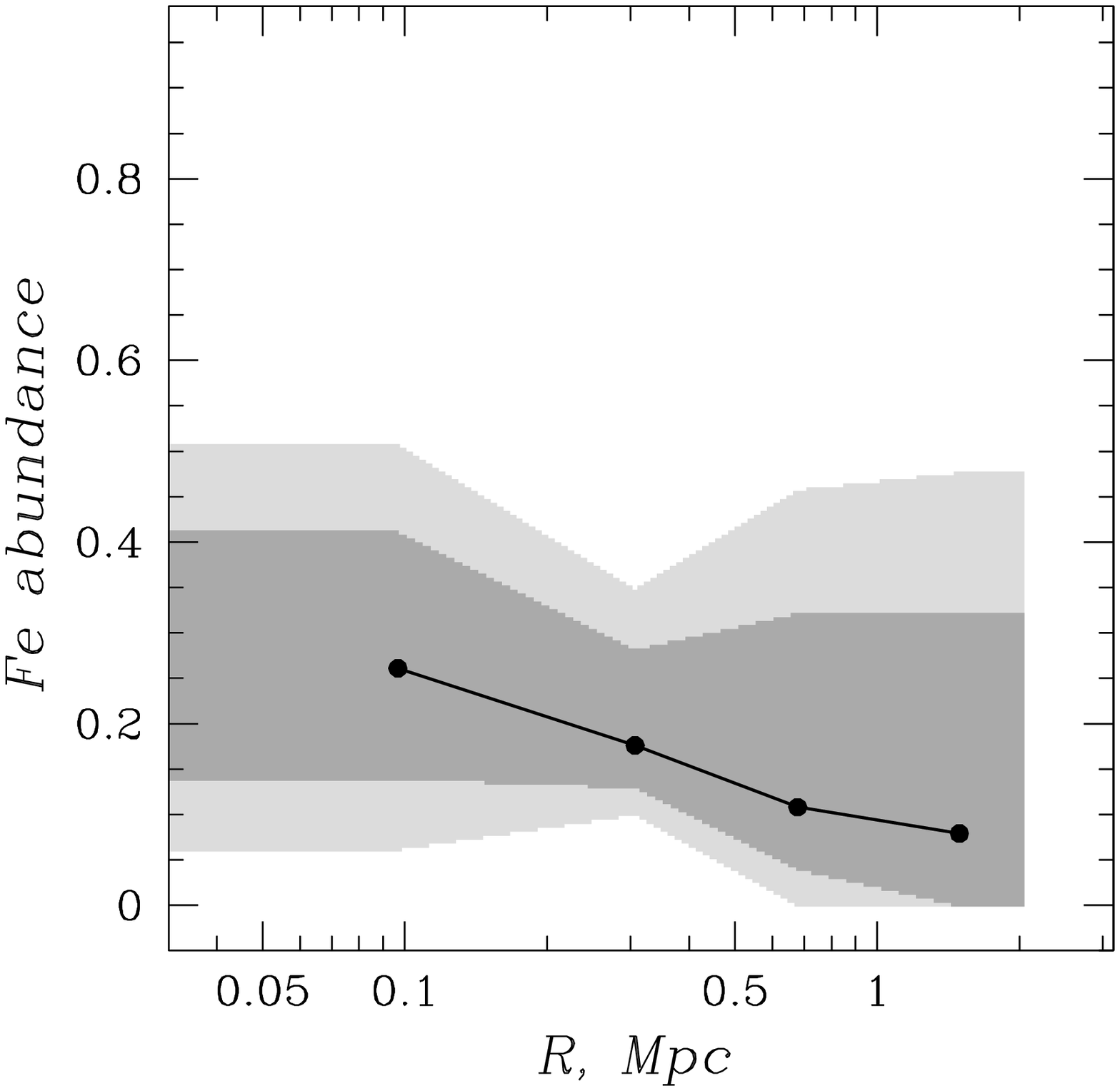} \hfill \includegraphics[width=1.8in]{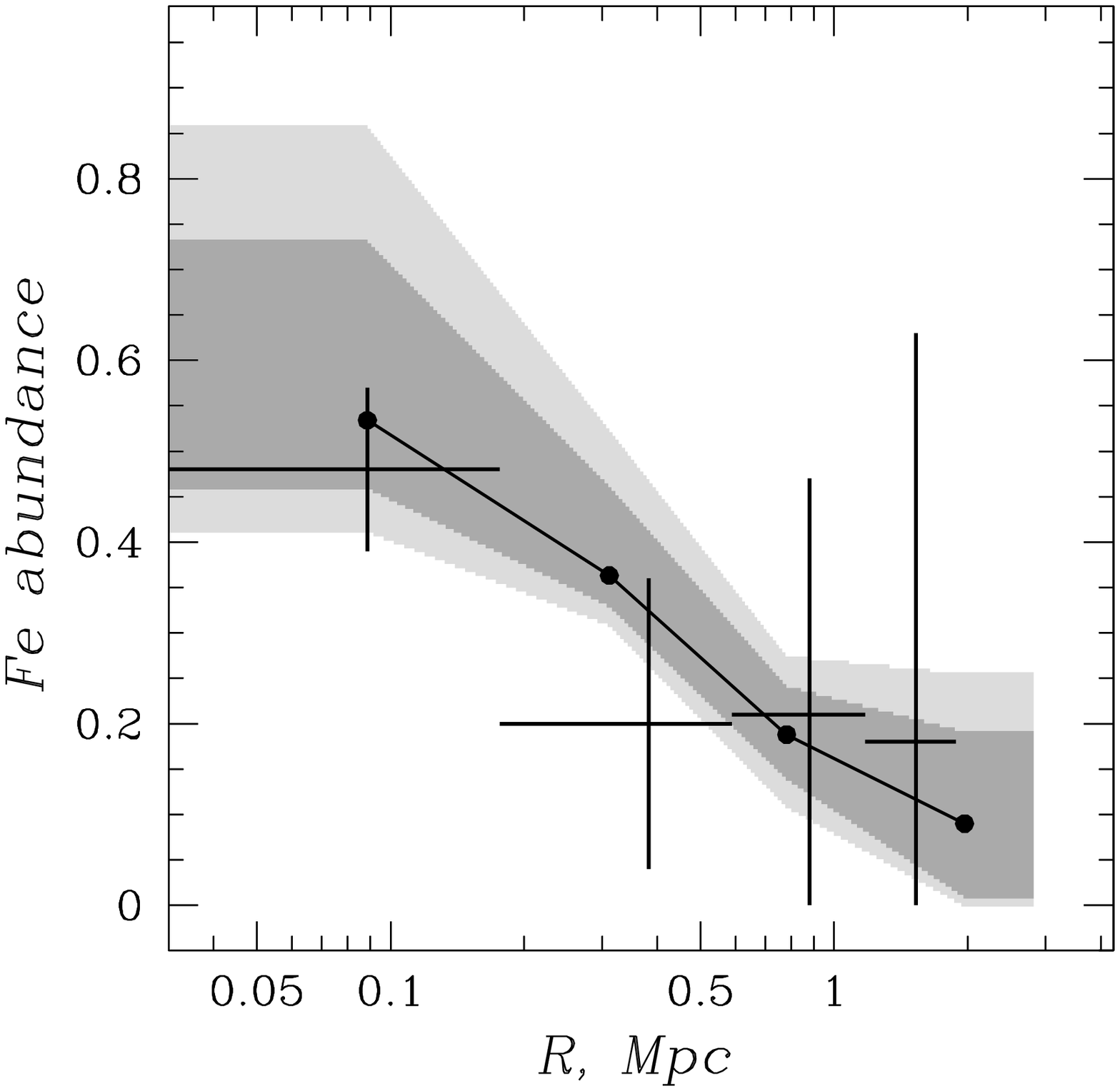} \hfill \includegraphics[width=1.8in]{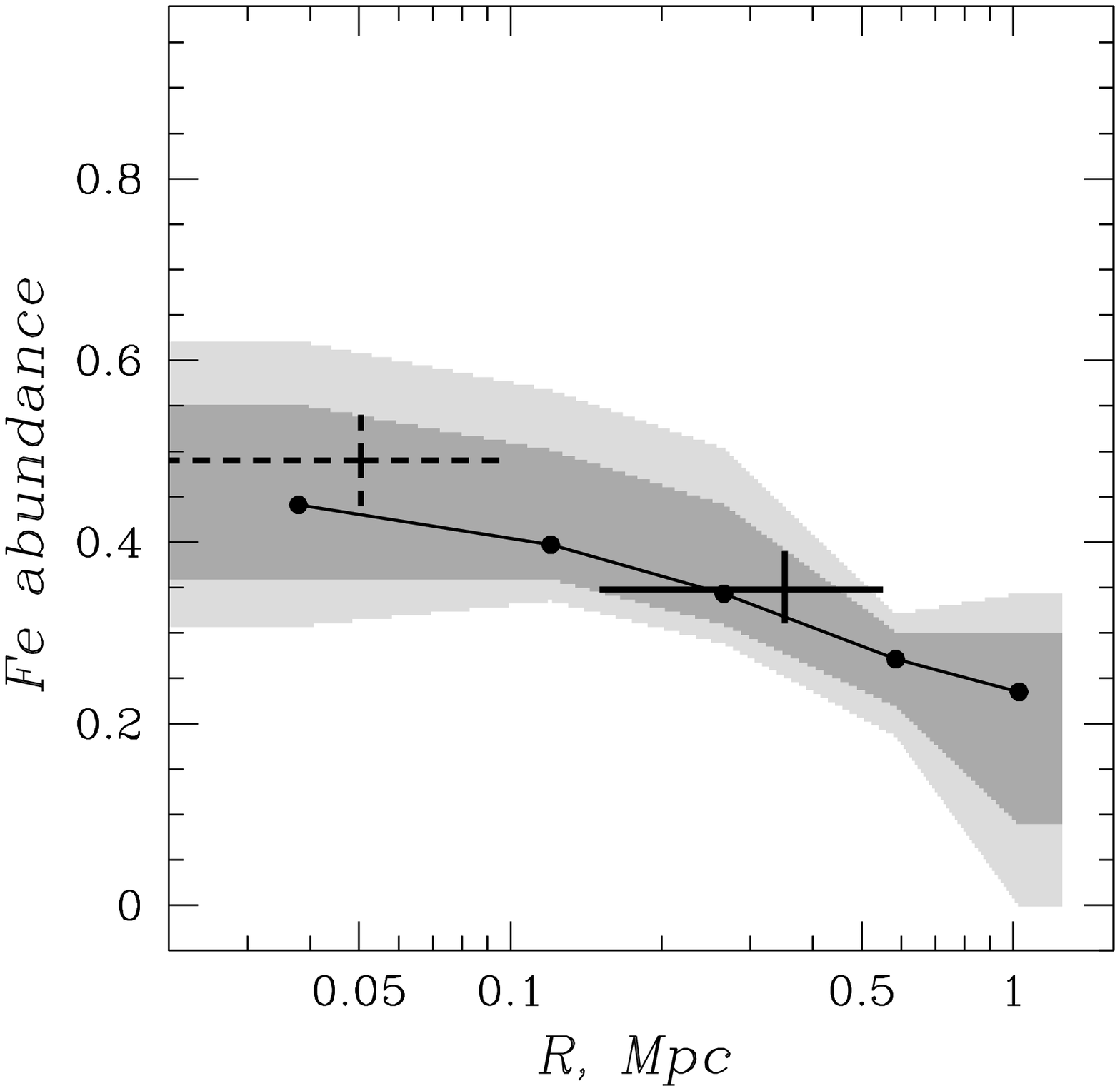} \hfill \includegraphics[width=1.8in]{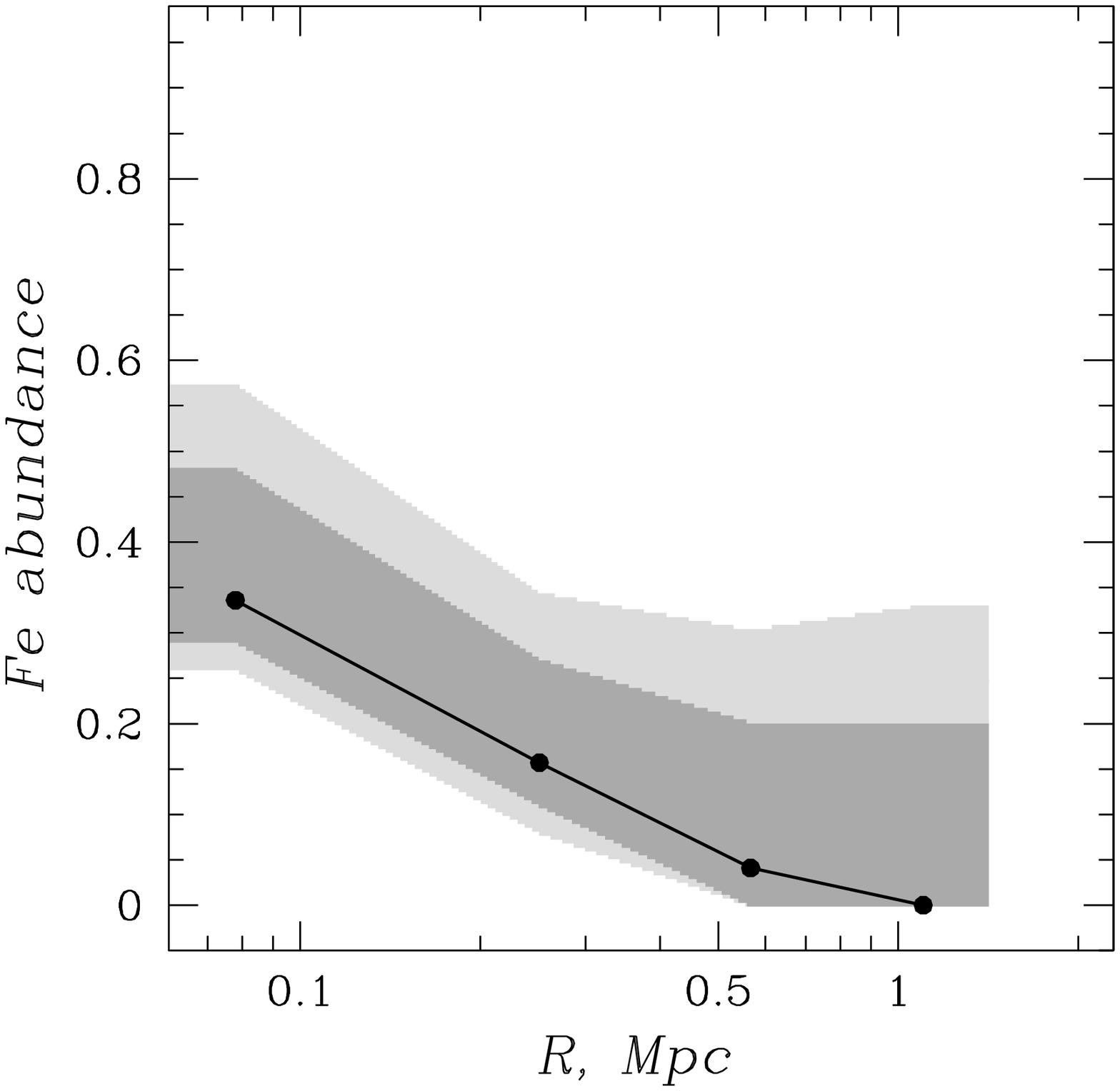}

\includegraphics[width=1.8in]{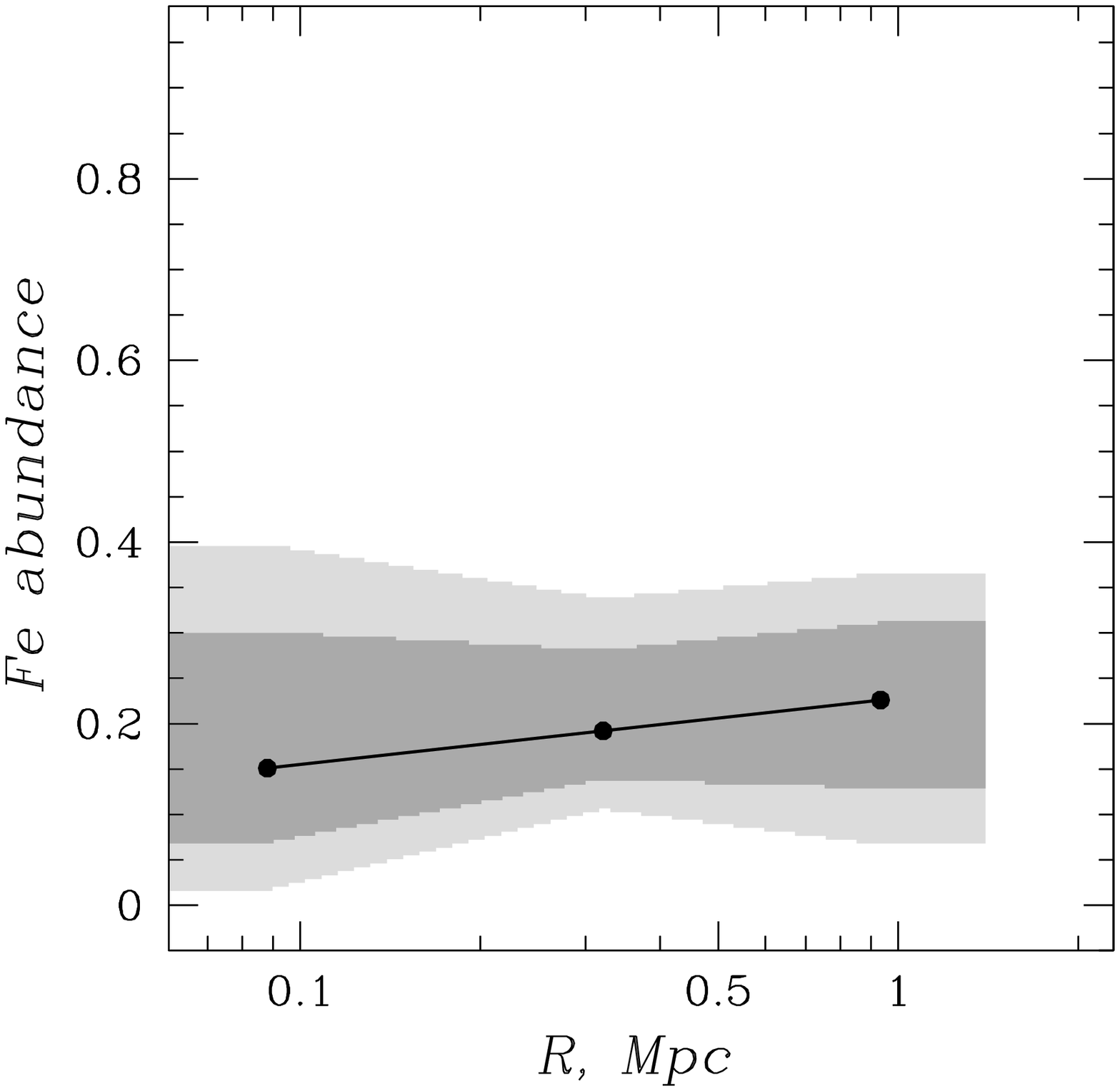} \hfill \includegraphics[width=1.8in]{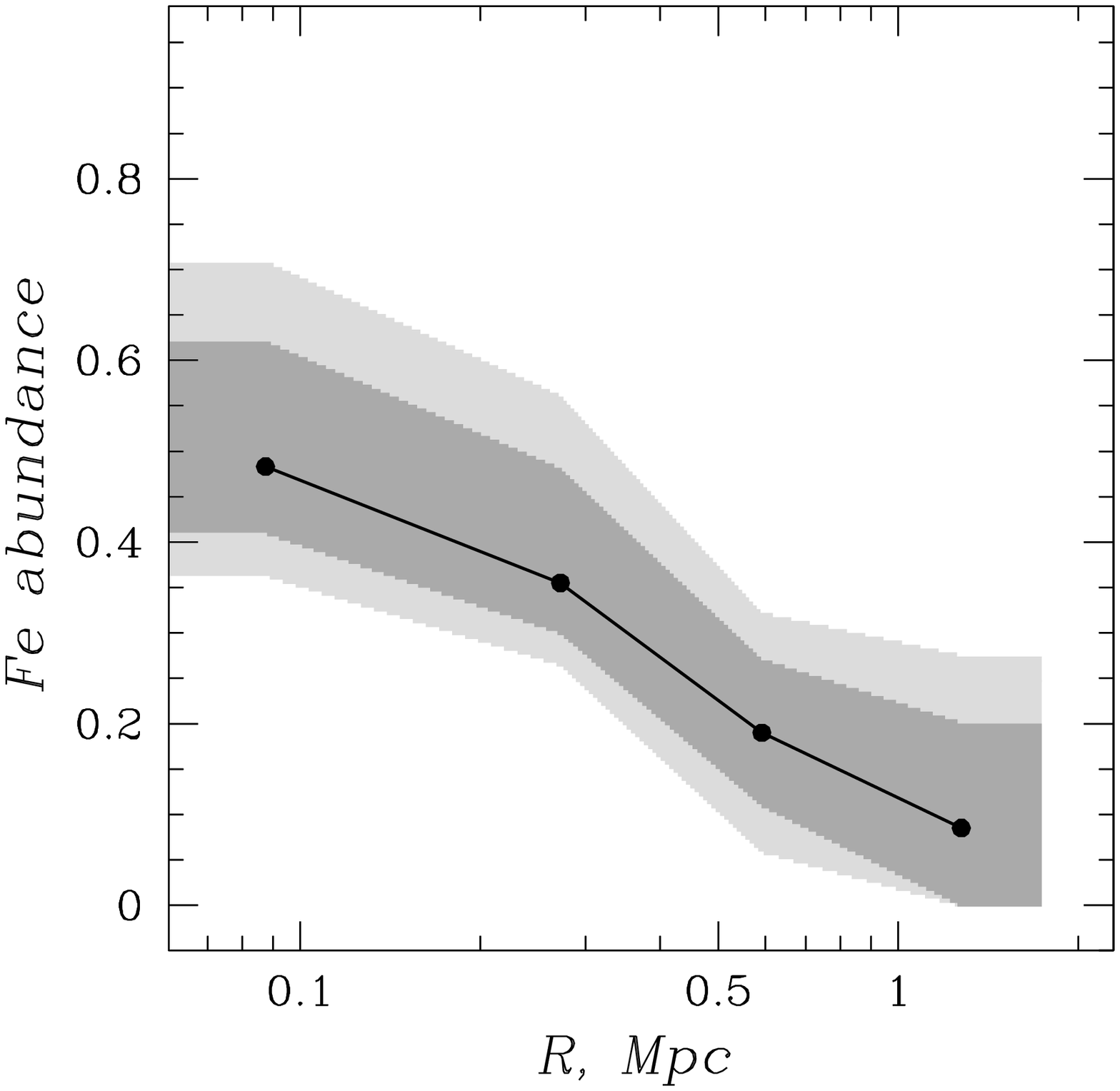} \hfill \includegraphics[width=1.8in]{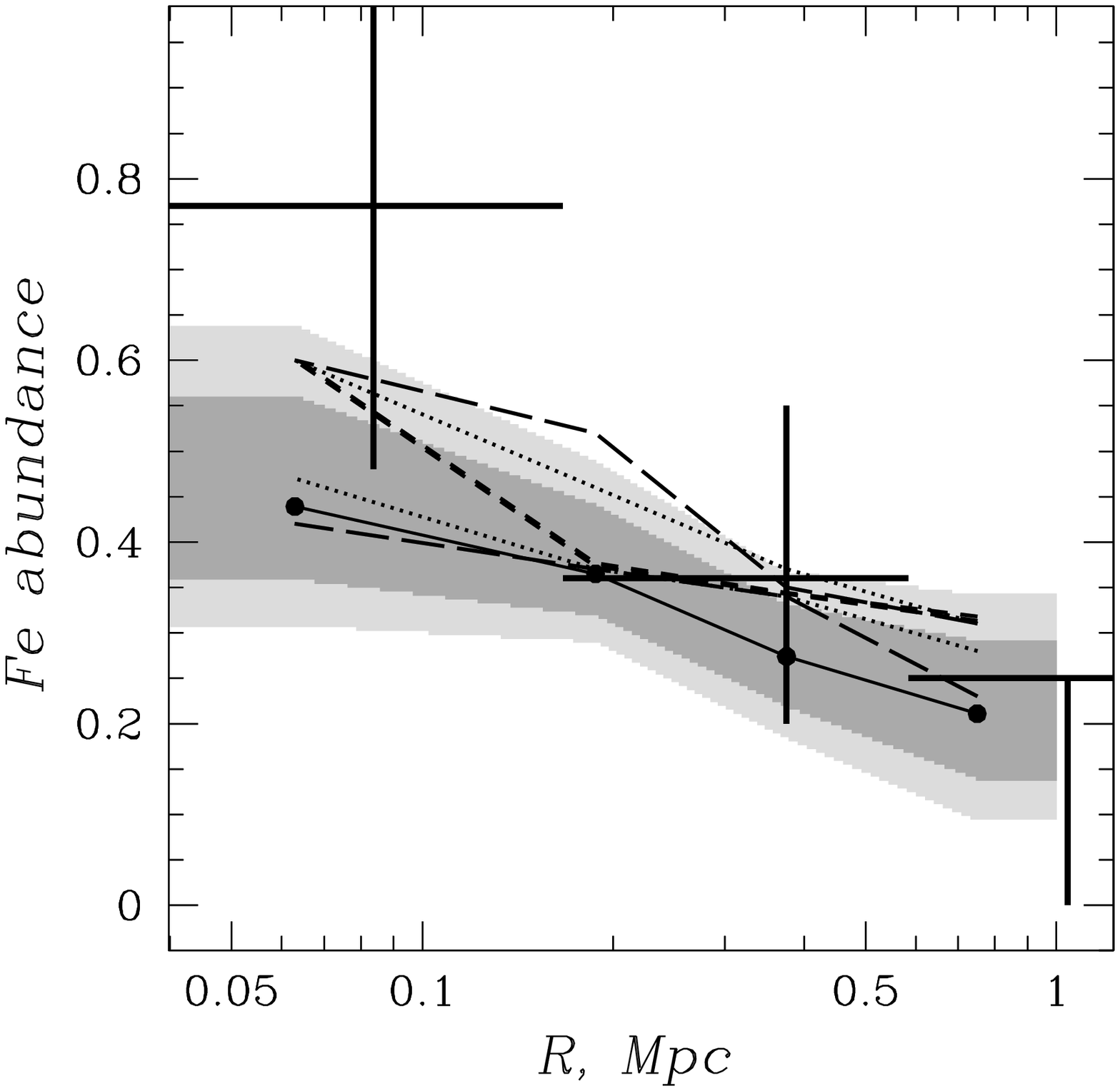} \hfill \includegraphics[width=1.8in]{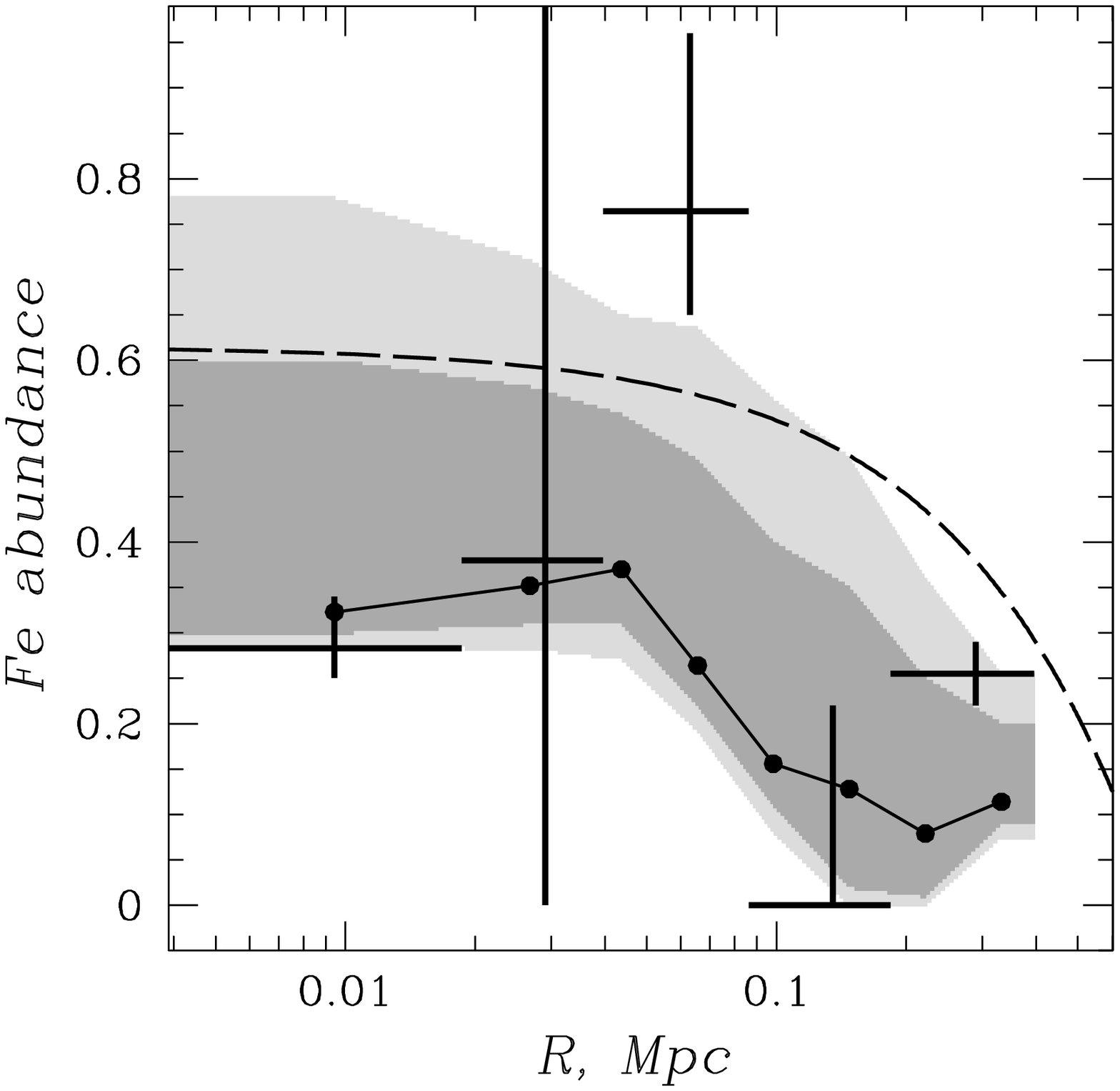}

\figcaption{Derived Fe abundances (in units of 4.68 $10^{-5}$ for iron
number abundances relative to H). The solid lines correspond to the best-fit
Fe abundances derived from the ASCA data. The filled circles indicate the
spatial binning used in the analysis. Dark and light shaded zones around the
best fit curves denote the 68 and 90 per cent confidence areas.  Crosses on
the A2029 panel represent previous ASCA SIS+GIS results from Sarazin \etal
(1998). Solid crosses on the A496, A1060 and A2199 panels show the results
from Mushotzky \etal (1996). A cross on the MKW4 panel shows the results
from Fukazawa \etal (1998) with radii of measurement from Fukazawa (private
communication). Crosses on the A780 plot denote the results of modeling of
GIS data from Ikebe \etal (1997) and similarly on the A4059 plot from
Kikuchi \etal (1999). Dashed crosses on the A496 and A2199 panels show
abundance determinations from Dupke \& White (1999). On the A4059 panel we
illustrate the effect of the inclusion of a different cooling flow model
(short-dashed line), ASCA PSF systematics (dotted lines) and usage of the
ROSAT image as an input (long-dashed lines for two extreme solutions). On
the HCG62 panel we also show the results obtained when the regularization in
spectral fitting is turned off (solid crosses). This is discussed in the
Appendix. The dashed line indicates the results of an independent analysis
of ROSAT PSPC data, as discussed in Finoguenov \& Ponman (1999).
\label{fe-fig}}
\vspace*{-15.8cm}

{\it \hfill MKW4\hspace*{0.5cm} \hfill A496\hspace*{0.5cm} \hfill A780\hspace*{0.3cm} \hfill A1060\hspace*{0.2cm}  }
\vspace*{4cm}

{\it \hfill A1651\hspace*{0.3cm} \hfill A2029\hspace*{0.3cm} \hfill A2199\hspace*{0.3cm} \hfill A2597\hspace*{0.3cm}}

\vspace*{4cm}

{\it \hfill A2670\hspace*{0.3cm} \hfill A3112\hspace*{0.3cm} \hfill A4059\hspace*{0.3cm} \hfill HCG62\hspace*{0.3cm}}

\vspace*{4cm}

\end{figure*}

Variations in the surface brightness representation were investigated by
using the results of Mohr \etal (1999), which were derived using a surface
brightness fitting technique different from our own. Uncertainties in the
PSF and ROSAT profiles were found to have significant effects on
normalization and somewhat on the temperature, but little impact on the
element abundance determinations. We have included the effects of
systematics in the parameter's uncertainties, presented in this paper. We
have also checked the effect of the elliptical geometry of some clusters by
taking the actual ROSAT image (as an alternative to its radial profile) as
the basis for modeling the flux distribution in A4059, which is the most
elliptical of our clusters. The results of this, of variations in the
treatment of the central cooling flow, and of uncertainties in the ASCA PSF,
are all shown for A4059 in Fig.\ref{fe-fig}.  It can be seen that the
abundance distribution is quite robust to such differences in the analysis,
with systematic variations being confined within the 90\% statistical error
region.

As discussed above, any errors in our analysis resulting from differences
between X-ray profiles in the ROSAT and ASCA energy bands should reduce as
smaller spatial bins are used for the analysis. Within the limits set by the
increasing effects of ASCA PSF systematics, when going to smaller spatial
bins (Markevitch private communication) we have checked whether our
assumption of the spectral-spatial separability of the emission (Finoguenov
\etal 1999) affects our final results, by increasing the number of spatial
bins. Since no significant changes in element distributions were obtained,
we conclude that this assumption is reasonable.

Table \ref{tab:opt} contains the optical and X-ray properties of our sample.
Column (1) identifies the system, (2) adopted luminosity distance, (3)
corresponding scale length, (4--5) total blue light and radius for this
measurement, (6) estimated blue luminosity within 0.5 Mpc based on the
Bahcall (1981) central galaxy density within 0.5 Mpc, (7) luminosity of
dominant central galaxy (referred to throughout the present
paper as a `cD'), and (8) core radius of the galaxy
distribution. Cols. (9--10) give the results of the cluster surface
brightness fitting with a $\beta$ model, taken from Vikhlinin, Forman, Jones
(1999), where regions contaminated by the cooling flow were excluded from
the analysis, supplemented by data from Finoguenov and Ponman (1999),
Finoguenov and Jones (2000), (11) gives the best fit cluster
emission-weighted temperature, (12) the corresponding 90\% error, (13) the
virial radius of the system ($r_{180}=1.23 T_{keV}^{0.5}h_{50}^{-1}$ Mpc;
Evrard \etal 1996).

The observation log for the cluster sample is given in Table
\ref{tab:obslog}. The SIS exposure times are based on applied standard
screening.  For the new data, analyzed here, distances were taken from
compilation of Vikhlinin \etal (1999). Optical luminosities are obtained
from the compilation in Arnaud \etal (1992), converting from V to B using
$B=V+0.94$. Optical measurements for MKW4 and AWM7 are taken from Beers
\etal (1984). For A780 and A1651 no data on luminosity of the cD galaxy was
located. Core radii are obtained from Girardi \etal (1995).  For clusters
without a measured core radius, we use the mean value of 0.24~Mpc from
Girardi \etal (1995). For MKW4 we derived the core radius from data in
Dell'Antonio, Geller \& Fabricant (1995). For A496, a measurement of the
core radius was taken from Whitmore \etal (1993). Cluster data in
cols. (11-12) is taken from GIS temperature estimates by Markevitch \etal
(1998). The emission weighted temperatures in cols. (11-12) for the
low-temperature systems are derived from our SIS spectral analysis and are
corrected for the presence of cooling flows.

The new temperature data on MKW4 and A1060, derived in this paper are
presented in Tab.\ref{table-te}. We see evidence for a declining temperature
in both MKW4 and A1060. The later system is often cited as an example of an
isothermal cluster (Tamura \etal 1996), however their recent reanalysis of
the data agrees with our findings (Tamura \etal 2000).

%\begin{table}[H]
{
\footnotesize
\centering
\tabcaption{\footnotesize
\centerline{New ASCA SIS temperature measurements$^{\dag}$$^{\ddag}$}
\label{table-te}}

\begin{tabular}{lclc}
\hline
\hline
Annulus (\amin) & $kT_e$ & Annulus (\amin) & $kT_e$ \\
\hline
 & & & \\
\multicolumn{2}{c}{MKW4 \hspace*{0.8cm}}  &  \multicolumn{2}{c}{A1060 \hspace*{0.8cm}}    \\                                          
0.0---1.5  & 1.770 (1.50:1.80)  &  0.0---0.8  &  4.36 (3.0:6.0) \\ 
1.5---2.9  & 1.781 (1.72:1.85)  &  0.8---1.5  &  4.19 (3.3:5.3) \\ 
2.9---5.5  & 1.779 (1.72:1.85)  &  1.5---2.4  &  3.83 (3.4:4.3) \\ 
5.5---10.5 & 1.650 (1.59:1.70)  &  2.4---3.9  &  3.27 (2.9:3.6) \\ 
10.5---20. & 1.478 (1.39:1.58)  &  3.9---6.2  &  3.05 (2.8:3.3)  \\
                    & &           6.2---10.  &  3.34 (3.0:3.6) \\ 
                     & &          10.---16.  &  2.69 (2.5:2.8)  \\
\hline
\end{tabular}                                        

\begin{enumerate}
\item[{$^{\dag}$}]{\footnotesize ~ Errors are given at 68\% confidence level
    for one parameter of interest. MEKAL plasma code is used for spectral
    fitting.}
\item[{$^{\ddag}$}]{\footnotesize ~ For the rest of the clusters analyzed in
    this paper, the combined GIS+SIS+PSPC temperature measurements of
    Markevitch \etal (1998) supersede our data}
\end{enumerate}
}                                                    
%\end{table}                                         

For clusters without published optical luminosities, the luminosities are
derived from the correlation between the X-ray temperature and the Bahcall
number density, $N_{0.5}$ (Edge \& Stewart 1991), the latter is directly
related to optical luminosity within 0.5~Mpc (cf Edge \& Stewart 1991).
Since the correlation found in Edge \& Stewart (1991) has such a large
scatter, we recalibrated it using ASCA measured temperatures (Markevitch
\etal 1998 and this paper) and find a best fit relation given by:

\centerline{\psfig{file=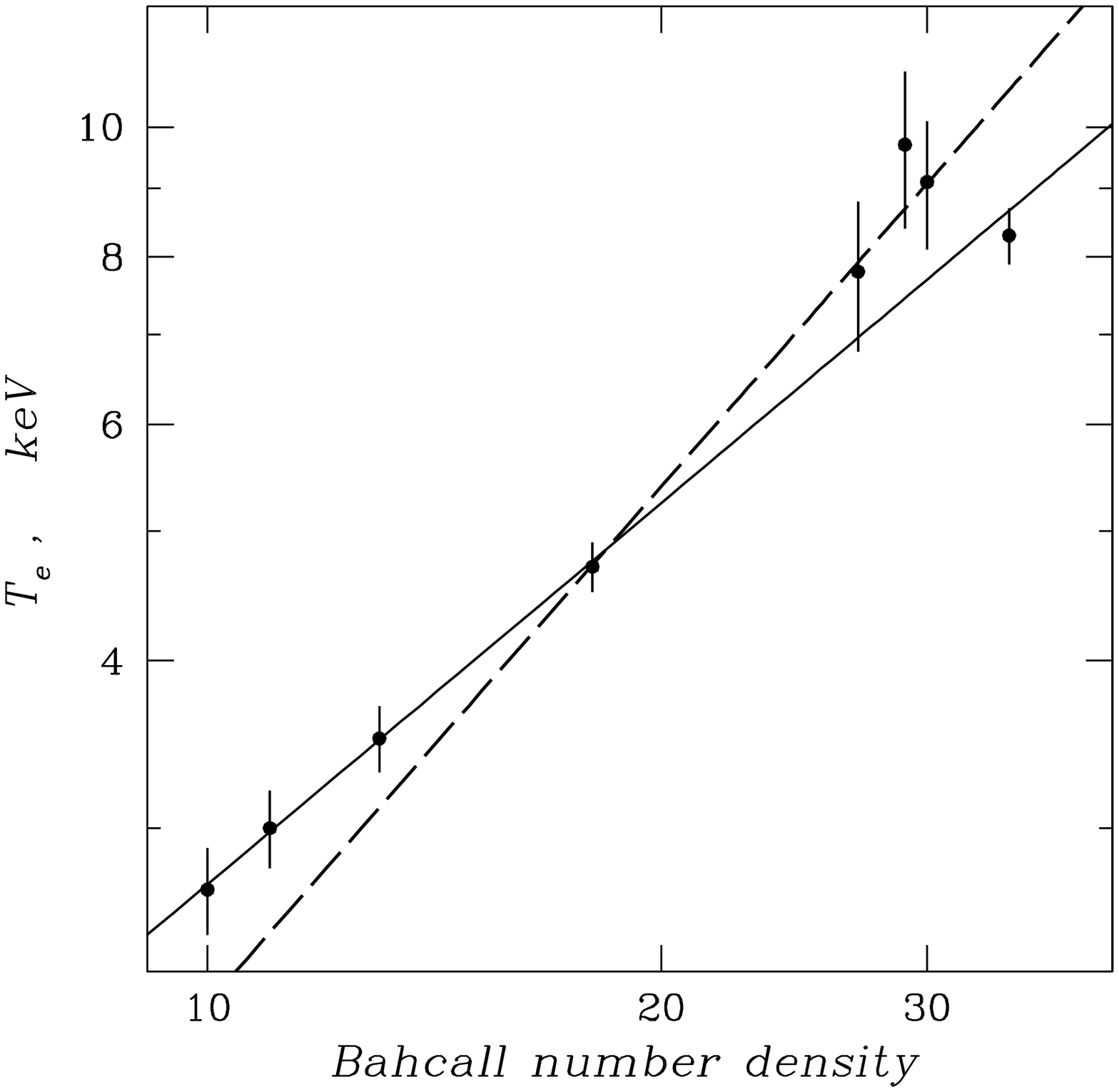,width=3.25in}}

\figcaption{ Cluster gas temperature vs galaxy number density (the Bahcall
  number) within 0.5~Mpc. The solid line indicates our best-fit relation,
and the dashed line indicates the best-fit relation obtained by Edge and
Stewart (1991). Error bars are shown at the 90\% confidence level. In order
of increasing Bahcall number, the points correspond to Virgo, A1060, AWM7,
A2199, A1795, A2141, A2029 and A401. 
\label{fig:nb}}
\medskip

\begin{equation}\label{nb-kt}
N_{0.5} = 3.46\pm0.6 \; \times \; T^{1.058\pm 0.06}
\end{equation}

\noindent

where $T$ is in units of keV.  The errors are given at the 90\% confidence
level. As can be seen in the figure, the scatter about this line is
remarkably small, corresponding to 20\% in $N_{0.5}$.  While our best-fit
relation is within the uncertainties of Edge and Stewart (1991), their
best-fit prediction deviates somewhat from the data on the cool clusters
Virgo, AWM7, and A1060 (see Fig.\ref{fig:nb}). However, since in practice we
only use Eq.\ref{nb-kt} to estimate the cluster blue luminosities for
clusters in the range 4--6 keV, no significant difference would occur 
if their relation were used.

\bigskip

\section{Radial Distribution of Heavy Elements}\label{sec:radial}

The abundance data, derived in this work, are tabulated in
Table \ref{table-ab}.

\begin{figure*}

\includegraphics[width=1.8in]{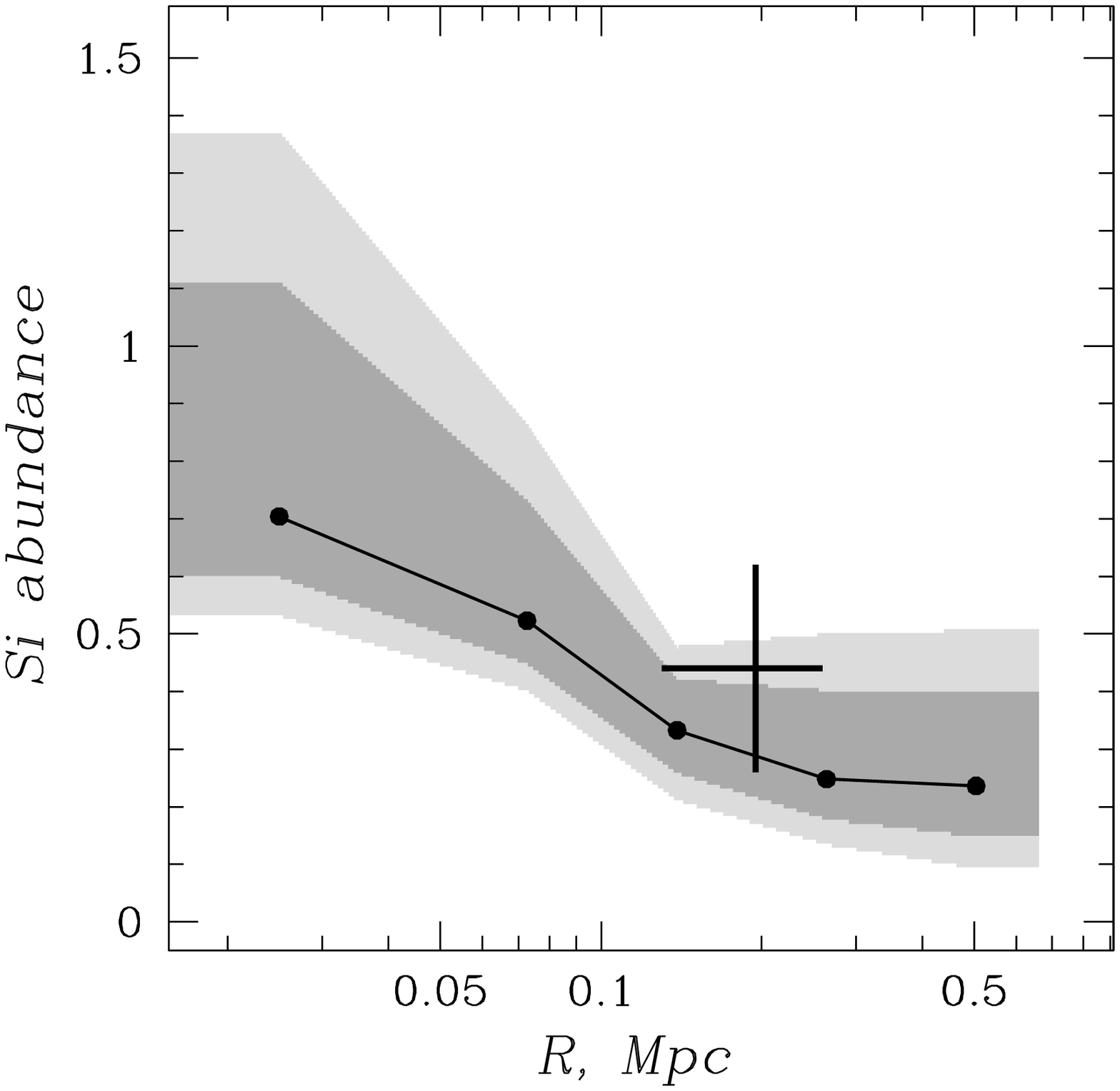} \hfill \includegraphics[width=1.8in]{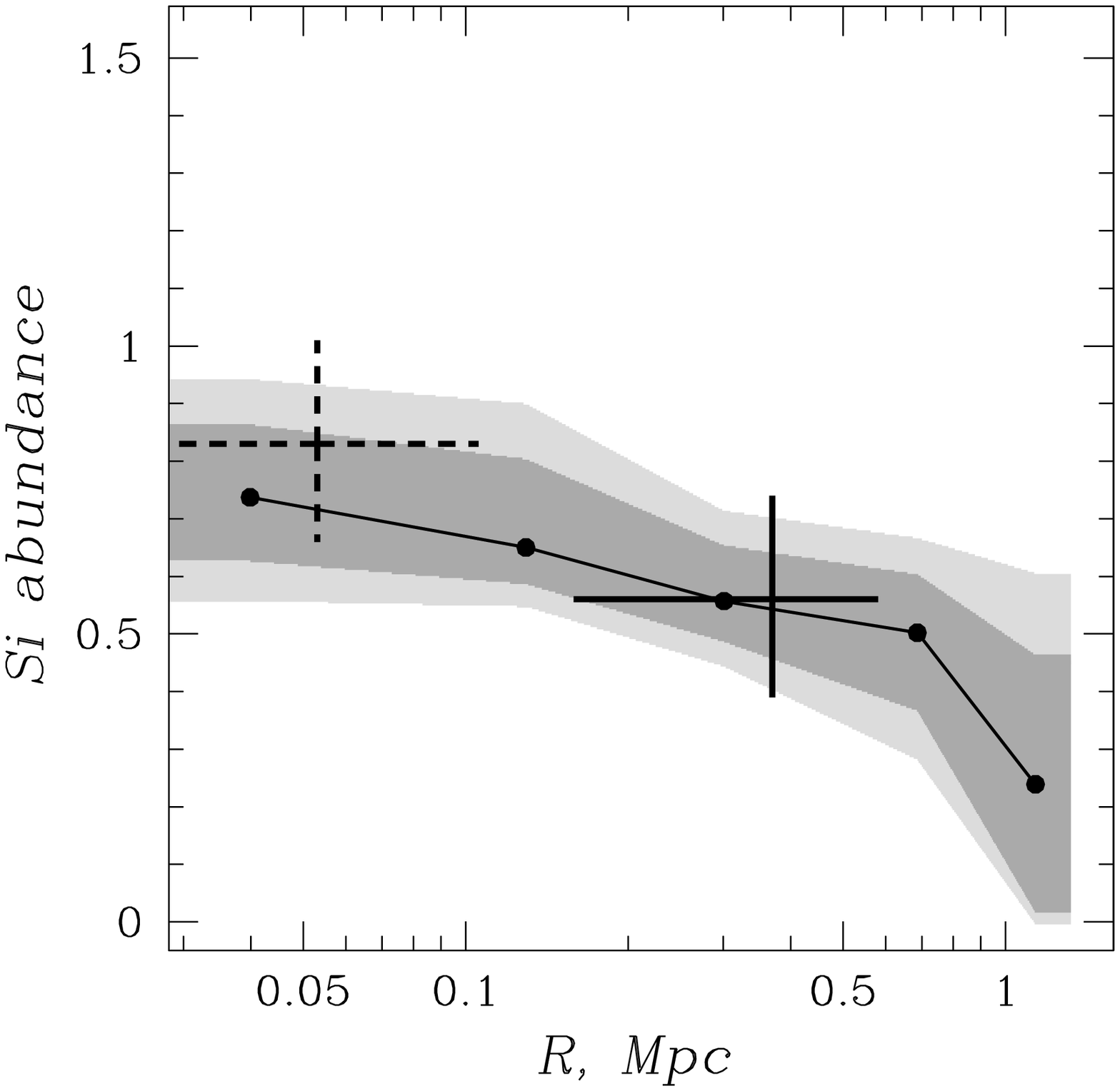} \hfill \includegraphics[width=1.8in]{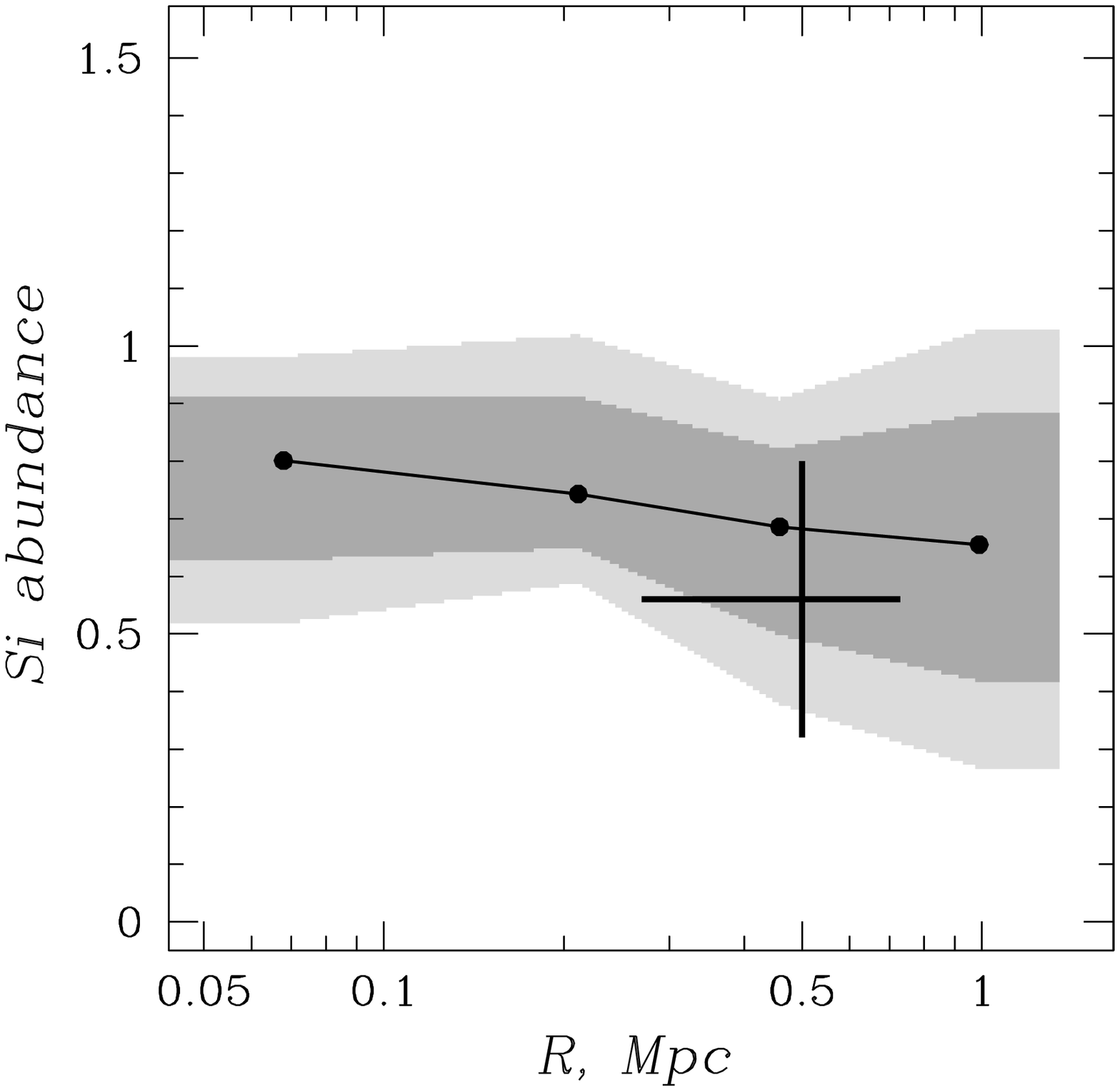} \hfill \includegraphics[width=1.8in]{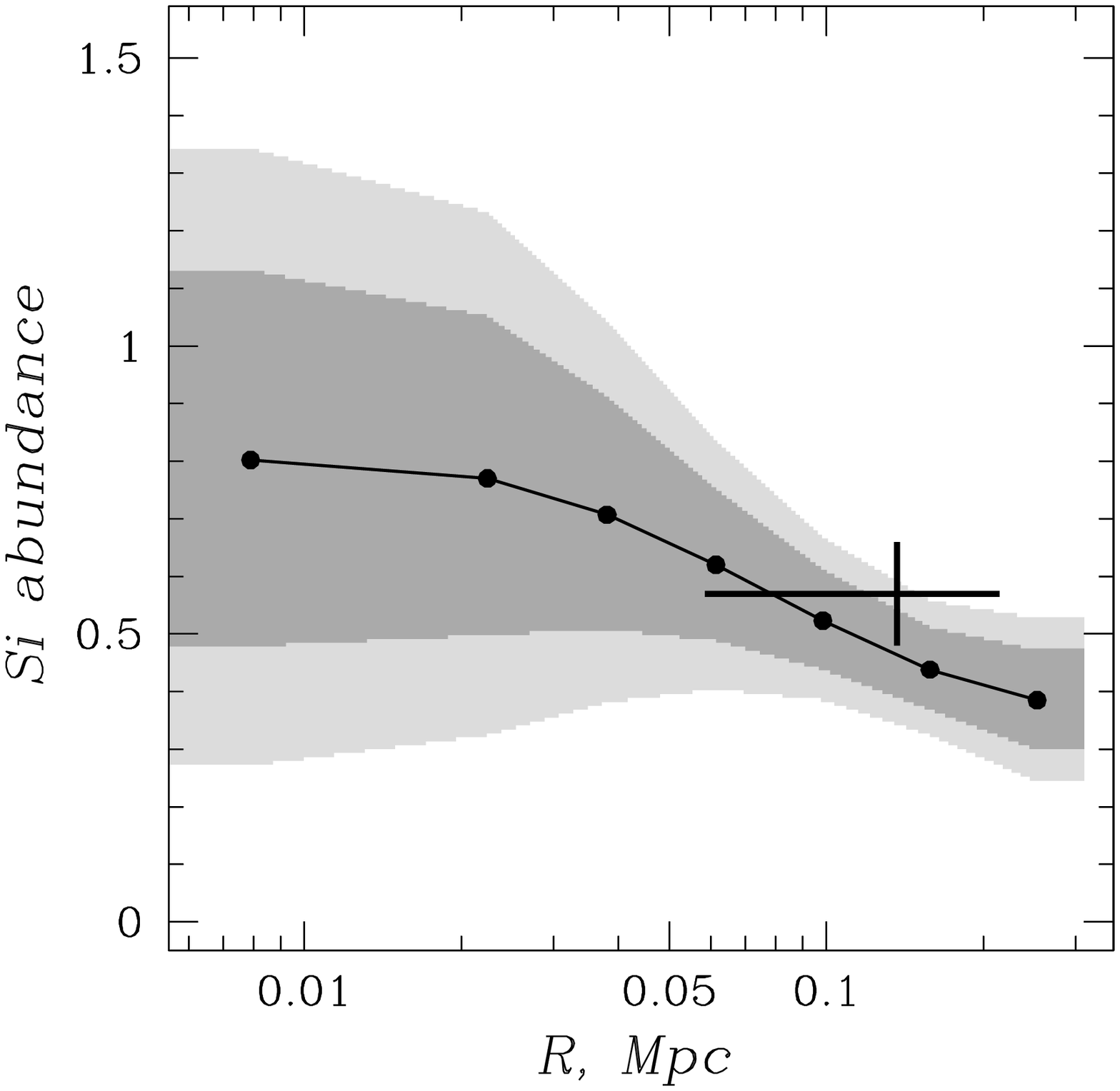}

\includegraphics[width=1.8in]{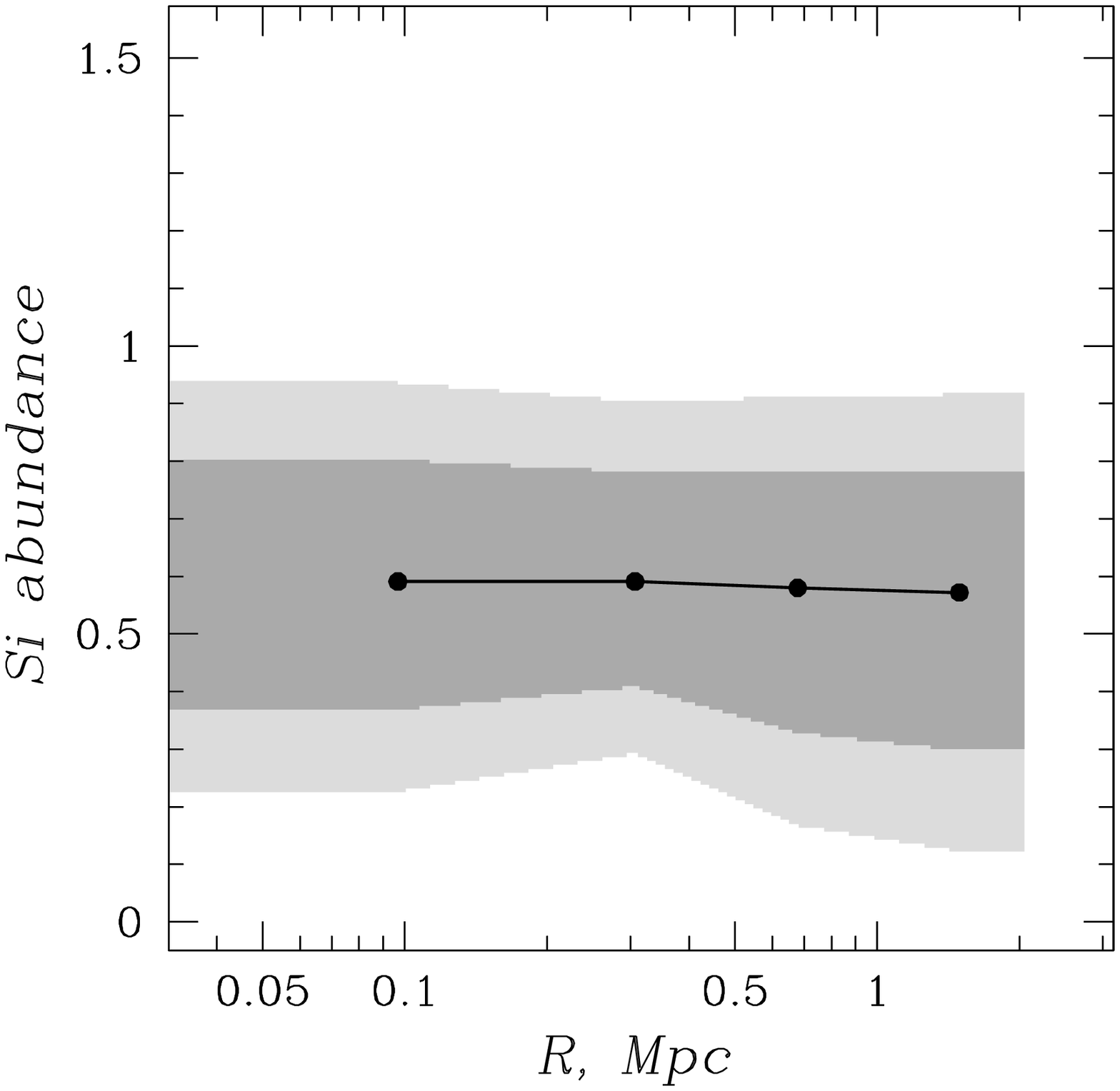} \hfill \includegraphics[width=1.8in]{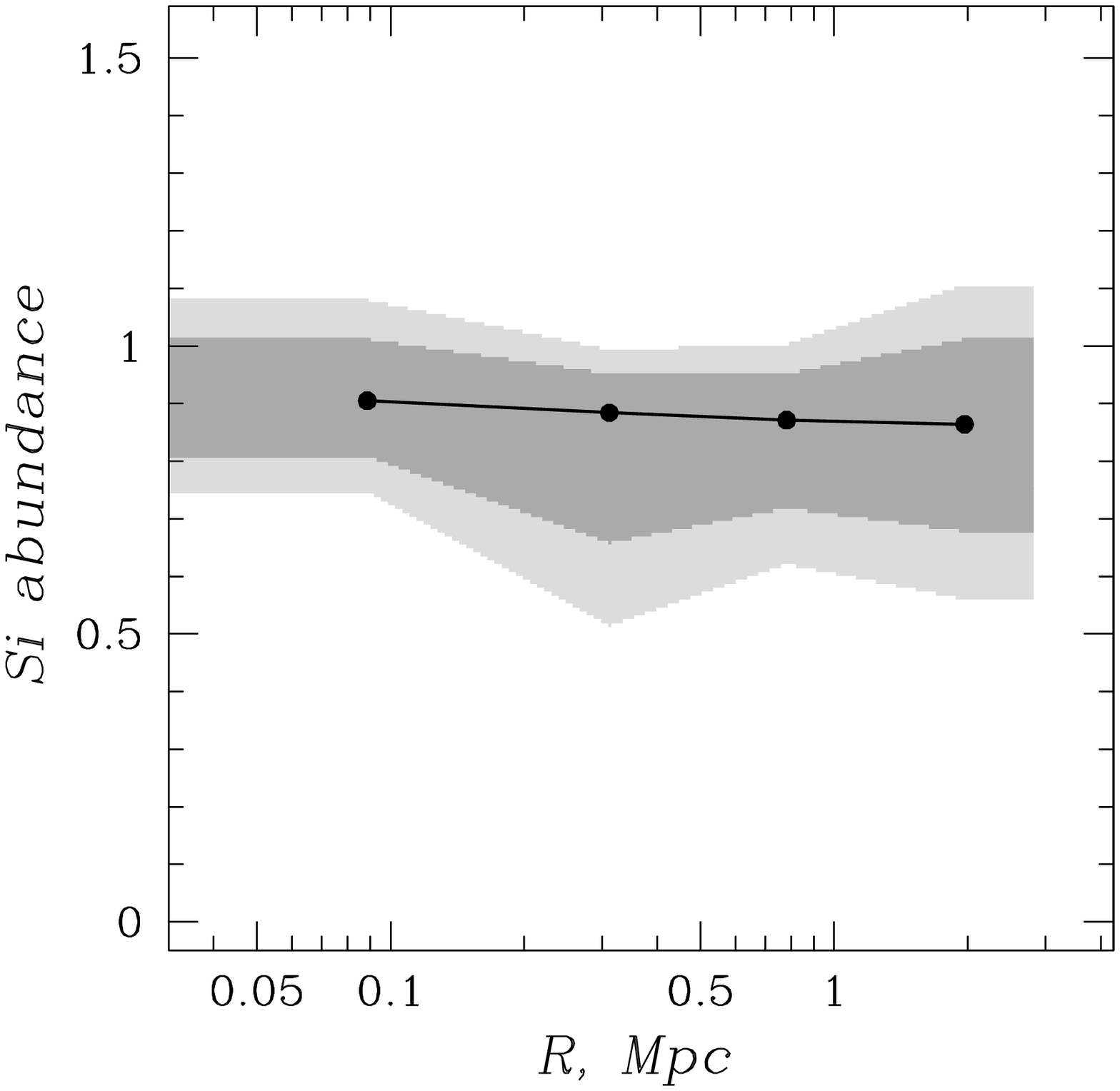} \hfill \includegraphics[width=1.8in]{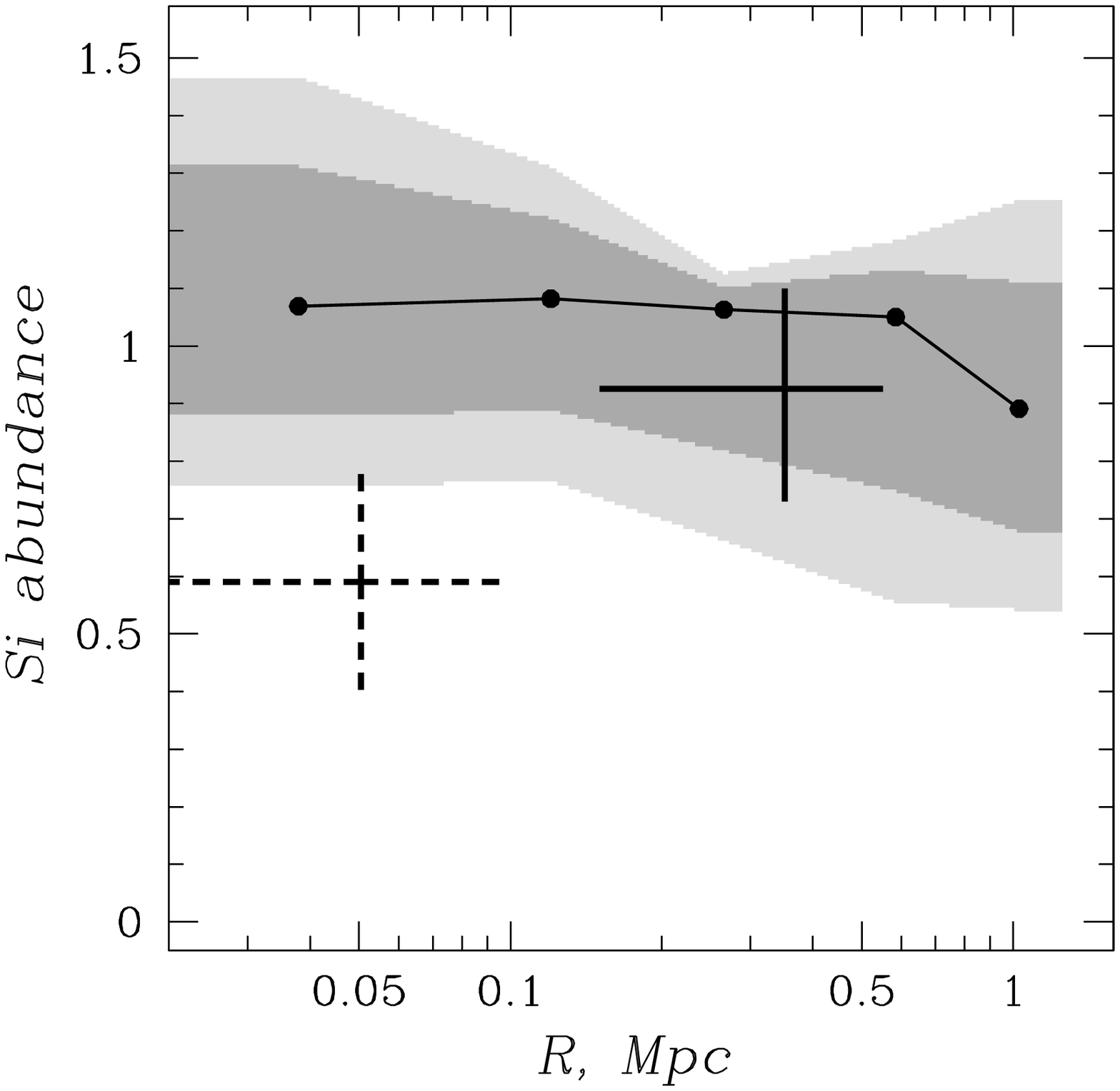} \hfill \includegraphics[width=1.8in]{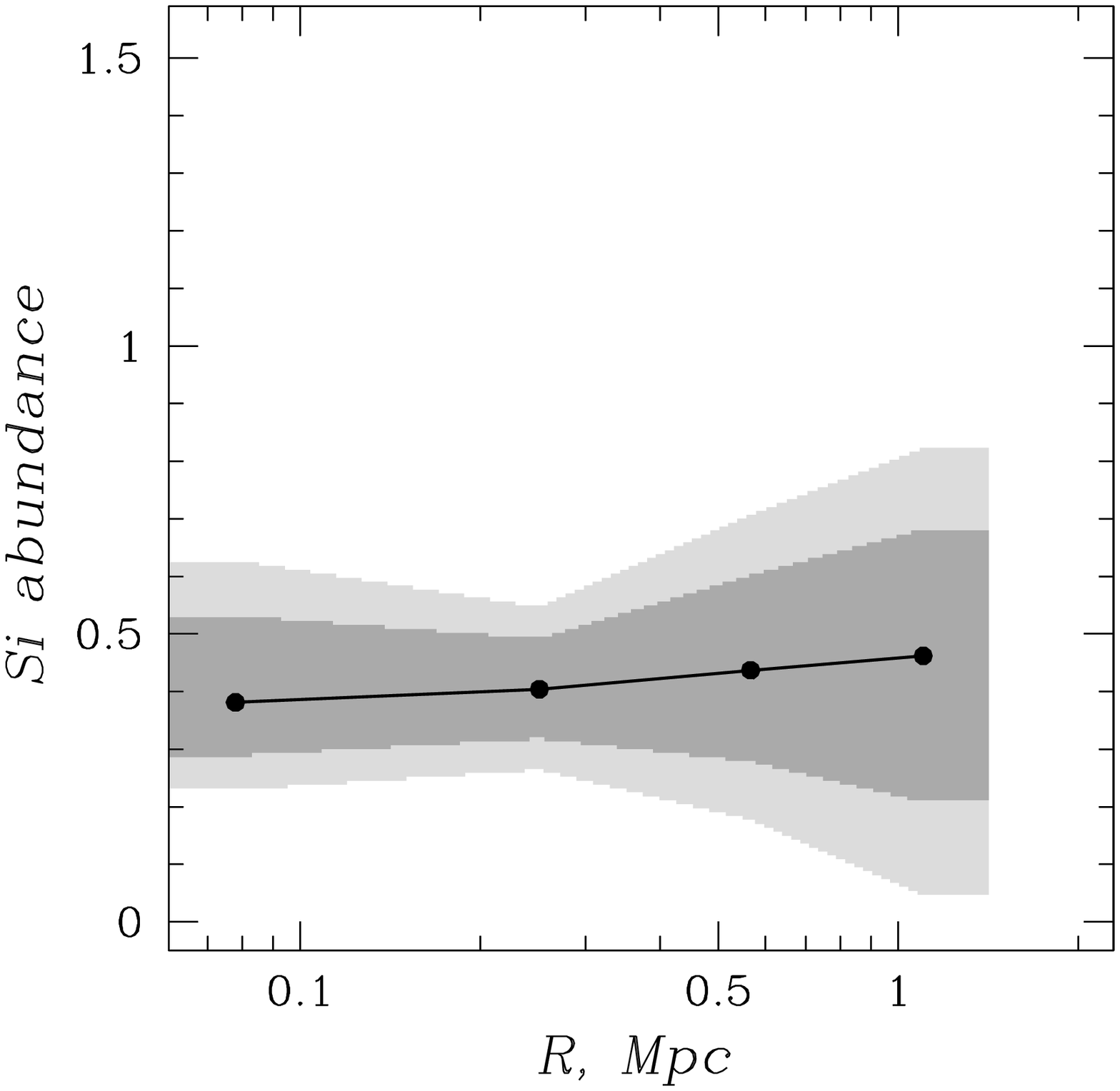}

\includegraphics[width=1.8in]{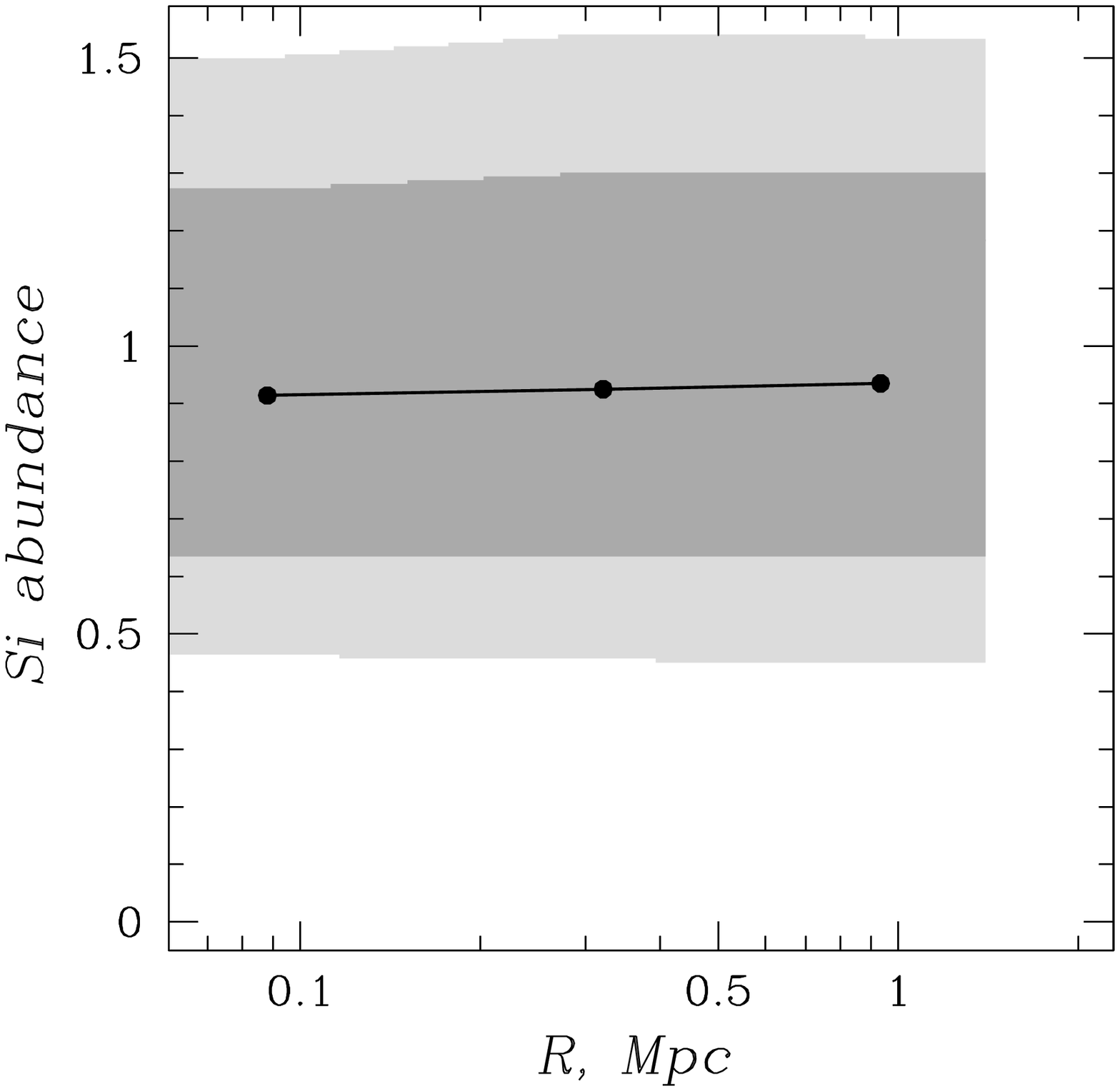} \hfill \includegraphics[width=1.8in]{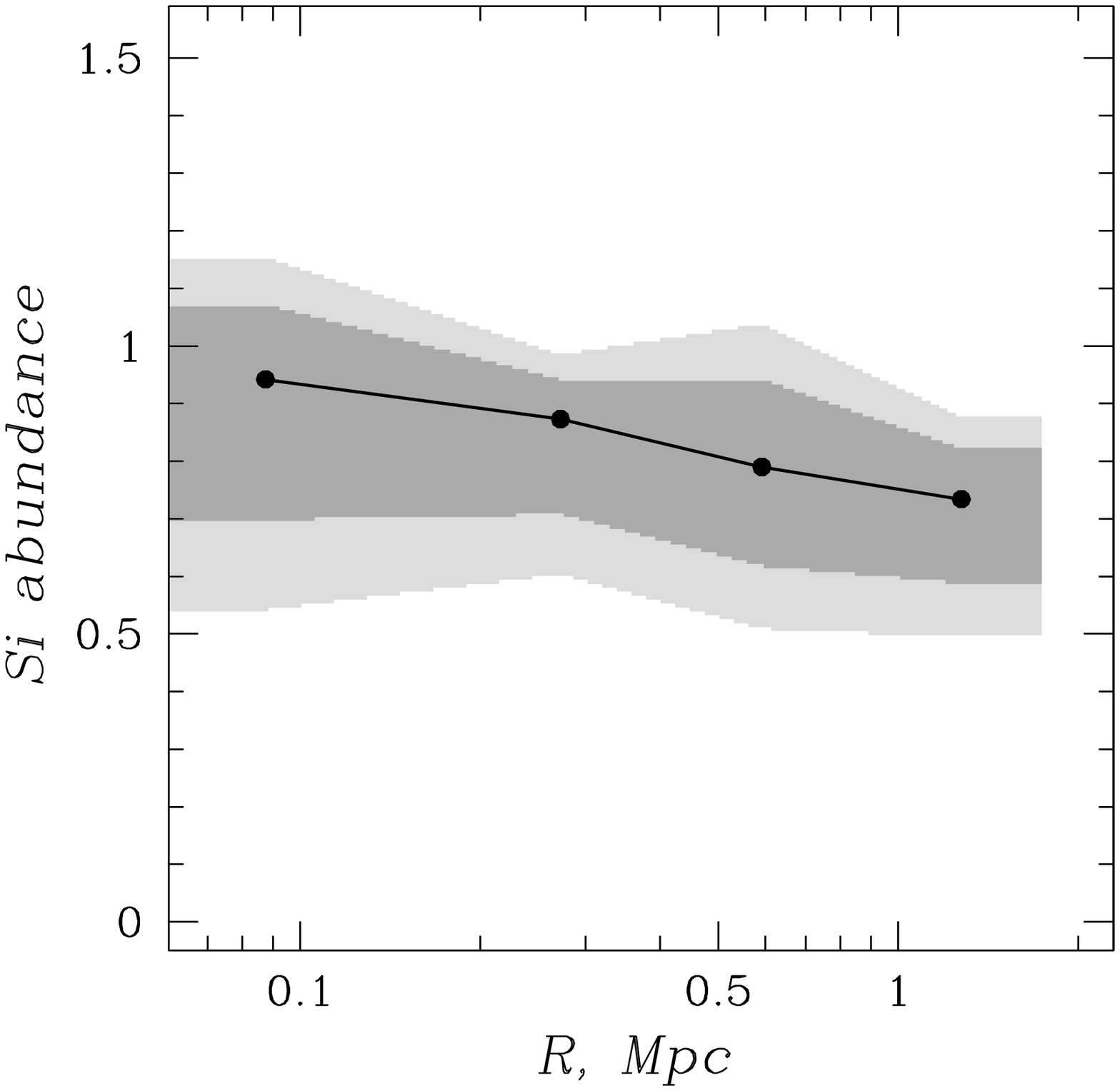} \hfill \includegraphics[width=1.8in]{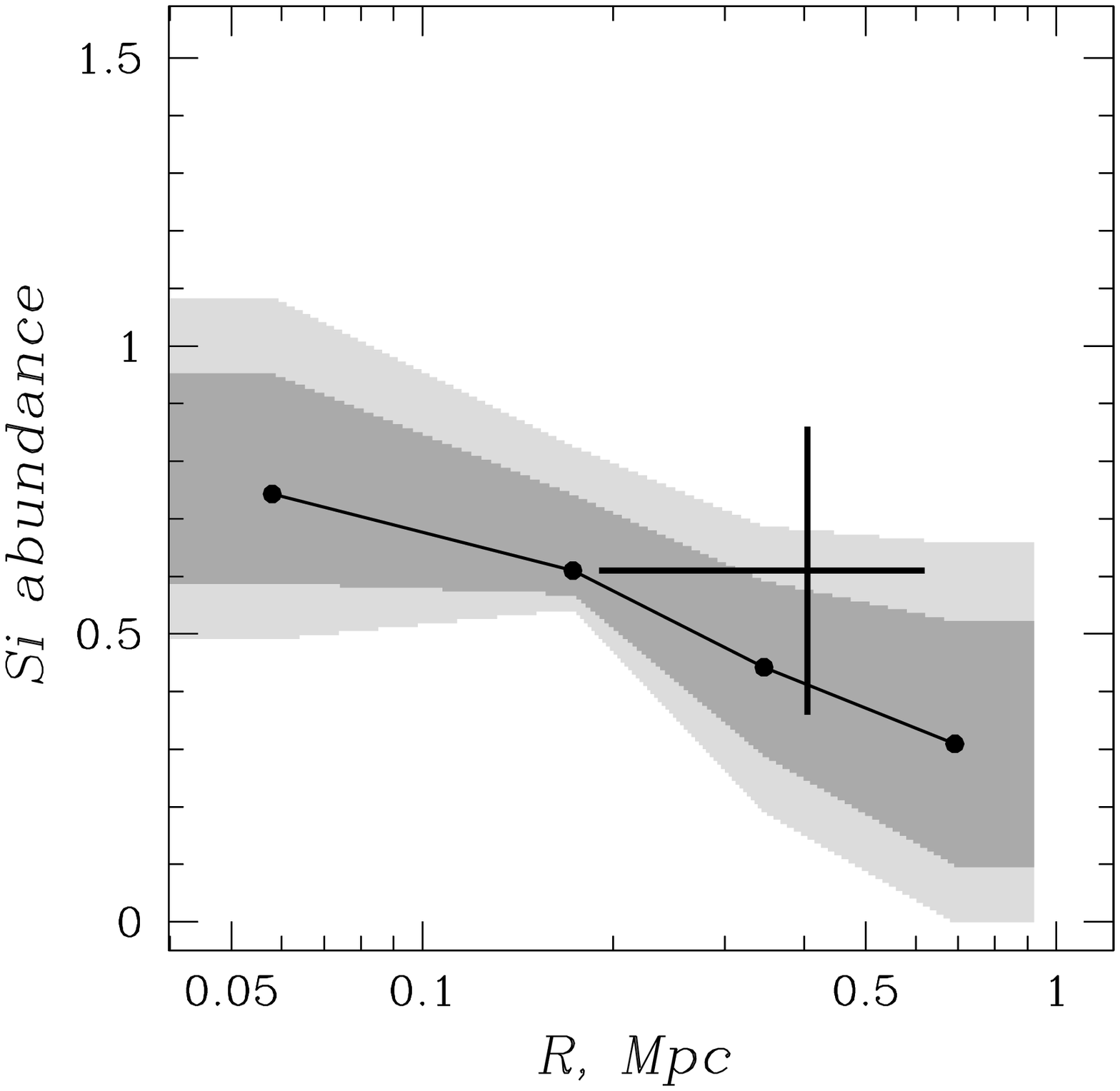}

\figcaption{Derived Si abundances.  The solid lines correspond to the
best-fit Si abundances derived from the ASCA data. The filled circles
indicate the spatial binning used in the analysis.  Dark and light shaded
zones around the best fit curve denote the 68 and 90 per cent confidence
areas, respectively. Solid crosses on the A496, A1060 and A2199 panels
represent ASCA values from Mushotzky \etal (1996). Crosses on MKW4, A780 and
A4059 panels represent ASCA values from Fukazawa \etal (1998), with radii of
measurements from Fukazawa (private communication). Dashed crosses on
the A496 and A2199 panels show abundance determinations from Dupke \& White
(1999).
\label{si-fig}}
\vspace*{-14.4cm}

{\it \hfill MKW4\hspace*{0.5cm} \hfill A496\hspace*{0.5cm} \hfill A780\hspace*{0.3cm} \hfill A1060\hspace*{0.2cm}  }
\vspace*{4cm}

{\it \hfill A1651\hspace*{0.3cm} \hfill A2029\hspace*{0.3cm} \hfill A2199\hspace*{0.3cm} \hfill A2597\hspace*{0.3cm}}

\vspace*{4cm}

{\it \hfill A2670\hspace*{2.7cm} \hfill A3112\hspace*{2.6cm} \hfill A4059\hspace*{0.3cm}}

\vspace*{4cm}

\end{figure*}

\begin{figure*}

\includegraphics[width=1.8in]{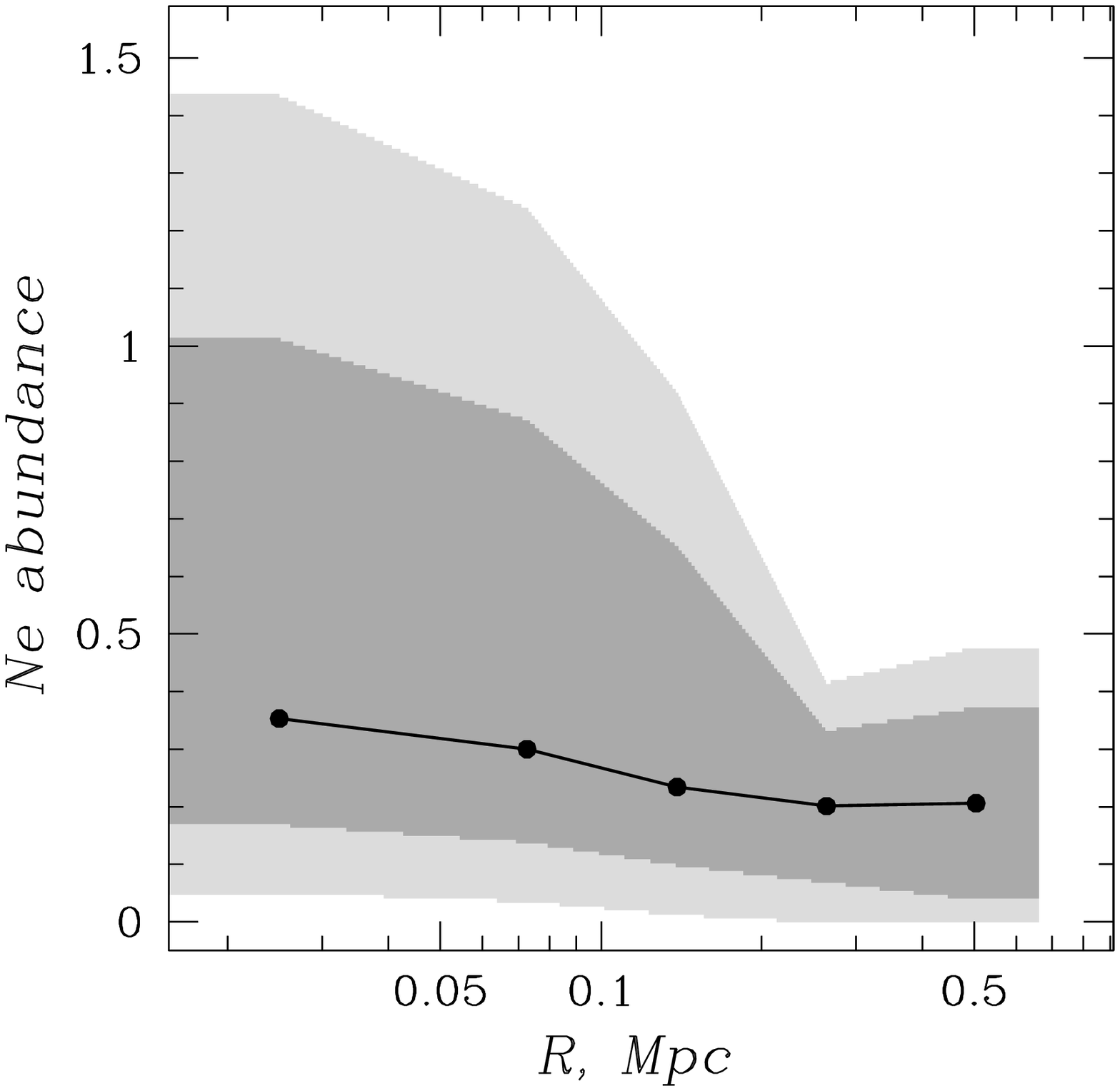} \hfill \includegraphics[width=1.8in]{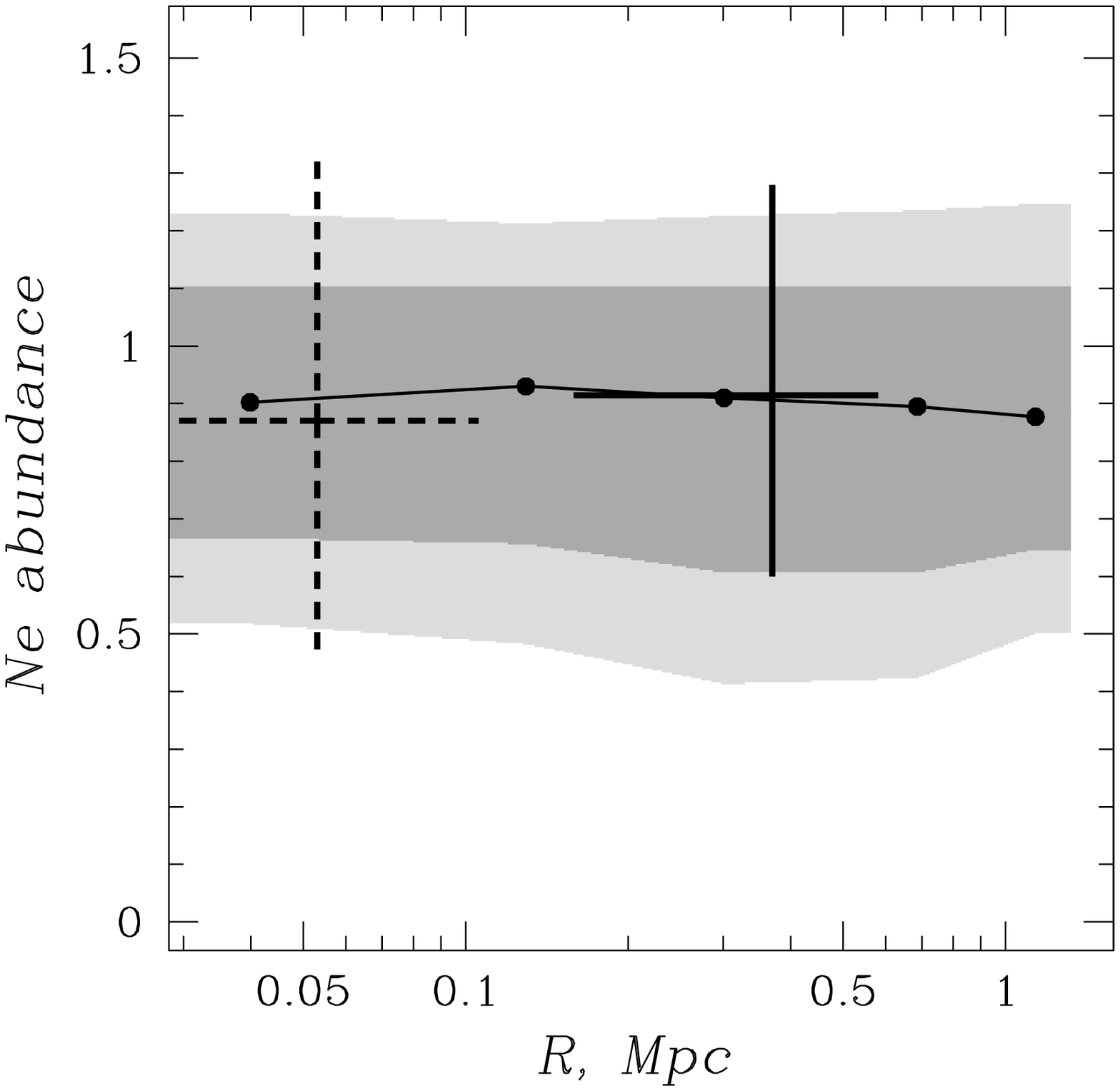} \hfill \includegraphics[width=1.8in]{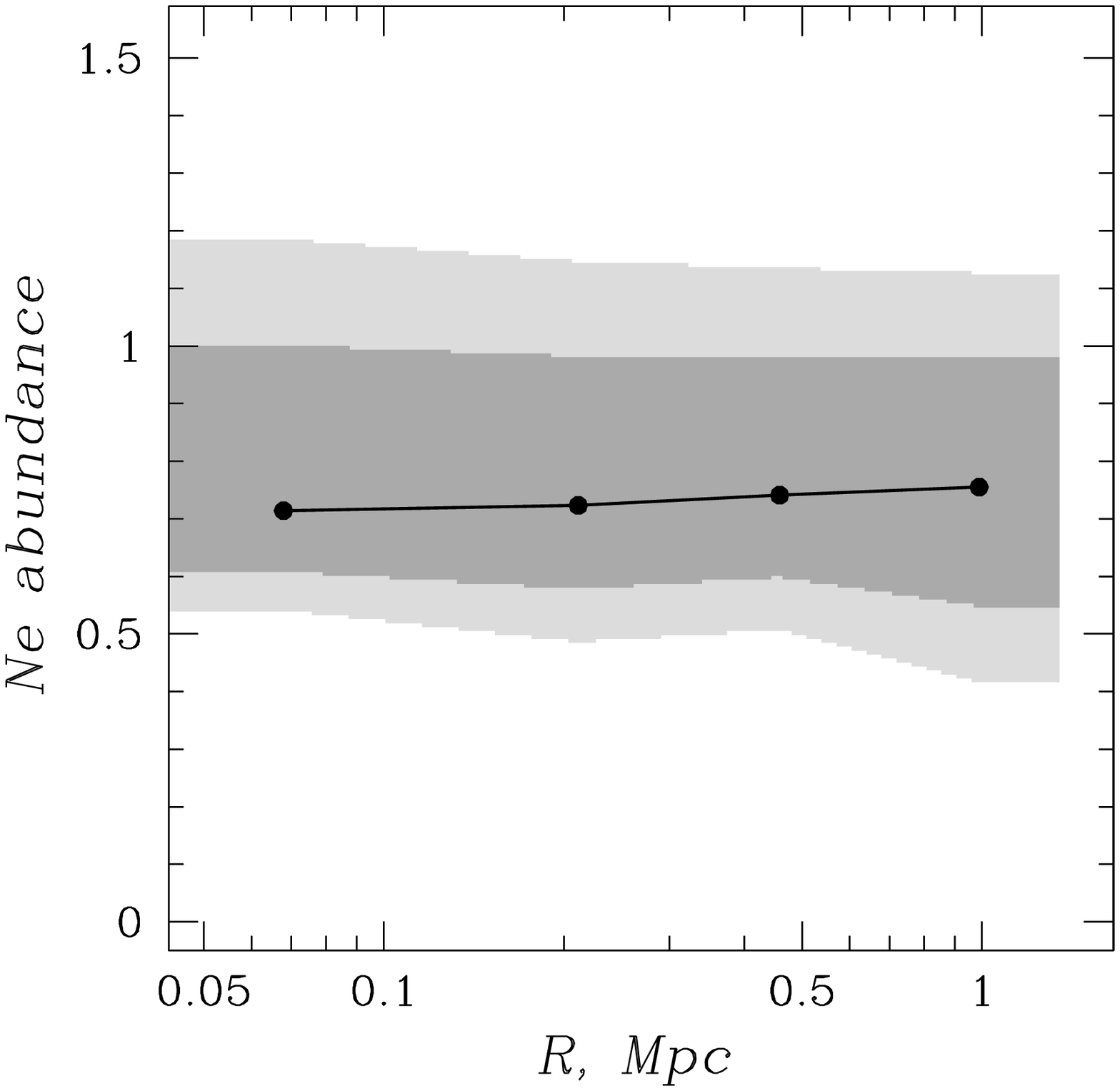}

\includegraphics[width=1.8in]{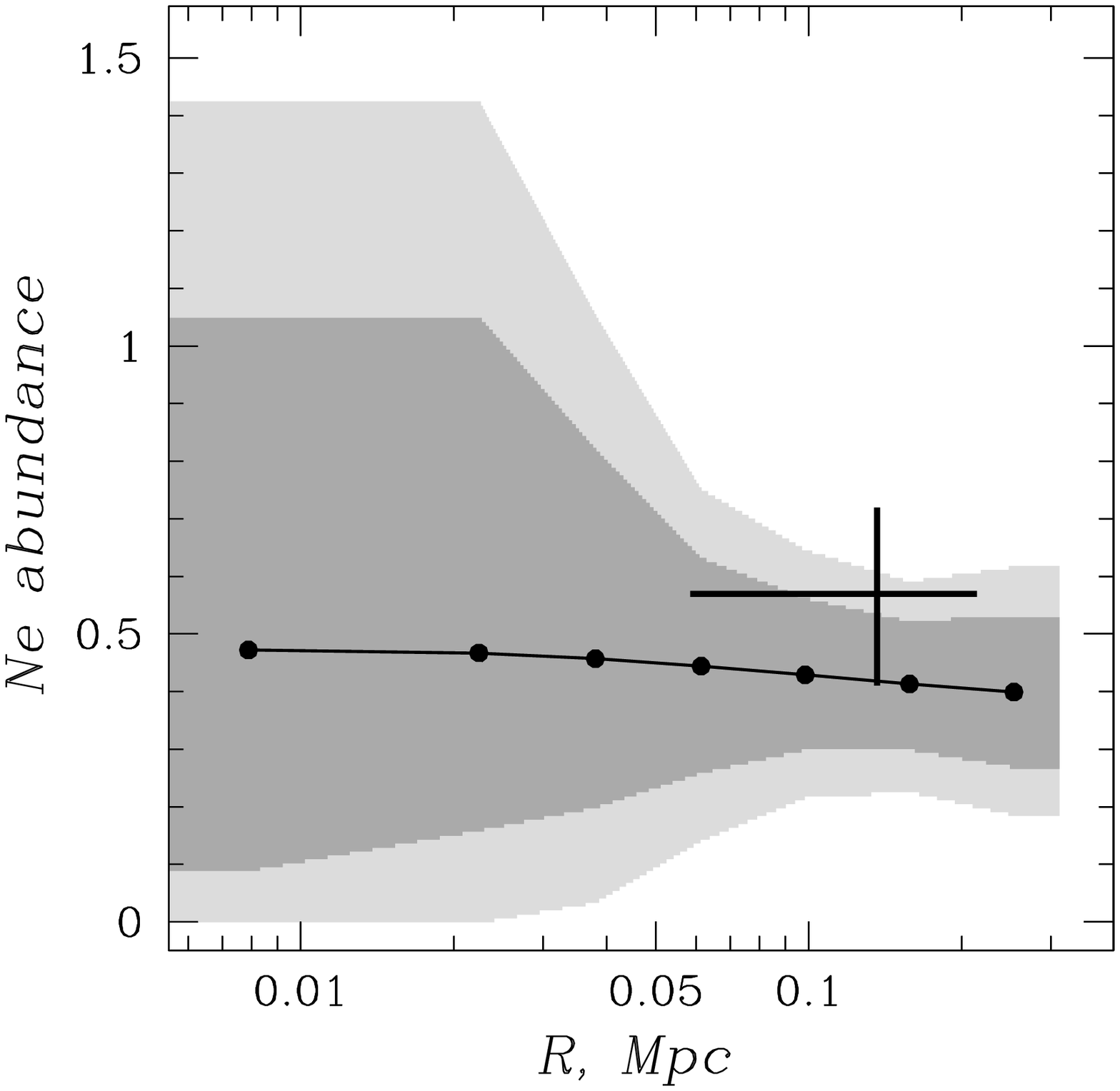} \hfill \includegraphics[width=1.8in]{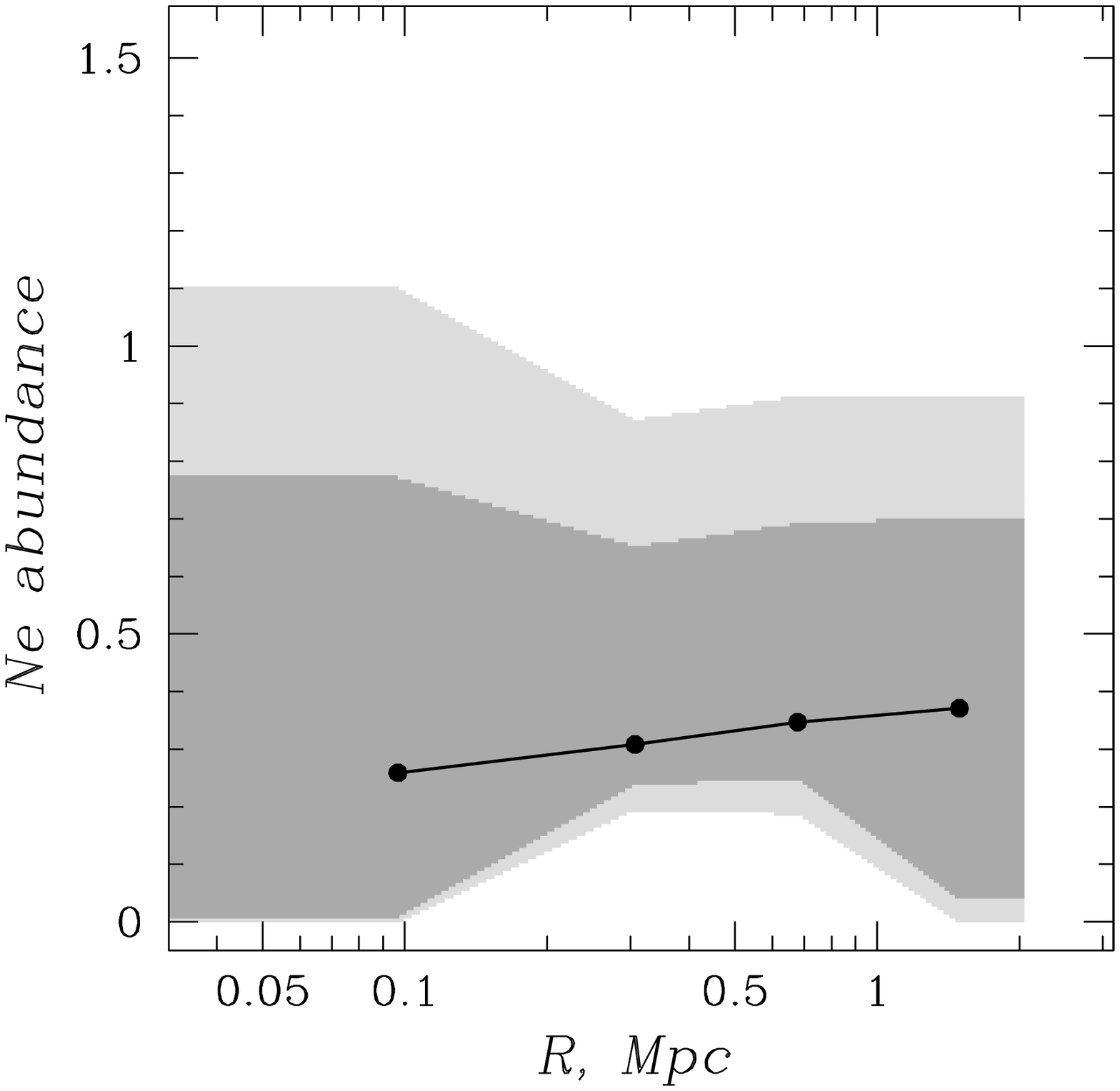}  \hfill \includegraphics[width=1.8in]{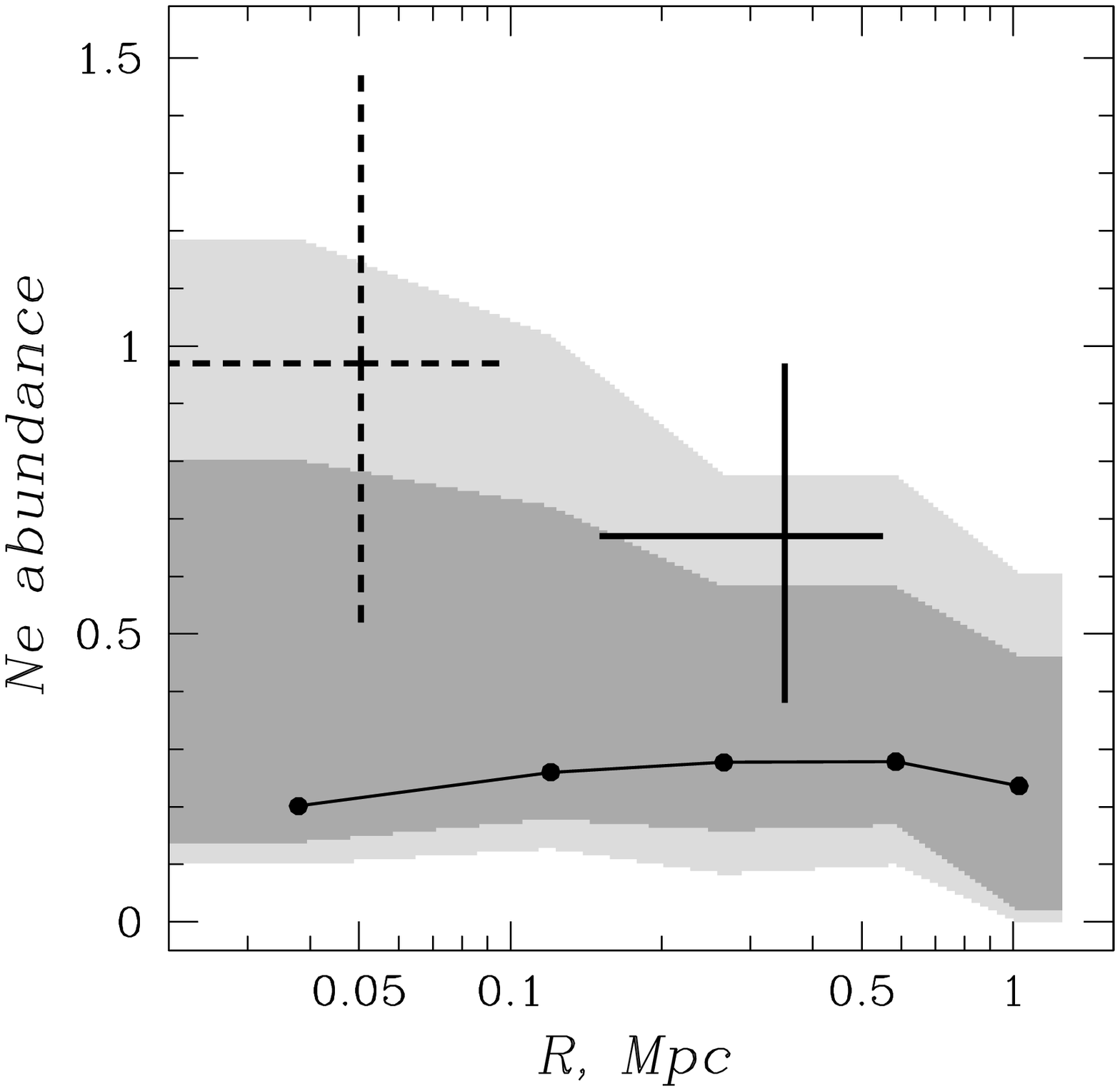}

\includegraphics[width=1.8in]{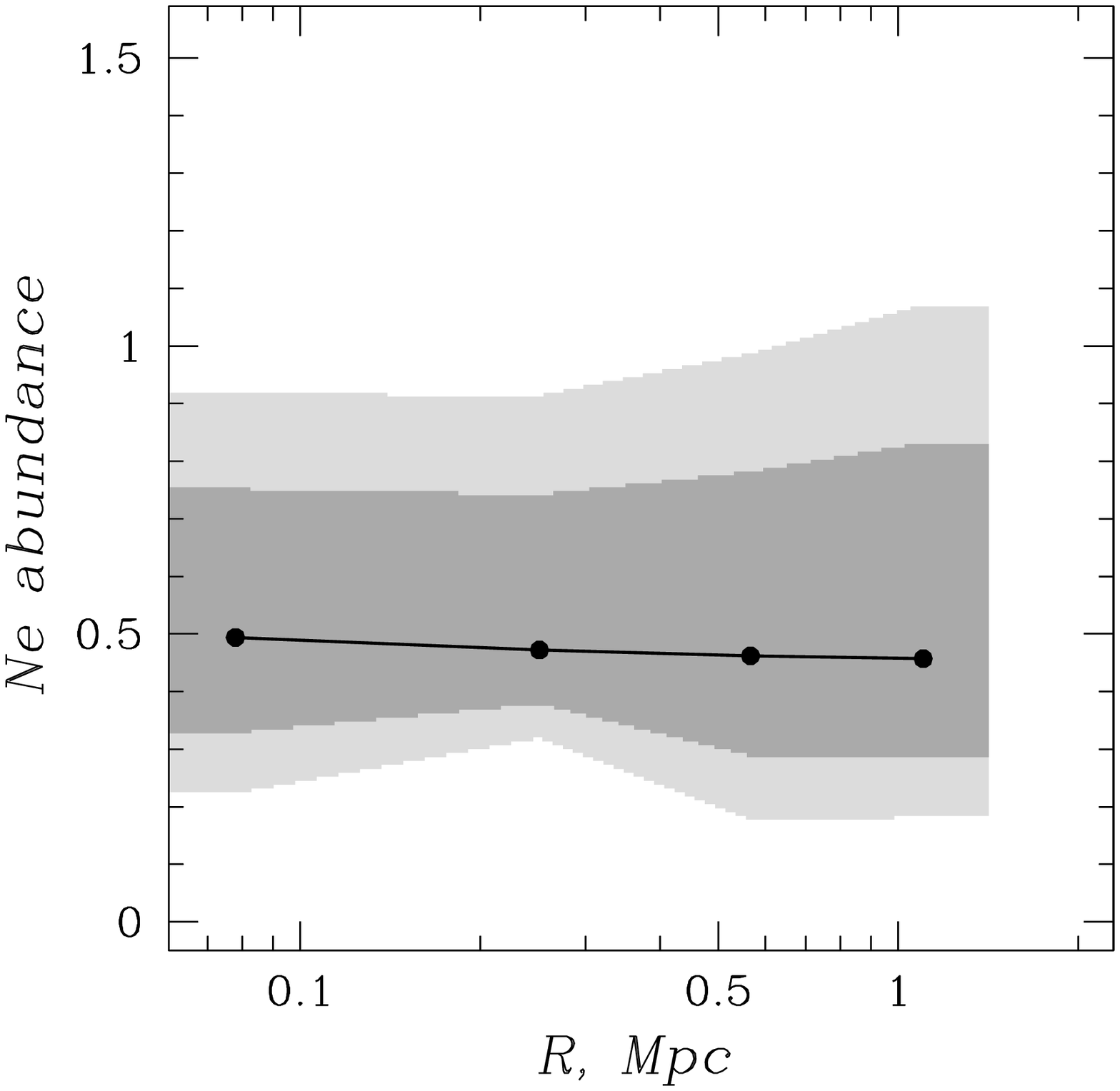} \hfill \includegraphics[width=1.8in]{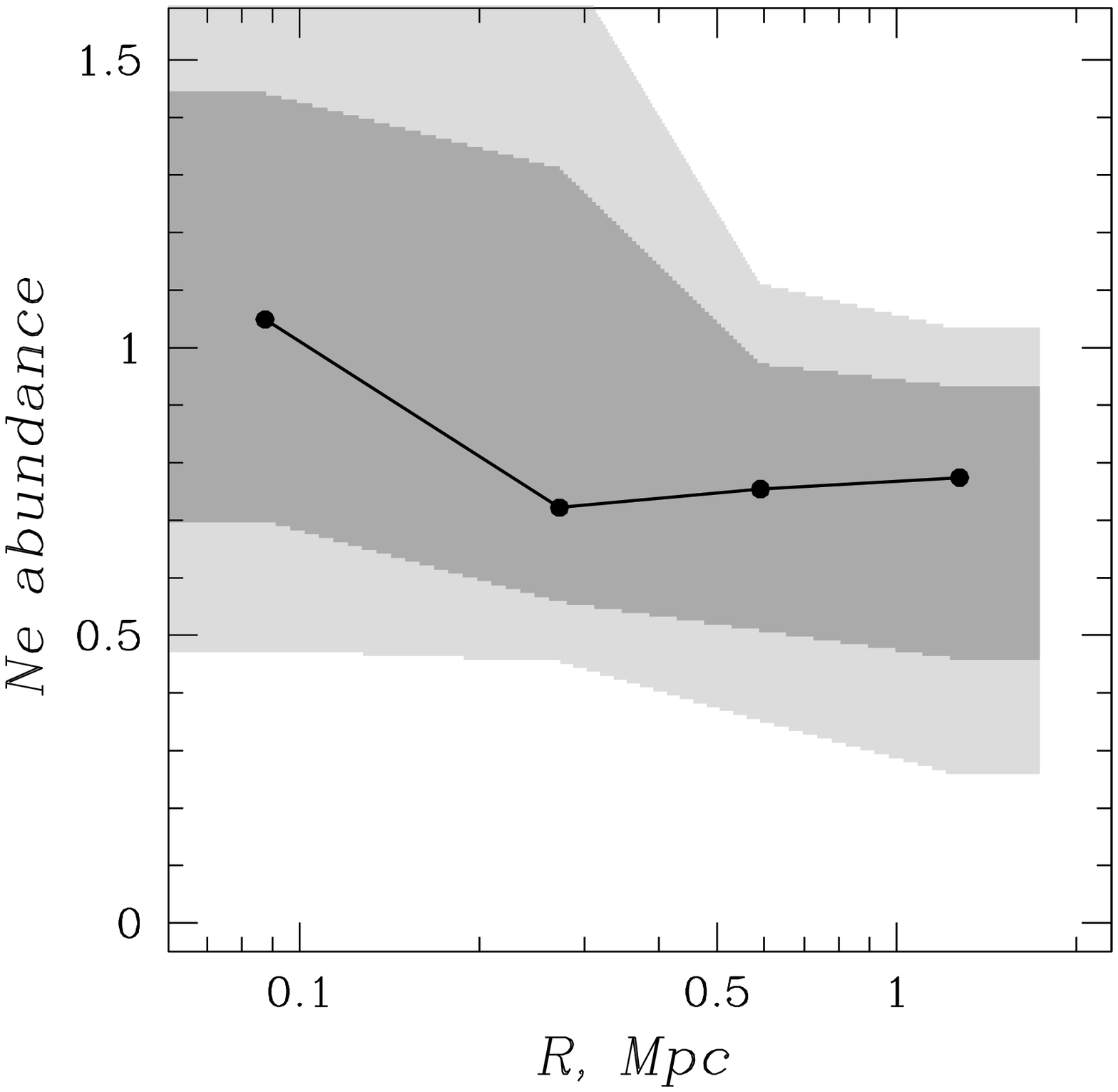} \hfill \includegraphics[width=1.8in]{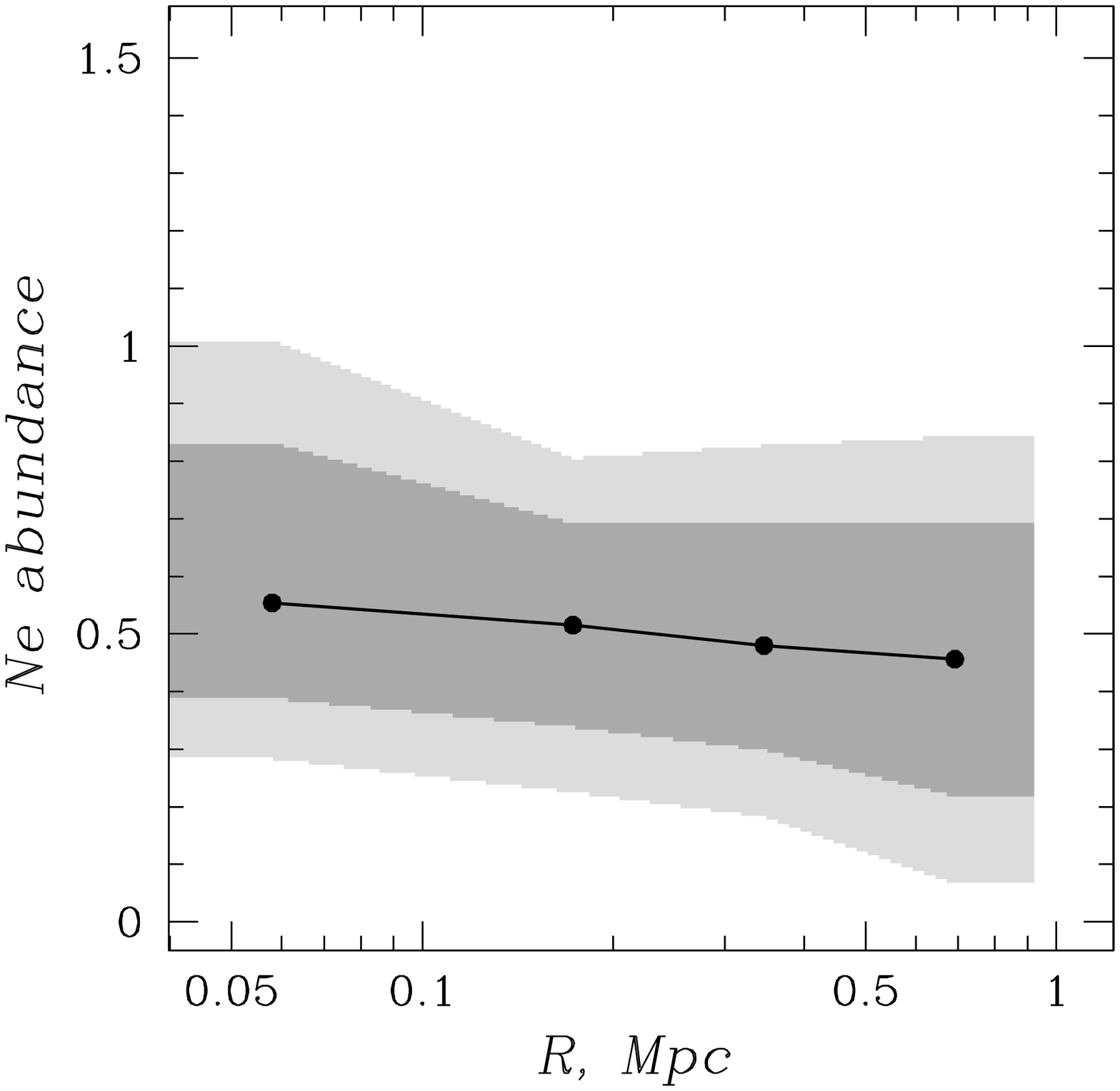}

\figcaption{Derived Ne abundances. Solid line represents the best-fit curve
describing ASCA results, with filled circles indicating the spatial binning
used. Dark and light shaded zones around the best fit curve denote the 68
and 90 per cent confidence areas, respectively. Solid crosses on the A496,
A1060 and A2199 panels represent ASCA values from Mushotzky \etal
(1996). Dashed crosses on the A496 and A2199 panels show abundance
determinations from Dupke \& White (1999).
\label{ne-fig}}
\vspace*{-14.0cm}

{\it \hfill MKW4\hspace*{2.8cm} \hfill A496\hspace*{2.7cm} \hfill A780\hspace*{0.3cm}}
\vspace*{4cm}

{\it \hfill A1060\hspace*{2.7cm} \hfill A1651\hspace*{2.6cm} \hfill A2199\hspace*{0.3cm}}

\vspace*{4cm}

{\it \hfill A2597\hspace*{2.7cm} \hfill A3112\hspace*{2.6cm} \hfill A4059\hspace*{0.3cm}}

\vspace*{4cm}

\end{figure*}

\begin{figure*}

\includegraphics[width=1.8in]{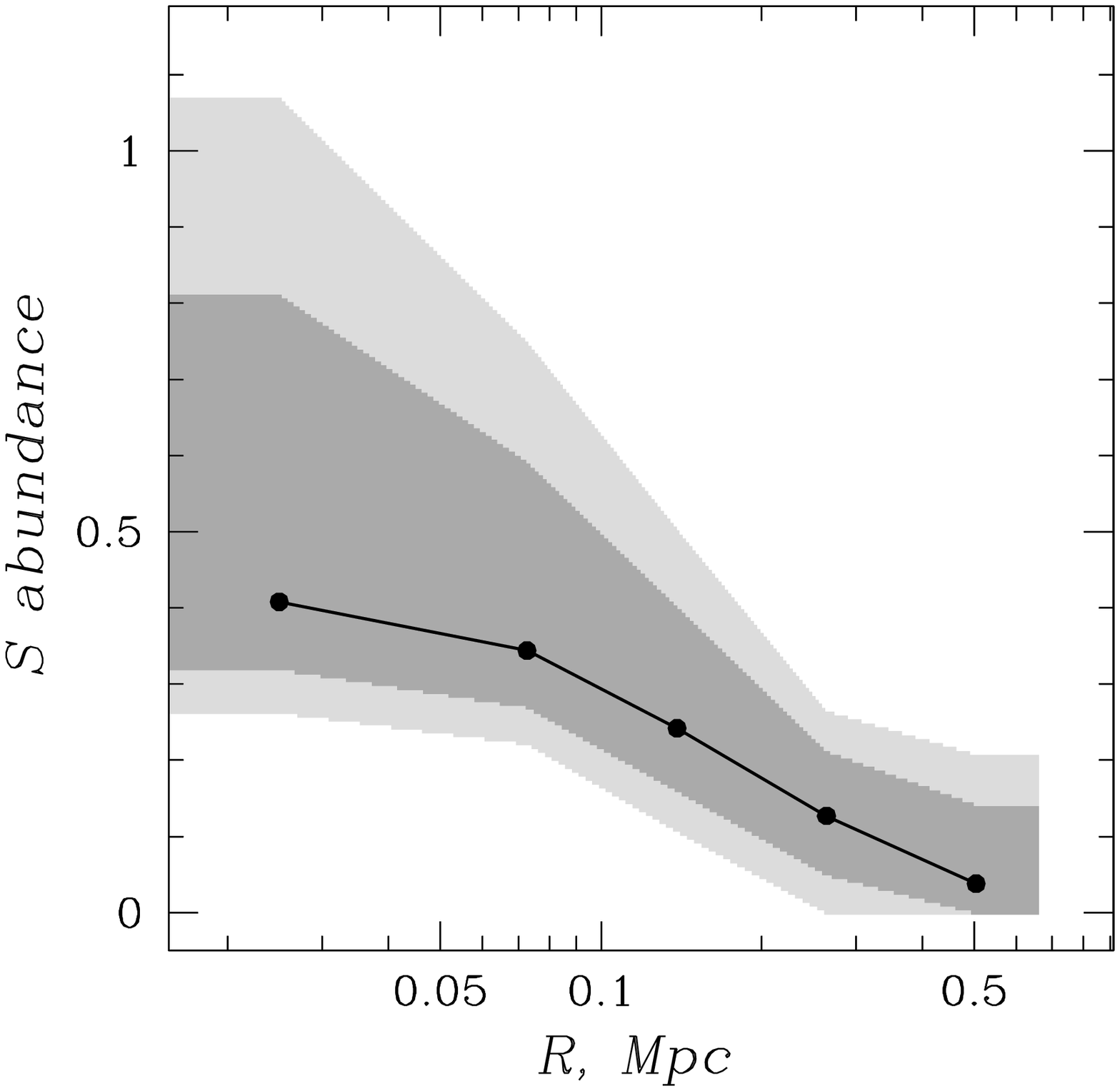} \hfill \includegraphics[width=1.8in]{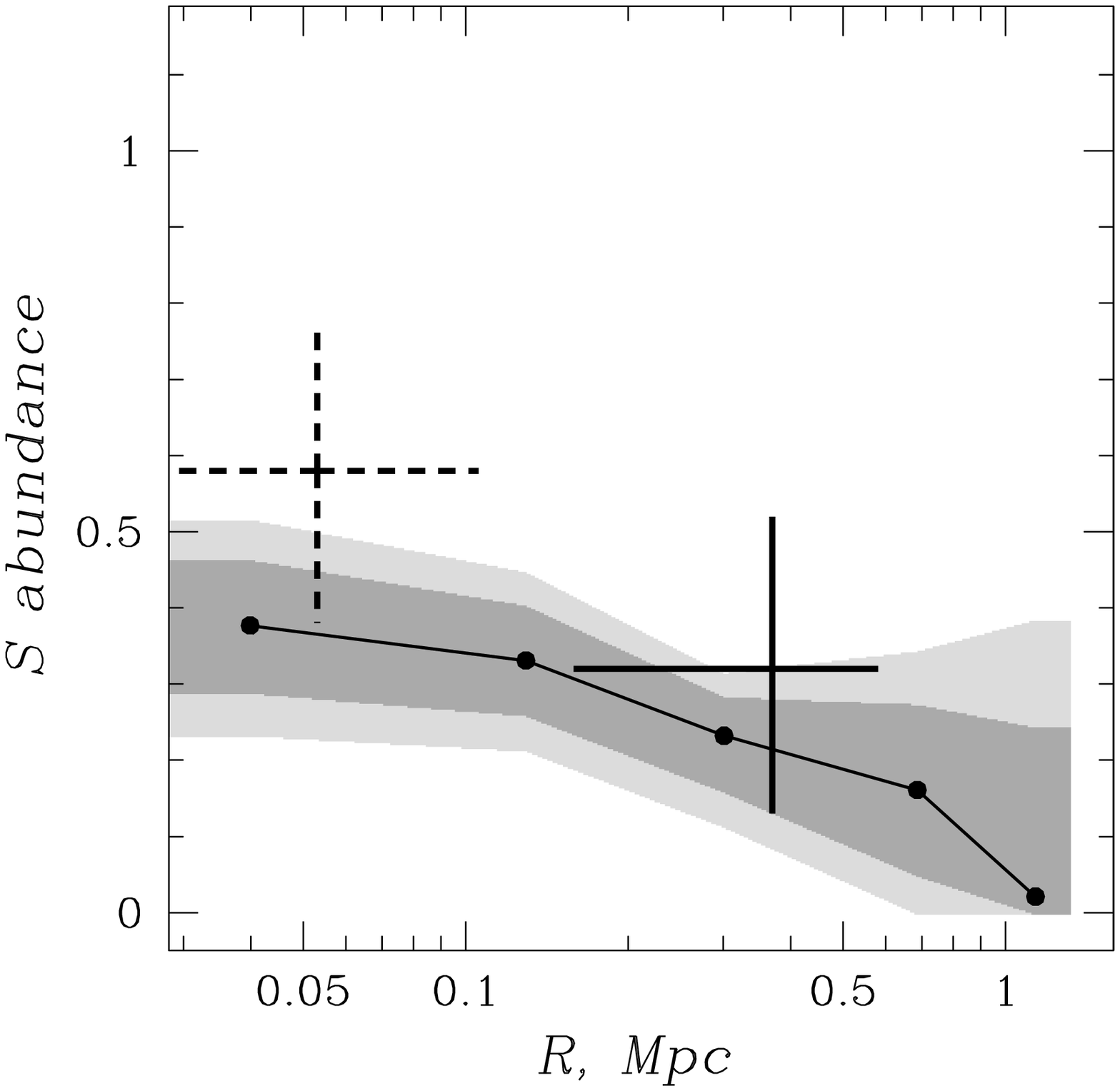} \hfill \includegraphics[width=1.8in]{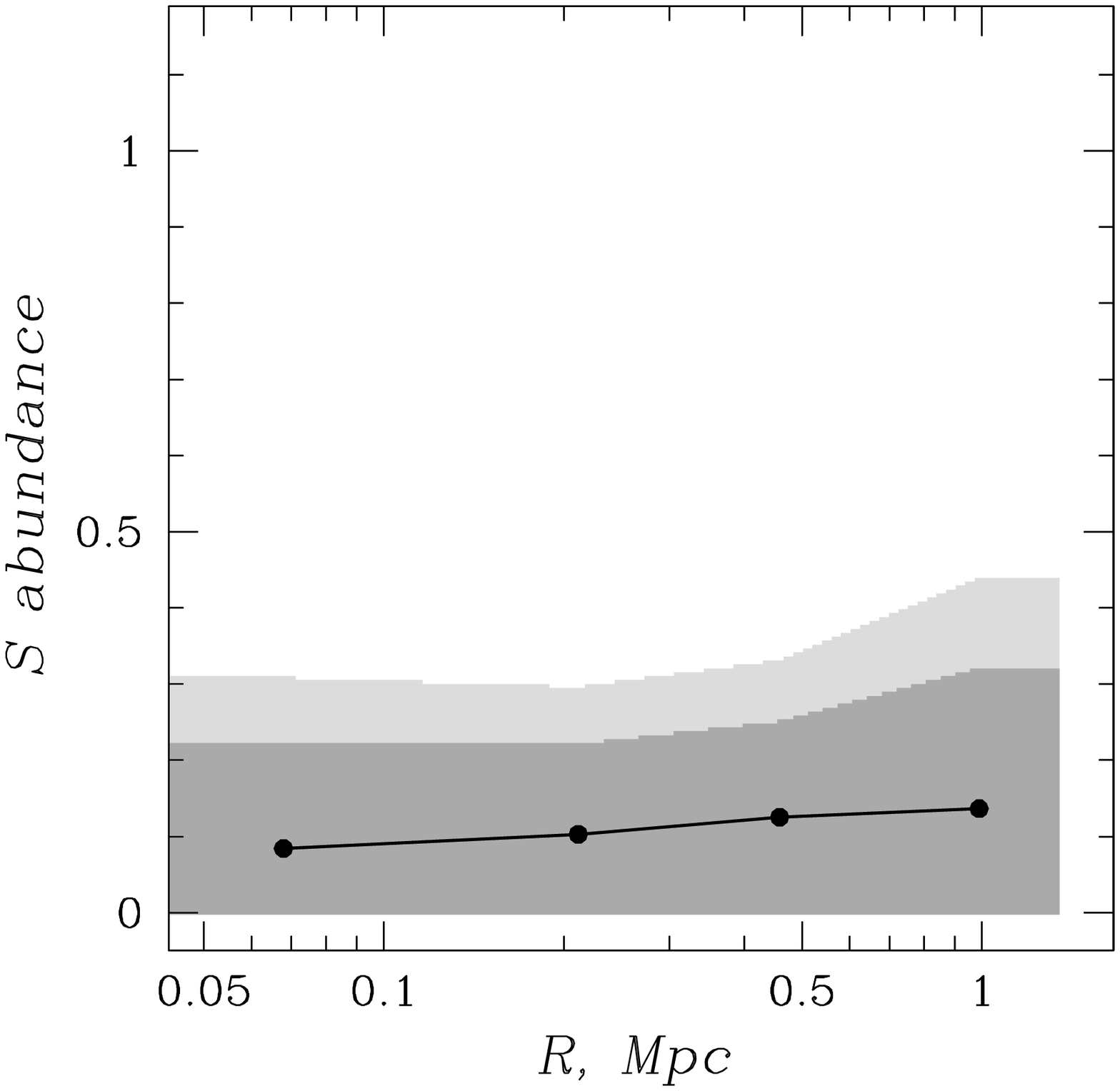}

\includegraphics[width=1.8in]{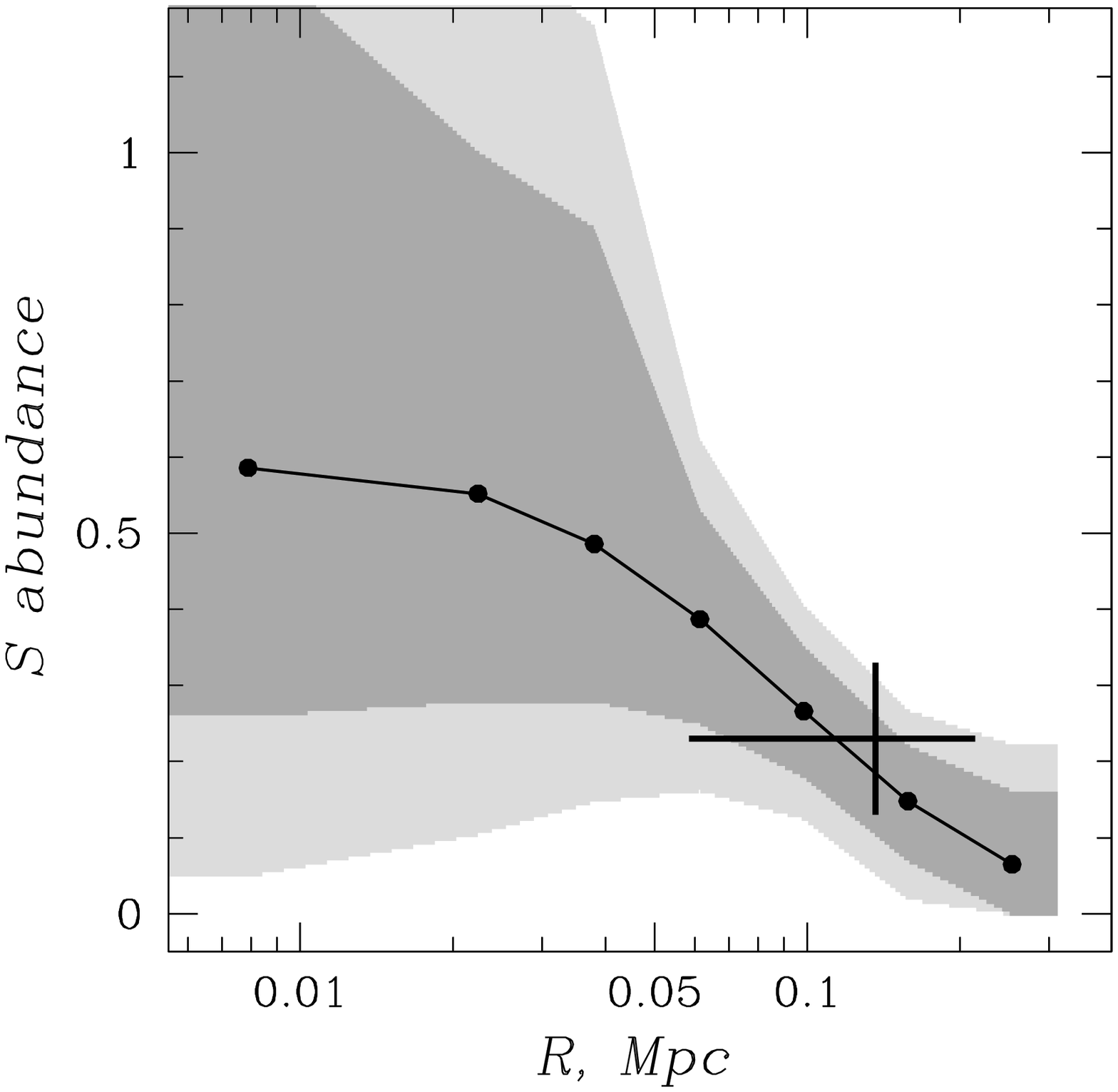} \hfill \includegraphics[width=1.8in]{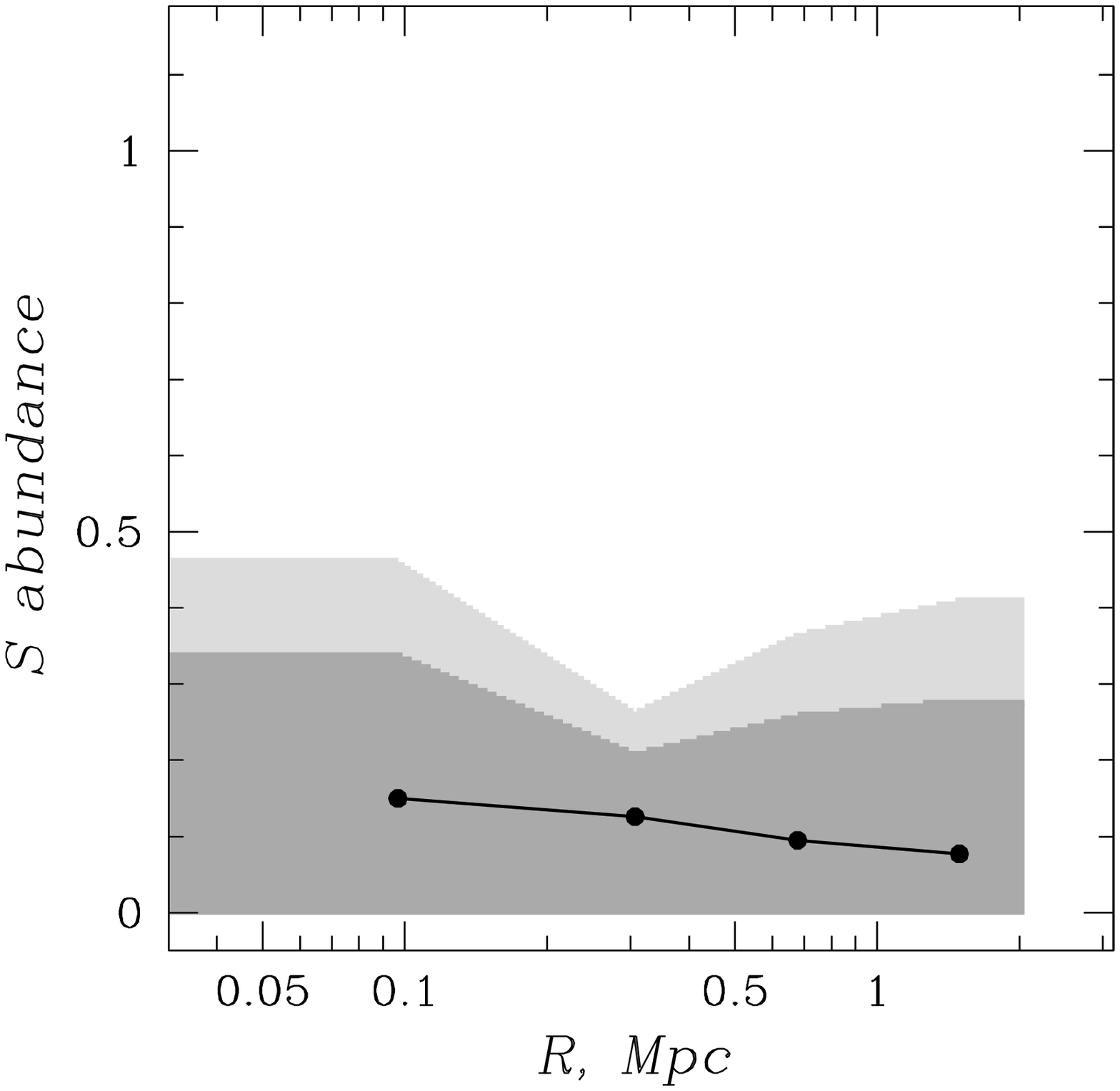}  \hfill \includegraphics[width=1.8in]{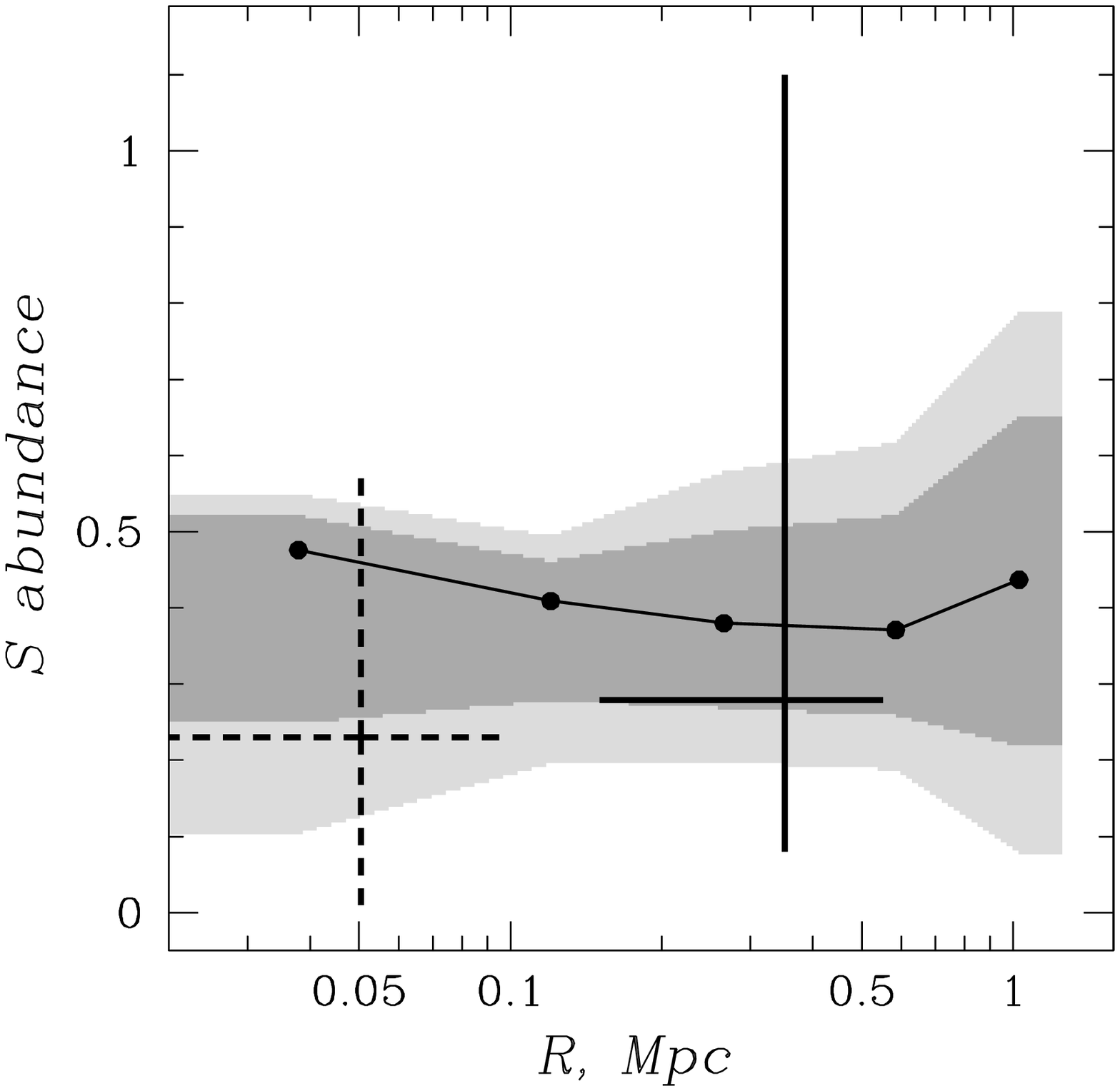}

\includegraphics[width=1.8in]{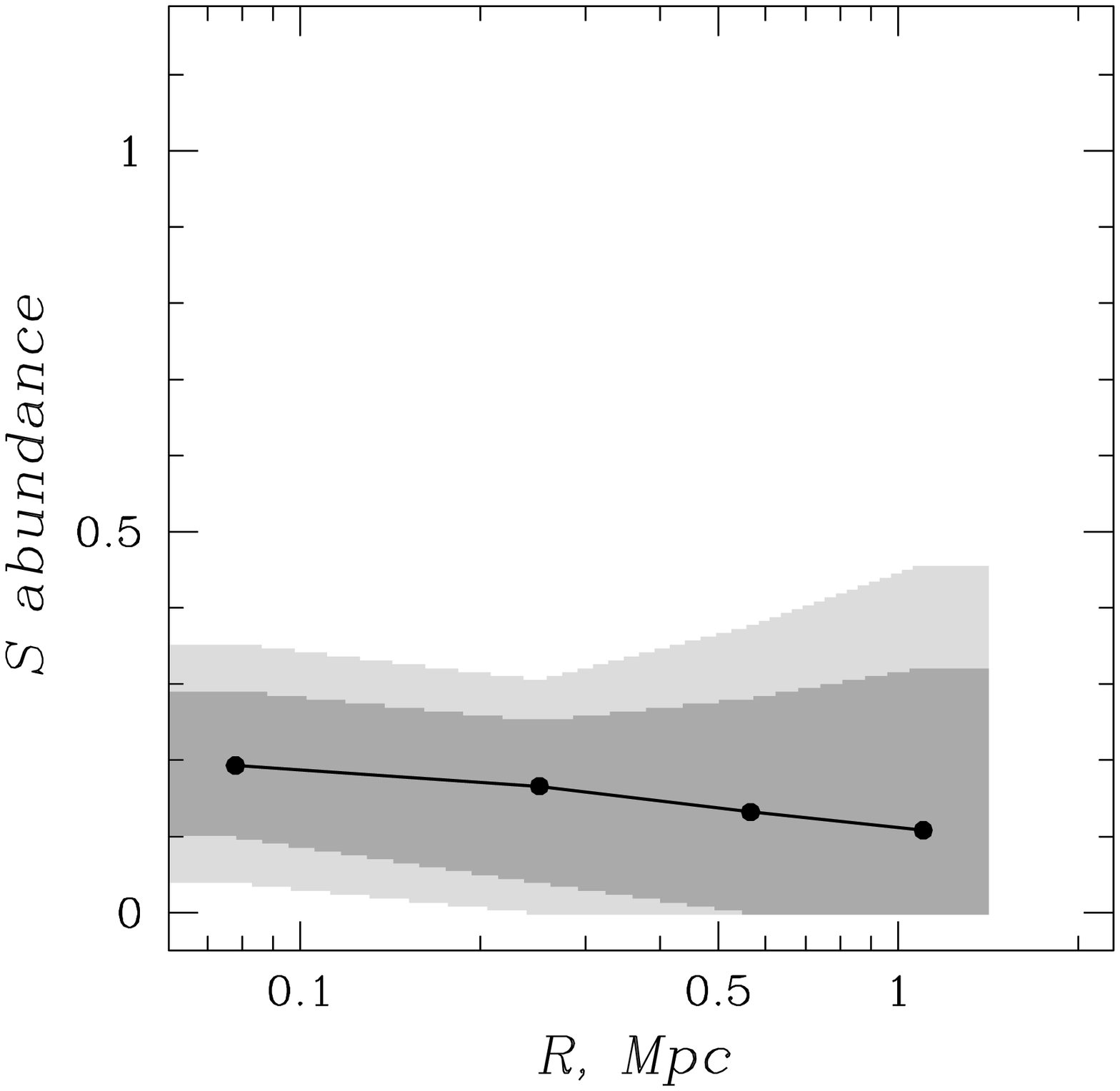} \hfill \includegraphics[width=1.8in]{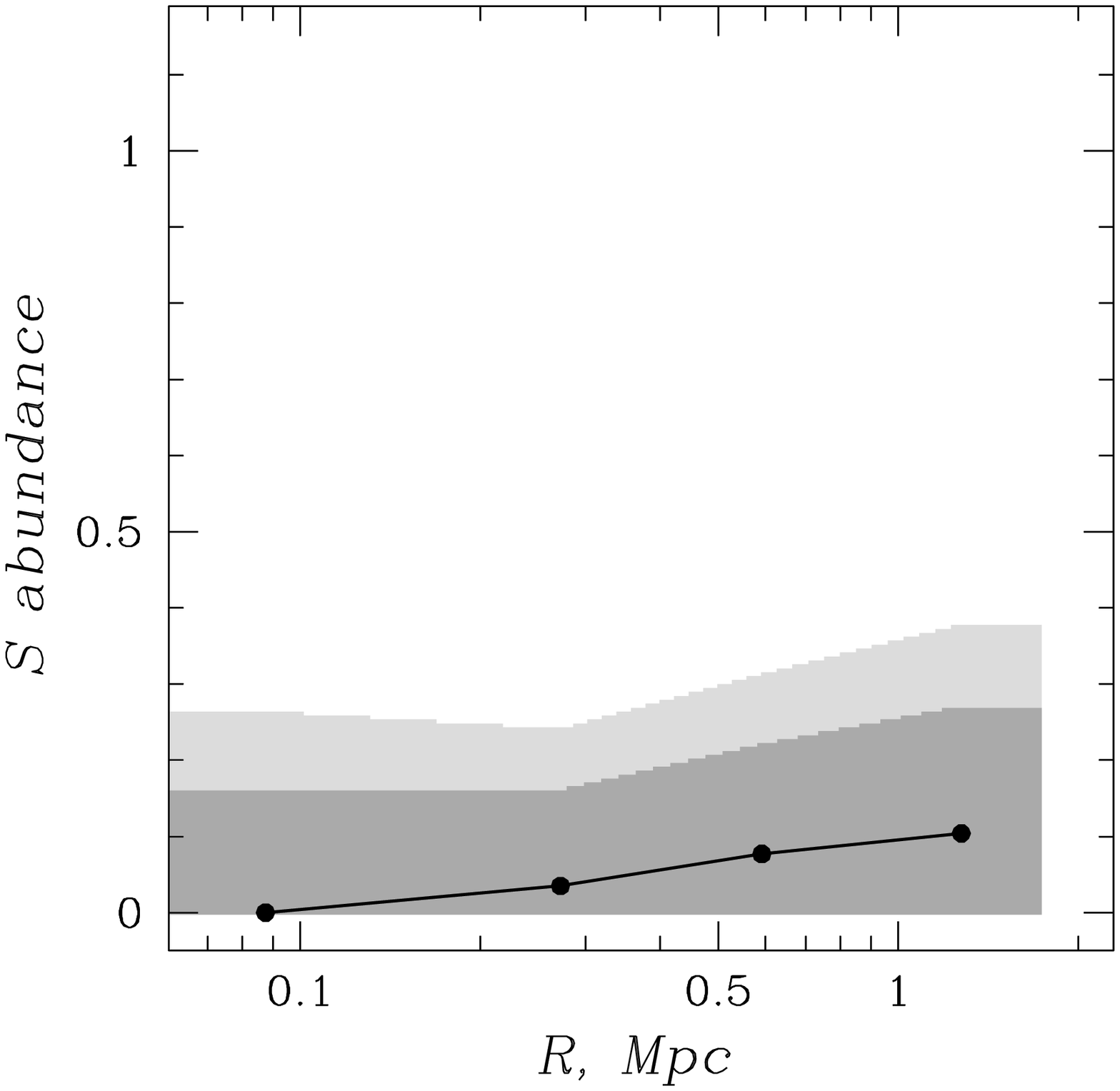} \hfill \includegraphics[width=1.8in]{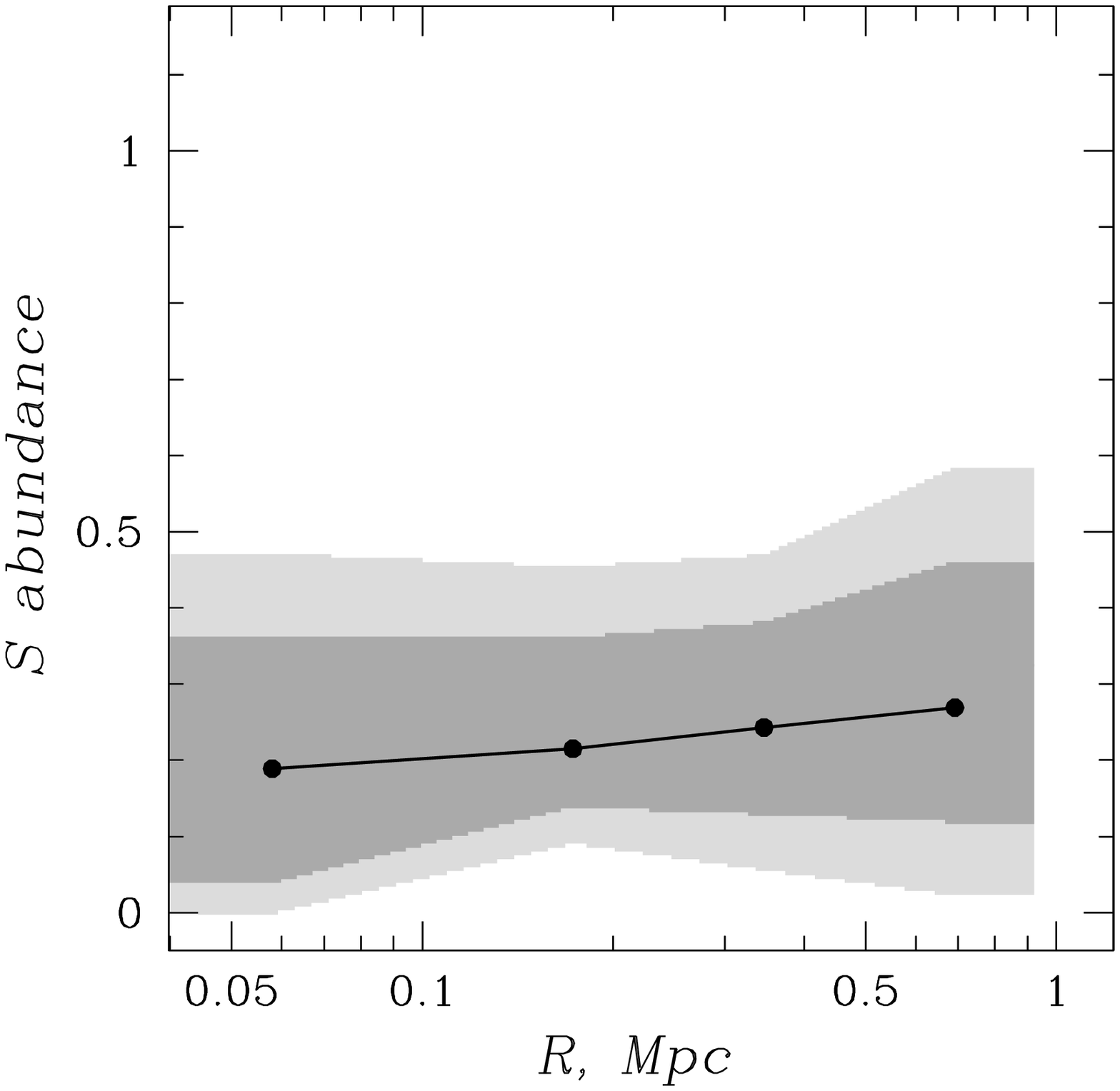}

\figcaption{ Derived S abundances. Solid line represents the best-fit curve
describing ASCA results, with filled circles indicating the spatial binning
used. Dark and light shaded zones around the best fit curve denote the 68
and 90 per cent confidence areas, respectively. Solid crosses on the A496,
A1060 and A2199 panels represent ASCA values from Mushotzky \etal (1996).  S
is coupled with Ar in our fitting, so our limits are sometimes more strict in
comparison to the values of Mushotzky \etal (1996). Dashed crosses on the A496
and A2199 panels show abundance determinations from Dupke \& White (1999).
\label{s-fig}}
\vspace*{-14.0cm}

{\it \hfill MKW4\hspace*{2.8cm} \hfill A496\hspace*{2.7cm} \hfill A780\hspace*{0.3cm}}
\vspace*{4cm}

{\it \hfill A1060\hspace*{2.7cm} \hfill A1651\hspace*{2.6cm} \hfill A2199\hspace*{0.3cm}}

\vspace*{4cm}

{\it \hfill A2597\hspace*{2.7cm} \hfill A3112\hspace*{2.6cm} \hfill A4059\hspace*{0.3cm}}

\vspace*{4cm}

\end{figure*}

%\clearpage

\begin{table*}
{
\footnotesize
\centering
\tabcaption{\footnotesize
\centerline{ASCA SIS heavy element abundance measurements$^{\dag}$}
\label{table-ab}}

\begin{tabular}{lllll}
\hline
\hline
Annulus (\amin) & $Ne/Ne_{\odot}$ & $Si/Si_{\odot}$ & $S/S_{\odot}$ & $Fe/Fe_{\odot}$ \\
\hline
 & & & & \\
\multicolumn{2}{c}{MKW4} & & & \\
 0.0---1.5 & 0.35 (0.17:1.01) & 0.70 (0.60:1.11) & 0.41 (0.32:0.81) & 0.62 (0.58:0.89)\\
 1.5---2.9 & 0.30 (0.14:0.87) & 0.52 (0.45:0.73) & 0.34 (0.27:0.59) & 0.45 (0.41:0.62)\\
 2.9---5.5 & 0.23 (0.10:0.65) & 0.33 (0.26:0.42) & 0.24 (0.16:0.40) & 0.26 (0.22:0.44)\\
 5.5---10.5 & 0.20 (0.07:0.33) & 0.25 (0.18:0.40) & 0.13 (0.05:0.21) & 0.13 (0.10:0.24)\\
10.5---20.0 & 0.21 (0.04:0.37) & 0.24 (0.15:0.40) & 0.04 (0.00:0.14) & 0.06 (0.03:0.11)\\
 & & & & \\
\multicolumn{2}{c}{A496} & & & \\
 0.0---1.5 & 0.90 (0.67:1.10) & 0.74 (0.63:0.86) & 0.38 (0.29:0.46) & 0.47 (0.43:0.55)\\
 1.5---3.4 & 0.93 (0.66:1.10) & 0.65 (0.59:0.80) & 0.33 (0.26:0.40) & 0.39 (0.34:0.48)\\
 3.4---8.0 & 0.91 (0.61:1.10) & 0.56 (0.49:0.65) & 0.23 (0.16:0.28) & 0.31 (0.26:0.36)\\
 8.0---18.0 & 0.90 (0.61:1.10) & 0.50 (0.37:0.60) & 0.16 (0.05:0.27) & 0.27 (0.22:0.31)\\
18.0---25.0 & 0.88 (0.65:1.10) & 0.24 (0.02:0.46) & 0.02 (0.00:0.24) & 0.13 (0.00:0.31)\\
 & & & & \\
\multicolumn{2}{c}{A780} & & & \\
 0.0---1.5 & 0.71 (0.61:1.00) & 0.80 (0.63:0.91) & 0.08 (0.00:0.22) & 0.39 (0.33:0.55)\\
 1.5---3.2 & 0.72 (0.58:0.98) & 0.74 (0.65:0.91) & 0.10 (0.00:0.22) & 0.29 (0.26:0.35)\\
 3.2---7.0 & 0.74 (0.60:0.98) & 0.69 (0.50:0.82) & 0.13 (0.00:0.25) & 0.16 (0.05:0.21)\\
 7.0---15.0 & 0.76 (0.55:0.98) & 0.66 (0.42:0.88) & 0.14 (0.00:0.32) & 0.08 (0.00:0.16)\\
 & & & & \\
\multicolumn{2}{c}{A1060} & & & \\
 0.0---0.8 & 0.47 (0.09:1.05) & 0.80 (0.48:1.13) & 0.59 (0.26:1.27) & 0.44 (0.24:0.65)\\
 0.8---1.5 & 0.47 (0.16:1.05) & 0.77 (0.50:1.05) & 0.55 (0.28:1.00) & 0.42 (0.27:0.58)\\
 1.5---2.4 & 0.46 (0.20:0.82) & 0.71 (0.51:0.91) & 0.49 (0.28:0.90) & 0.39 (0.30:0.48)\\
 2.4---3.9 & 0.44 (0.26:0.63) & 0.62 (0.49:0.75) & 0.39 (0.25:0.53) & 0.34 (0.29:0.39)\\
 3.9---6.2 & 0.43 (0.30:0.56) & 0.52 (0.44:0.61) & 0.27 (0.18:0.35) & 0.30 (0.26:0.34)\\
 6.2---10.0 & 0.41 (0.30:0.52) & 0.44 (0.37:0.51) & 0.15 (0.07:0.22) & 0.26 (0.22:0.30)\\
10.0---16.0 & 0.40 (0.27:0.53) & 0.39 (0.30:0.47) & 0.07 (0.00:0.16) & 0.22 (0.19:0.25)\\
 & & & & \\
\multicolumn{2}{c}{A1651} & & & \\
 0.0---1.5 & 0.26 (0.01:0.77) & 0.59 (0.37:0.80) & 0.15 (0.00:0.34) & 0.26 (0.14:0.41)\\
 1.5---3.3 & 0.31 (0.24:0.65) & 0.59 (0.41:0.78) & 0.13 (0.00:0.21) & 0.18 (0.13:0.28)\\
 3.3---7.3 & 0.35 (0.25:0.69) & 0.58 (0.33:0.78) & 0.10 (0.00:0.26) & 0.11 (0.04:0.32)\\
 7.3---16.0 & 0.37 (0.04:0.70) & 0.57 (0.30:0.78) & 0.08 (0.00:0.28) & 0.08 (0.00:0.32)\\
 & & & & \\
\multicolumn{2}{c}{A2029} & & & \\
0.0---1.5   &  & 0.91 (0.81:1.01) &  & 0.53 (0.46:0.73)  \\
1.5---3.8   &  & 0.88 (0.66:0.95) &  & 0.36 (0.33:0.46)  \\
3.8---9.5   &  & 0.87 (0.72:0.95) &  & 0.19 (0.14:0.24)  \\
9.5---24.0  &  & 0.86 (0.68:1.01) &  & 0.09 (0.01:0.19)  \\
 & & & & \\
\multicolumn{2}{c}{A2199} & & & \\
 0.0---1.5 & 0.20 (0.14:0.80) & 1.07 (0.88:1.31) & 0.48 (0.25:0.52) & 0.44 (0.36:0.55)\\
 1.5---3.3 & 0.26 (0.18:0.72) & 1.08 (0.89:1.22) & 0.41 (0.28:0.46) & 0.40 (0.36:0.50)\\
 3.3---7.3 & 0.28 (0.16:0.58) & 1.06 (0.82:1.10) & 0.38 (0.27:0.50) & 0.34 (0.31:0.44)\\
 7.3---16.0 & 0.28 (0.17:0.58) & 1.05 (0.75:1.13) & 0.37 (0.26:0.52) & 0.27 (0.22:0.30)\\
16.0---25.0 & 0.24 (0.02:0.46) & 0.89 (0.68:1.11) & 0.44 (0.22:0.65) & 0.23 (0.09:0.30)\\
 & & & & \\
\multicolumn{2}{c}{A2597} & & & \\
 0.0---1.2 & 0.49 (0.33:0.75) & 0.38 (0.29:0.53) & 0.19 (0.10:0.29) & 0.34 (0.29:0.48)\\
 1.2---2.7 & 0.47 (0.38:0.74) & 0.40 (0.32:0.49) & 0.17 (0.04:0.25) & 0.16 (0.11:0.27)\\
 2.7---6.1 & 0.46 (0.29:0.78) & 0.44 (0.28:0.60) & 0.13 (0.00:0.28) & 0.04 (0.00:0.20)\\
 6.1---11.0 & 0.46 (0.29:0.83) & 0.46 (0.21:0.68) & 0.11 (0.00:0.32) & 0.00 (0.00:0.20)\\
 & & & & \\
\multicolumn{2}{c}{A2670} & & & \\
0.0---1.5   &  & 0.91 (0.64:1.27) &  & 0.15 (0.07:0.30)  \\
1.5---4.0   &  & 0.93 (0.64:1.30) &  & 0.19 (0.14:0.28)  \\
4.0---12.0  &  & 0.94 (0.64:1.30) &  & 0.23 (0.13:0.31)  \\
 & & & & \\
\multicolumn{2}{c}{A3112} & & & \\
 0.0---1.5 & 1.05 (0.70:1.44) & 0.94 (0.70:1.07) & 0.00 (0.00:0.16) & 0.48 (0.41:0.62)\\
 1.5---3.2 & 0.72 (0.56:1.31) & 0.87 (0.71:0.94) & 0.04 (0.00:0.16) & 0.36 (0.30:0.48)\\
 3.2---7.0 & 0.75 (0.51:0.97) & 0.79 (0.62:0.94) & 0.08 (0.00:0.22) & 0.19 (0.11:0.27)\\
 7.0---15.0 & 0.77 (0.46:0.93) & 0.73 (0.59:0.82) & 0.10 (0.00:0.27) & 0.09 (0.00:0.20)\\
 & & & & \\
\multicolumn{2}{c}{A4059} & & & \\
 0.0---1.5 & 0.55 (0.39:0.83) & 0.74 (0.59:0.95) & 0.19 (0.04:0.36) & 0.44 (0.36:0.60)\\
 1.5---3.0 & 0.52 (0.34:0.69) & 0.61 (0.57:0.74) & 0.22 (0.14:0.36) & 0.37 (0.32:0.44)\\
 3.0---6.0 & 0.48 (0.30:0.69) & 0.44 (0.29:0.59) & 0.24 (0.13:0.38) & 0.27 (0.22:0.34)\\
 6.0---12.0 & 0.46 (0.22:0.69) & 0.31 (0.10:0.52) & 0.27 (0.12:0.46) & 0.21 (0.14:0.31)\\
\hline  
\end{tabular}

\begin{enumerate}
\item[{$^{\dag}$}]{\footnotesize ~ Definition of the solar abundance units
is 12.3, 3.55, 1.62 and 4.68 $\times10^{-5}$ for the number abundance of Ne,
Si, S and Fe relative to H, respectively. Errors are given at 68\%
confidence level (see text for further details). MEKAL plasma code is used
for spectral fitting.}
\end{enumerate}
}
\end{table*}

%\clearpage

\subsection{Iron}

Significant Fe abundance gradients are detected in MKW4, A496, A780, A2029,
A2199, A2597 and A3112 (see Fig. \ref{fe-fig}).  An Fe abundance gradient
was previously reported for A780 by Ikebe \etal (1997) using GIS
data. Detection of an Fe abundance gradient in A4059 using the wider field of
view of the GIS (Kikuchi \etal 1999) is in agreement with the trend seen in our
SIS analysis. Sarazin \etal (1998) reported a marginal Fe abundance gradient
in A2029 based on the central ASCA pointing alone. Beyond the cooling flow
region in the clusters, we find good agreement between the Fe abundances
derived from our three-dimensional modeling technique with those obtained
from previous estimates (\eg Mushotzky \etal 1996 and Fukazawa \etal 1998),
as well as the central abundances estimated in Dupke and White (1999).

At a radial distance of 1 Mpc, all clusters have an iron abundance below 0.3
solar. The mean Fe abundance at 1 Mpc is 0.20 solar with an rms of 0.10.
Due to the presence of Fe abundance gradients, the Fe abundance in the
central spatial bin is sensitive to the physical width of the bin, so nearby
systems have higher inferred central abundances. Most of the systems have
central Fe abundances in excess of 0.4 solar.  The exceptions are A2670 and
probably A1651, which do not have observed Fe abundance gradients, which
makes them candidates for recent mergers.  A2670 may be undergoing a merger
based on optical (Bird 1994) and ROSAT PSPC (Hobbs and Willmore 1997)
observations.

\subsection{Silicon}

The Si distribution within the clusters in our sample is, in general, more
spatially uniform than the Fe distribution.  Even in MKW4, where there is a
noticeable Si gradient, the decline in the Si abundance is not as strong as
the decline in the Fe abundance.  In most clusters, the Si abundance is
fairly uniform at 30--50\% of the solar value.  The only group in our
sample, MKW4, has a significantly lower Si abundance than the clusters. This
is in agreement with the general trend noted by Fukazawa \etal (1998).

Ponman, Cannon \& Navarro (1999) examined the density distribution in a
sample of clusters, and concluded that supernova-driven winds were probably
only efficient in reducing the content of gas and metals in clusters with
virial temperatures below approximately 4 keV.  The approximately flat
distribution of Si for clusters hotter than about 4 keV suggests that either
alpha-elements have been injected uniformly through the cluster by some
non-density dependent mechanism, or that the intergalactic gas has been well
mixed since the bulk of these elements were injected. The observed scatter
of Si abundances in our sample could be an indication of detailed
differences in the evolution of individual clusters (we return to this
subject below).

\subsection{Neon}

The results concerning the Ne distribution are similar to that of Si, except
that the errors are larger. As with Si, Ne gradients are not seen at a high
significance. The Ne abundance in A2670 and A2029 are highly uncertain, and
we do not present the results here.  The large uncertainties in the best fit
Ne abundances smear out the differences among the systems. In a few cases
(A496, A780, A2597 and A3112), the best fit Ne abundance at a radial
distance of 1 Mpc is constrained to be between $\sim 0.3$ and solar, while
for the remainder of the systems it is only constrained to be less 
than solar.

The derived Ne abundances are affected by the inclusion of a cooling flow
model, and also dependent on the model adopted for the cooling flow. This is 
due to the fact that the Ne lines lie at nearly the same energy as the L-shell 
Fe blend, so that emission from cooling gas can be confused with Ne emission.
In presenting the Ne results, we take the best fit values based on the
spectral analysis that includes a {\it mkcflow} model, while the errors
include the variance observed between different models, including simple
single temperature fits.

\subsection{Sulfur}

By coupling the abundances of S and Ar, we obtain errors smaller than the
errors listed in Mushotzky \etal (1996).  Even with this coupling, the
uncertainties in the S and Ar abundances in A2670 and A2029 are large, and
we do not present the results for these two clusters.  A central enhancement
in sulfur in detected in MKW4 and marginally in A1060 and A496 (see
Fig.\ref{s-fig}). For the remainder of the sample, the spectroscopic results
are consistent with a uniform sulfur distribution with values of 50--100\%
the solar value.

The sulfur abundance at a radius of 1 Mpc is less then 0.5 solar in all the
systems in our sample.  Definite non-zero values for the S abundance at 1
Mpc are only detected in A2199, and definite non-zero values at all radii
within a cluster could only be obtained in five systems (MKW4, A496, A1060,
A2199 and A4059).

\includegraphics[width=3.5in]{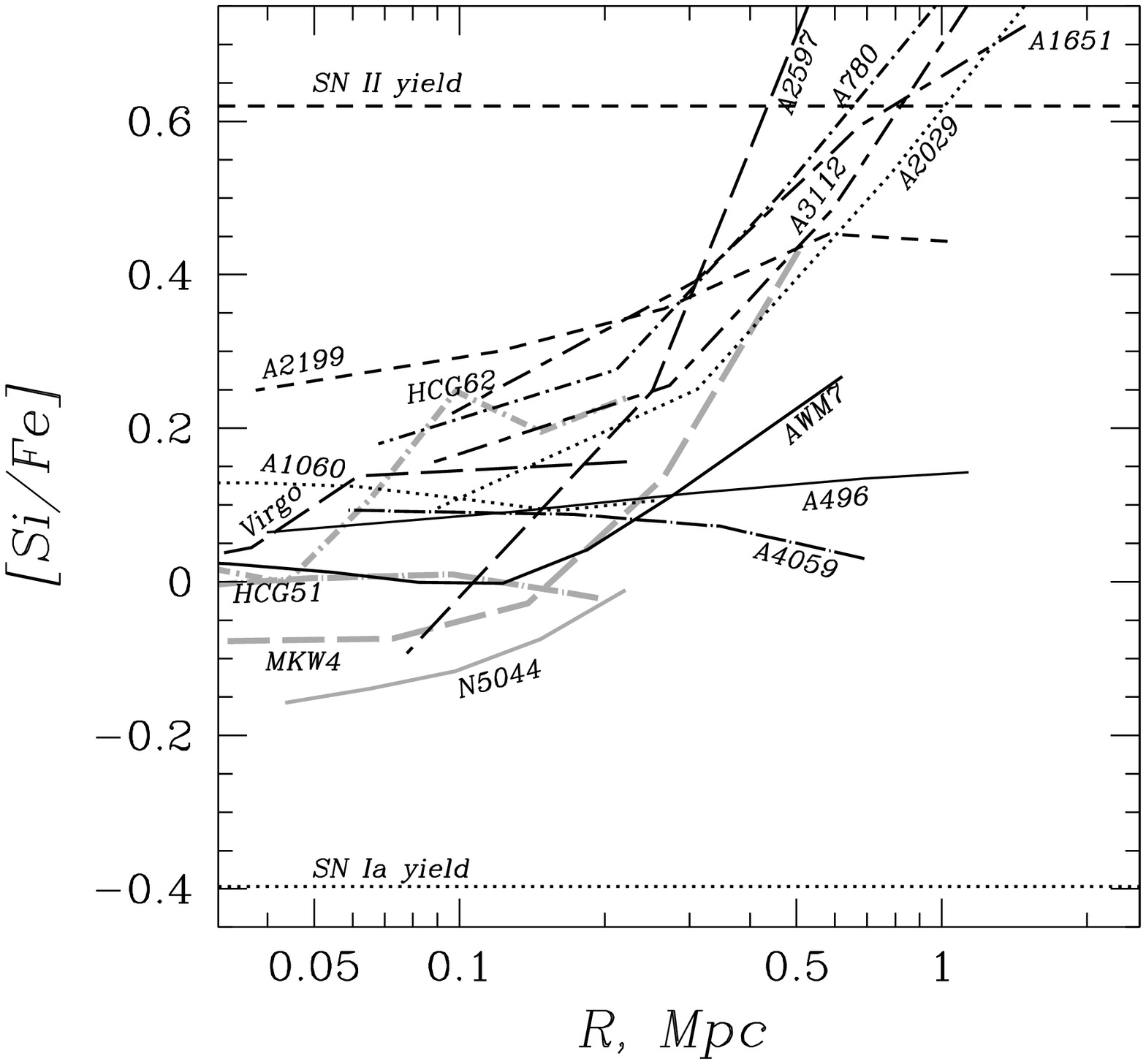} 
\figcaption{[Si/Fe] vs radius. Grey lines represent groups and black lines
represent clusters. The statistical uncertainty in the data increases with
radius from about 0.05 in the central regions to about 0.2 at 1 Mpc in terms
of [Si/Fe] at the 90\% confidence level. A2670 has a large uncertainty in
[Si/Fe] and is omitted from the plot.
\label{si2fe-rad}}

\begin{figure*}

\includegraphics[width=3.5in]{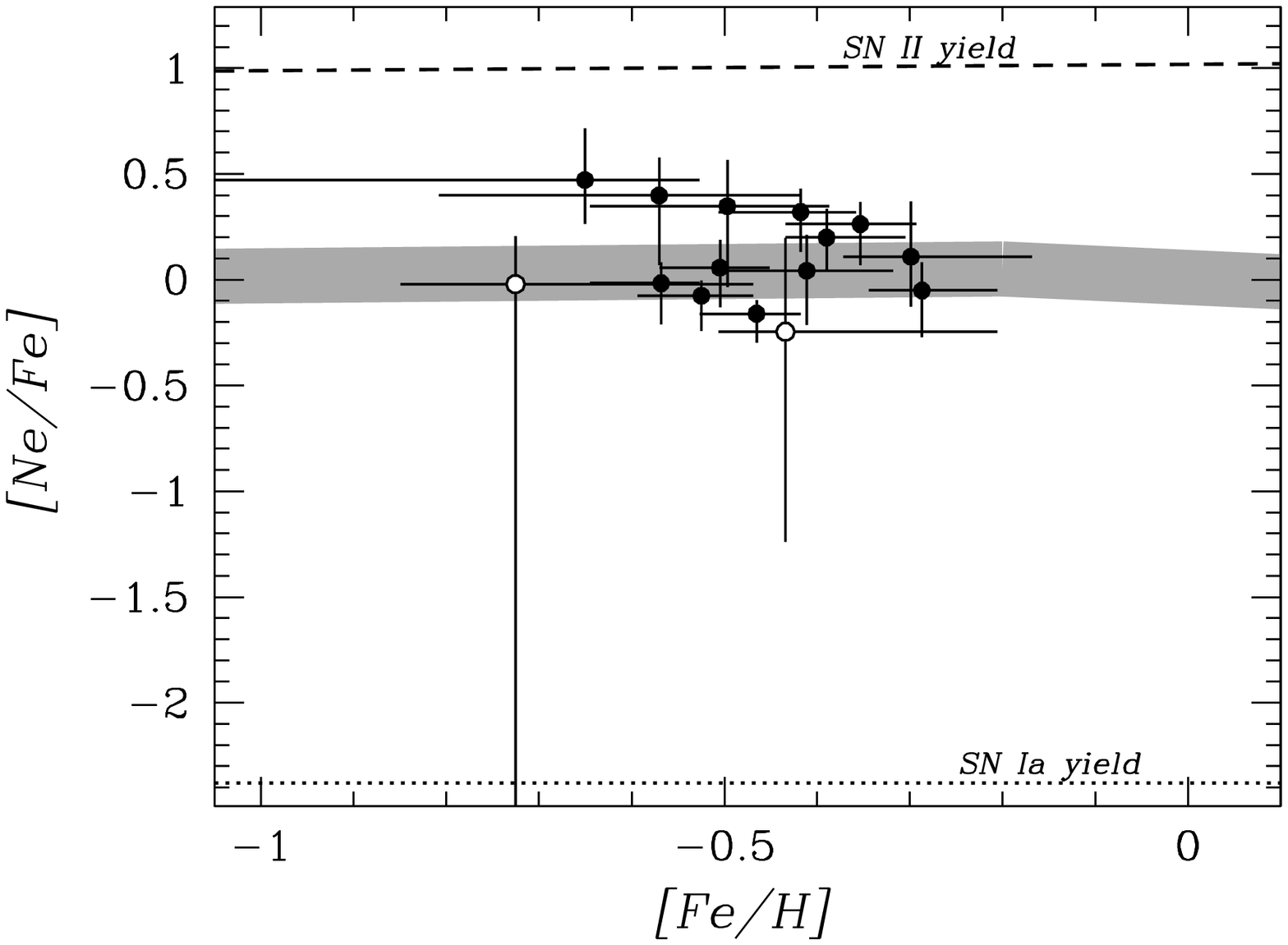} \hfill \includegraphics[width=3.5in]{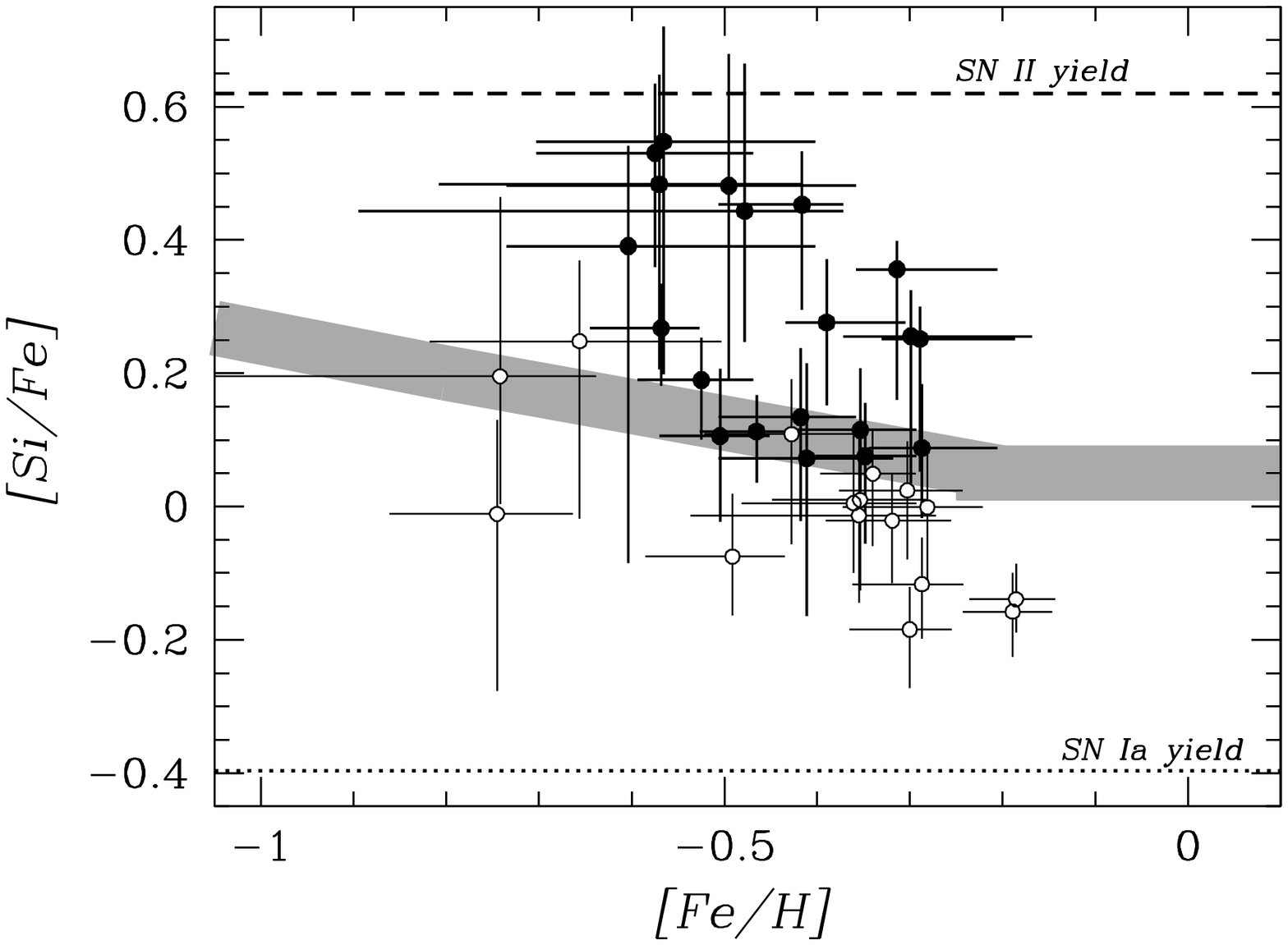}

\figcaption{ [Ne/Fe] {\it (left panel)} and [Si/Fe] {\it (right panel)} vs
[Fe/H] in clusters (filled circles) and groups (open circles).
The thick gray line in the right panel shows the observed abundance pattern
in Galactic stars and the thick gray line in the left panel shows the
corresponding theoretical prediction for Ne (Timmes, Woosley and Weaver
1995). The thickness of the lines reflect the intrinsic scatter. Abundance
units correspond to ``meteoritic'' values from Anders \& Grevesse (1989) and
[] indicates logarithmic values. Error bars are shown at the 68\% confidence
level. The dotted lines show our adopted SN Ia yield ratios and the dashed
lines our adopted SN II yield ratios. The plots contain several points 
per system at different radii.
\label{cl-plane-fig}}

\end{figure*}

While the results for the S distribution in clusters are affected by
uncertainties in the measurements, in general, global gradients in
$\alpha$-process elements in our sample appear to be considerably weaker than
the Fe abundance gradients. This has the direct implication that SN~II
products are more widely distributed in clusters, while SN~Ia products are
more centrally concentrated. According to the simulations of Metzler \&
Evrard (1994), much of the expected abundance gradient is due to different
distributions between galaxies and gas. Within this scenario, the different
distributions of SN~Ia and SN~II products can be explained if the radial
behavior of gas-to-galaxy ratio differs between the main epochs of release
of these products. SN II products should be released whilst galaxy and gas
distributions are similar (resulting in weak abundance gradients), whilst
later, when SN Ia products are released, the gas distribution should be more
extended than that of the galaxies in order to reproduce the observed
gradients. This picture is in general agreement with the findings of Ponman
\etal (1999), that SN~II-driven winds are mostly released before cluster
collapse and preheat the gas. The resulting flattening of the gas profile,
relative to the galaxy distribution, then leads naturally to the generation
of an abundance gradient if further processed gas is injected from galaxies
after the difference in density profiles has been established. We will
return to the role of the different SN types in the metal enrichment process
below.

\section{Element Abundance Ratios}\label{sec:ratios}

The radially increasing importance of SN~II in the enrichment process of
clusters is well illustrated in Fig.\ref{si2fe-rad}, which shows how the
Si/Fe abundance ratio varies with radius in groups and clusters. This figure
shows that most rich clusters are characterized by an overabundance of Si
relative to Fe at large radii relative to the solar ratio. At large radii,
groups can attain gas mass fractions comparable to rich clusters (see
below), but they still lack the metals in terms of {\it lower} Fe abundances
and IMLRs (see Figs.\ref{fe-fig} and \ref{fig:imlr}).

\includegraphics[width=3.25in]{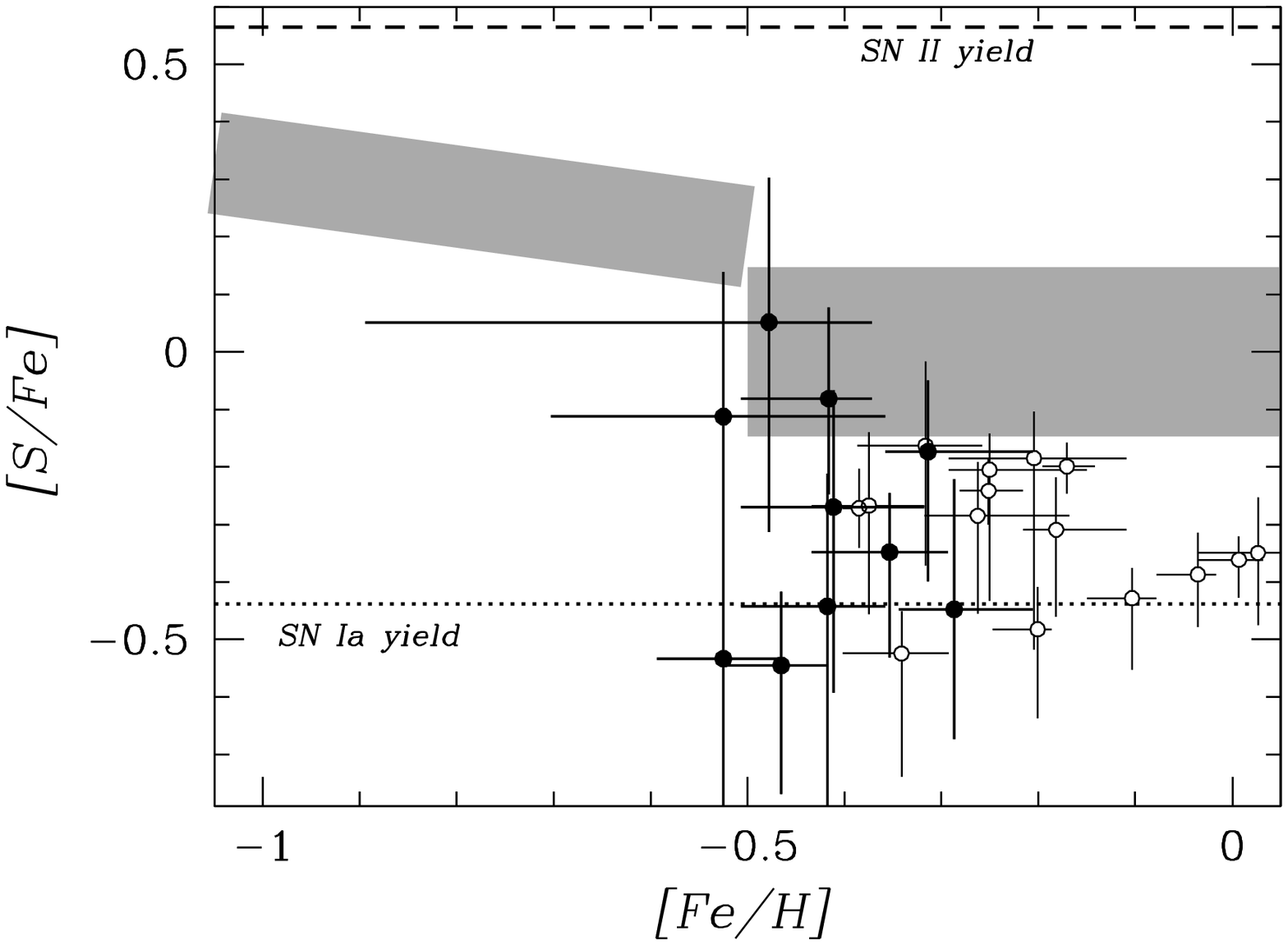} 
\figcaption{ [S/Fe] vs [Fe/H] in clusters of galaxies (filled circles) and
in cD galaxies (open circles).  The lines are the same as in
Fig.\ref{cl-plane-fig} (with the thick gray line representing the observed
abundance pattern in Galactic stars).
\label{cl-s2fe-fig}}

We compare our abundance determinations for the hot gas in groups and
clusters with stellar data in Fig.\ref{cl-plane-fig}.  To make a direct
comparison between the hot gas and stars, we convert our abundances to the
``meteoritic'' units in Anders \& Grevesse (1989). We omit the data within
the central 200 kpc of clusters where enrichment by cD galaxies may be
important. Fig.\ref{cl-plane-fig} shows that the gas in groups experienced a
greater relative enrichment from SN~Ia compared to the gas in rich clusters
($-0.2\pm0.05$ dex in terms of [Si/Fe] for a given [Fe/H]). This can arise
from either a greater release of SN~Ia elements or a greater loss of SN~II
ejecta in groups compared with that in rich clusters. As we show below, the
mass of heavy elements injected by SN~Ia per unit blue luminosity is similar
in groups and clusters; however, groups have a global deficit of SN~II
ejecta.  Examination of Fig.\ref{cl-plane-fig} shows that simply coadding
groups will not reproduce cluster abundance ratios even in the high
abundance central regions.  Groups also lack the high Si/Fe component, seen
in the outskirts of rich clusters.

Supersolar Si/Fe ratios found at the outskirts of clusters of galaxies can
be considered as an evidence for a top-heavy IMF (Loewenstein \& Mushotzky
1996; Wyse 1997). However, there is strong observational evidence in favor
of a universal IMF (of the form characterized by Kroupa, Trout and Gilmore
1993), based on several different studies (cf Proc. of 33 ESLAB Symposium;
Wyse 1997). So, as an alternative to suggestion of Loewenstein \& Mushotzky
(1996), we consider below a preferential loss of SNII ejecta. Our arguments
assume a delay in SN Ia enrichment relative to SN II enrichment. As long as
the escape of SN II ejecta occurs before SN Ia start to explode, than the
observed Si/Fe ratio in IGM can be reproduced.  Observations of our galaxy
imply a time lag in SN Ia enrichment of $\sim3$ Gyr (Yoshii \etal 1996).

According to theoretical modeling of SN Ia explosion, this delay results
from low-metallicity inhibition of SN Ia (Kobayashi \etal 1999), so a delay
in SN Ia enrichment in elliptical galaxies could be reduced to $\sim0.1$
Gyr. We thus conclude that the enrichment of the IGM originates from either
metal-poor galaxies or from short lived star bursts. Assuming the former, we
can reproduce the high iron abundance in the ICM only via metal-rich
galactic winds (but not e.g. by stripping of the gas), so star-bursts are
required in both scenaria. Observations of cluster galaxies indicate a
reduced star-formation activity (Balogh \etal 1998). Together with a flat
radial distribution of $\alpha$-process elements this argues in favor of SN
II enrichment occurring before cluster collapse.  Since escape of SN Ia
ejecta is enhanced at the end of the star burst (Recchi \etal 2000), detailed
simulations are needed in order to justify the short-lived starburst
scenario, proposed here.

As was shown by Mushotzky \etal (1996) and later confirmed by Finoguenov \&
Ponman (1999), the observed S/Fe ratios of clusters are consistent with pure
SN~Ia yields, despite the fact that S is an $\alpha$-process element.
Fig.\ref{cl-s2fe-fig} shows that our results reinforce this problem.
Applying the results of SN~II nucleosynthesis calculations to measurements
of [S/Fe] vs [Fe/H] for galactic stars, Timmes, Woosley and Weaver (1995)
conclude that sulfur is mainly produced by very massive progenitors
($M>30$\msun) of high metallicity (Fe abundance $>0.1$ solar).  Thus, lack
of S in ICM, argues in favor of production of bulk of elements in {\it
metal-poor} galaxies. As discussed in Timmes \etal (1995), sulfur is not
directly measured in stars or modeled with a high degree of certainty, but
may serve to discriminate the flat IMF required to explain both cluster and
elliptical enrichment.  In addition, [S/Fe] in the ICM can provides
constraints on the release of elements from spiral galaxies, which are
expected to have high [S/Fe].

Alternatively, a non-spherical SN II explosion was proposed to explain the
low S/Fe (Nagataki \& Sato 1998). In this scenario it is possible to have
the top-heavy IMF, and the production of sulfur in metal-rich galaxies
together.

%\medskip

\section{Iron Mass to Light ratios}\label{sec:disc}

The iron mass-to-light ratio (IMLR, in units of $h_{50}^{-1/2}$\msun/\lsun)
is a very useful diagnostic for characterizing the ability of a system to
synthesize and retain heavy elements. Previous calculations of the IMLR were
based on central iron abundance determinations, which can overestimate the
cumulative IMLR within 3 Mpc by a factor of 3, assuming that that the Fe
abundance drops to 0.1 solar at this radius. By resolving the spatial
distribution of heavy elements, we can make a comparison with the
distribution of optical light and not just the integrated values. The basic
optical properties of our cluster sample are summarized in Table
\ref{tab:opt}.  The optical light is normalized to the observed value $L_B$
(col. 3 or col. 5 in Tab.\ref{tab:opt}) within a radius $R_{L_B}$ (col. 4 in
Tab.\ref{tab:opt} or 0.5 Mpc, when $L_B$ values are taken from col. 5), and
a King profile is assumed for the galaxy light distribution. The cumulative
IMLR, including light from the cD galaxy (which we include as a central
point source) is shown in Fig.\ref{fig:imlr}. The radii in
Fig.\ref{fig:imlr} are expressed in units of the virial radius (see
Tab. \ref{tab:opt}, col. 13).

\centerline{\includegraphics[width=3.25in]{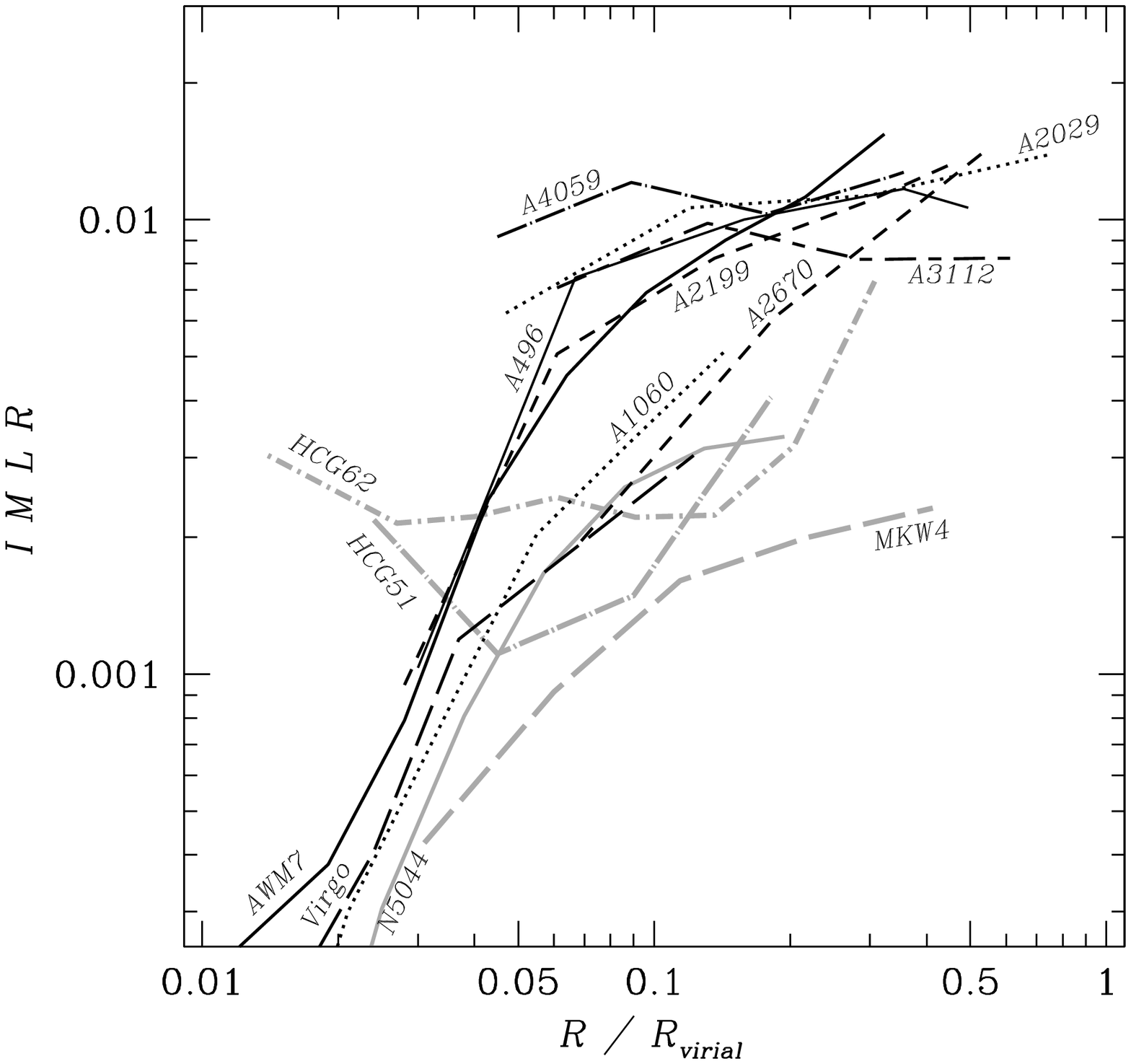}}

\figcaption{ Cumulative IMLR. The grey lines denote groups and black lines
denote clusters. A2597 is omitted from the plot for clarity,
however, due to the low optical luminosity of its cD galaxy,
the corresponding IMLR is similar to Fig.\ref{fig:imlr:nocd}.
\label{fig:imlr}}
\medskip

While the IMLR increases with radius in all systems, clusters have
higher IMLR values at a given radius compared with that in groups. The IMLR
is less than 0.02 at all radii in clusters and less than
0.006 at all radii in groups.  In the central $\sim 70$ kpc of cD groups and
clusters (this only excludes HCG51 and HCG62), the IMLR is less than
0.001. The Fe in this region can be produced by stellar mass loss from the
cD galaxies alone (cf Mathews 1989).

\begin{figure*}
\includegraphics[width=3.25in]{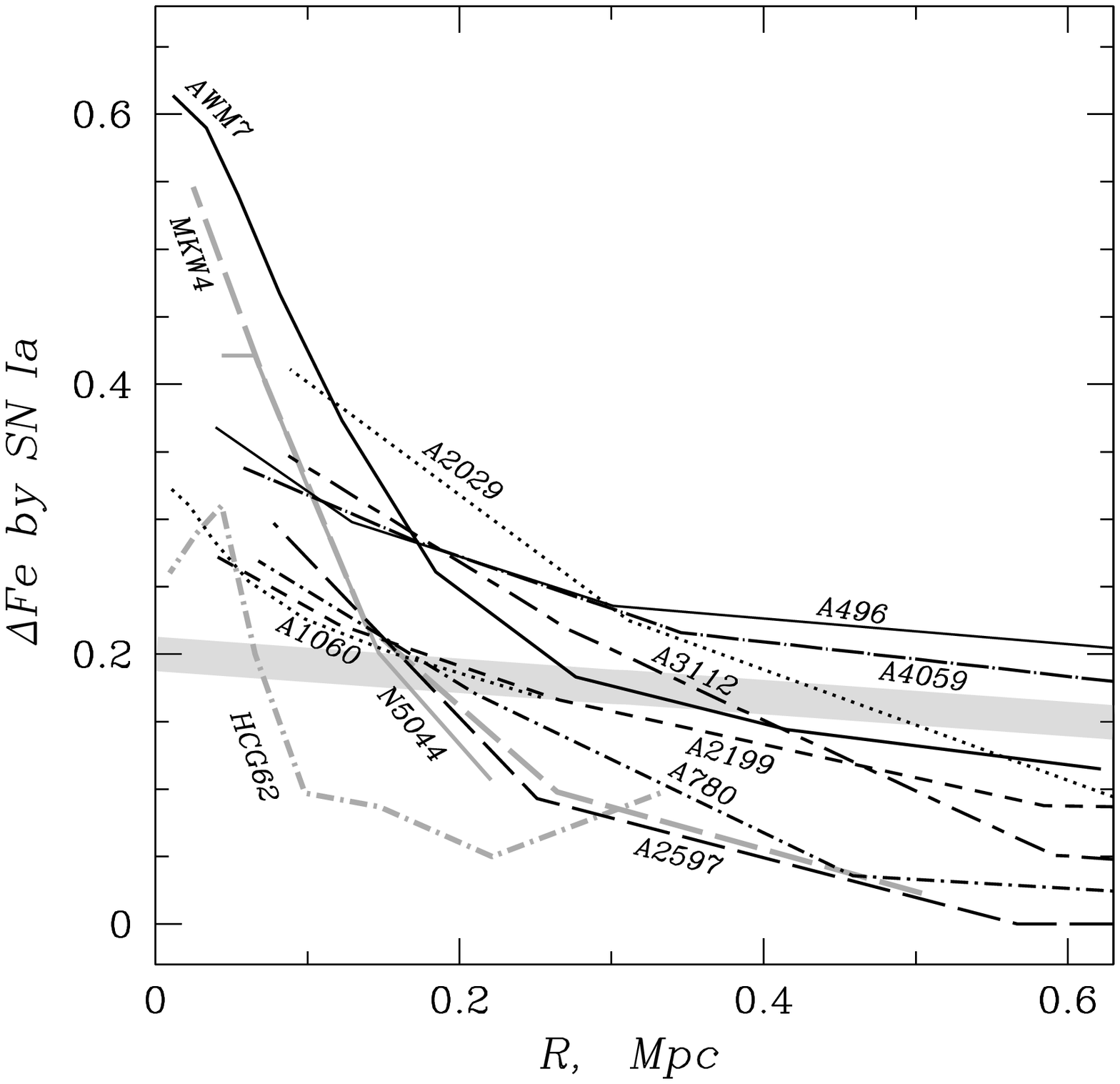}\includegraphics[width=3.25in]{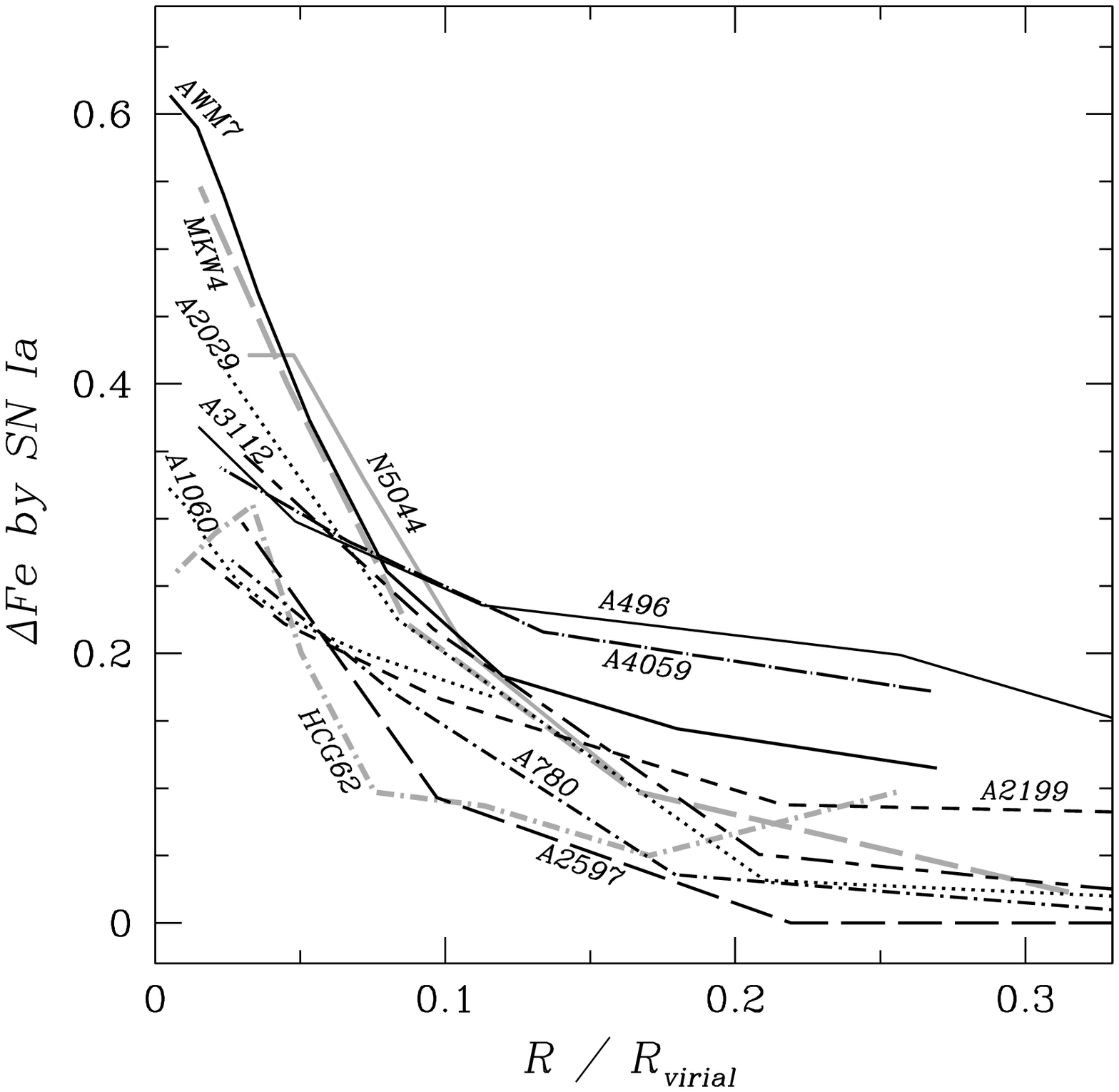}

\figcaption{ $\Delta Fe$ (in solar 'photospheric' units) injected by SN Ia
as a function of radius (in units of Mpc in the {\it left} panel and virial
radii in the {\it right} panel). Gray lines represent groups and black lines
represent clusters. Error bars at all radii are on average 0.1 at 68\%
confidence level. The thick gray line shows the variation in the early-type
galaxy fraction with radius in clusters (Whitmore, Gilmore and Jones 1993)
adjusted by a factor of 0.2 for illustrative purposes.  Comparing the two
figures one can see that the scatter in the data is diminished when virial
units are used.
\label{SNIa-light-fig}}
\medskip

\end{figure*}

\subsection{IMLRs of both Types of SNe}\label{sec:sne}

By resolving the spatial distribution of elements synthesized in different
types of SNe, we can determine the IMLR for each type. The primary method
for discriminating between enrichment by SN~Ia and SN~II lies in the
differences between their heavy element yields. This is primarily reflected
in differences between the $\alpha$-process elements and Fe. In order to
separate the input of different types of SN to the observed element
distributions, we have to make assumption about SN yields.  The heavy
element yields of SN~II were recently calculated by Woosley and Weaver
(1995). Since then a major effort has been carried out to compare their
yields with stellar data and observations of SN 1987A.  These studies (\eg\
Timmes, Woosley and Weaver 1995; Prantzos and Silk 1998) indicate that the
yields for the iron-peak elements in the Woosley and Weaver models should be
reduced by a factor of $\sim2$.  This reduction is primarily driven by the
observed O/Fe ratios in stars and observations of other $\alpha$-process
elements including Si (Prantzos private communication). One of the main
goals of this paper is to discriminate between the Fe that originated from
the two types of supernovae.  To accomplish this we adopt Si and Fe yields
of $y_{Si}=0.133$\msun\ and $y_{Fe}=0.07$\msun\ for SN II.  We note that
$y_{Fe}=0.07$\msun\ was also adopted in Renzini \etal (1993).  The yields of
SN Ia are obtained from Thielemann \etal (1993) with $y_{Si}=0.158$\msun\
and $y_{Fe}=0.744$\msun.  To decompose the Fe profiles into contributions
from the two types of SNe, we use the observed Si and Fe abundances in each
cluster and the adopted SN yields, as listed above.  This approach has the
advantage that Si and Fe abundance profiles have been determined for our
entire sample, and that the spread in the Si yields is low between different
modeling of SN element production (cf Gibson \etal 1997).

As can be seen in Fig.\ref{SNIa-light-fig}, the contribution from SN~Ia to
Fe abundance is centrally concentrated in all systems.  The [Fe/H]
distribution from SN~Ia depends on the spatial distribution of the injected
Fe relative to the distribution of the ambient gas when the enrichment
occurs.  For example, if we assume that during the period of enrichment the
gas and galaxies have the same distribution and that all galaxies contribute
equally to the enrichment, then [Fe/H] would be independent of radius. Since
a negative iron abundance gradient is observed, SN Ia enriched gas must be
preferentially released near the center of groups and clusters.

In Fig.\ref{SNIa-light-fig} we also show the radial behavior of the
early-type galaxy fraction in clusters (Whitmore \etal 1993). Whitmore \etal
find this behavior to be universal in all clusters in their sample. We note
however, that their sample only includes systems optically richer than
A1060. Beyond the central 300-400 kpc, the radial variation in the
early-type galaxy fraction is in good agreement with the declining iron
abundance.  In the central regions of the groups and clusters, however, the
abundance of iron increases much faster than the early-type galaxy fraction.
Fig.\ref{SNIa-light-fig} also shows that the scatter in SN Ia profiles
inside the central $0.1R_{virial}$ is diminished when virial units are used
instead of physical distances.  These results indicate that the central
concentration of SN Ia synthesized Fe cannot be explained by the increased
fraction of early-type galaxies at cluster centers. Density dependent
mechanisms (e.g. ram pressure stripping, galaxy harassment, and tidal
disruption) are therefore favored.

\begin{figure*}

\includegraphics[width=2.3in]{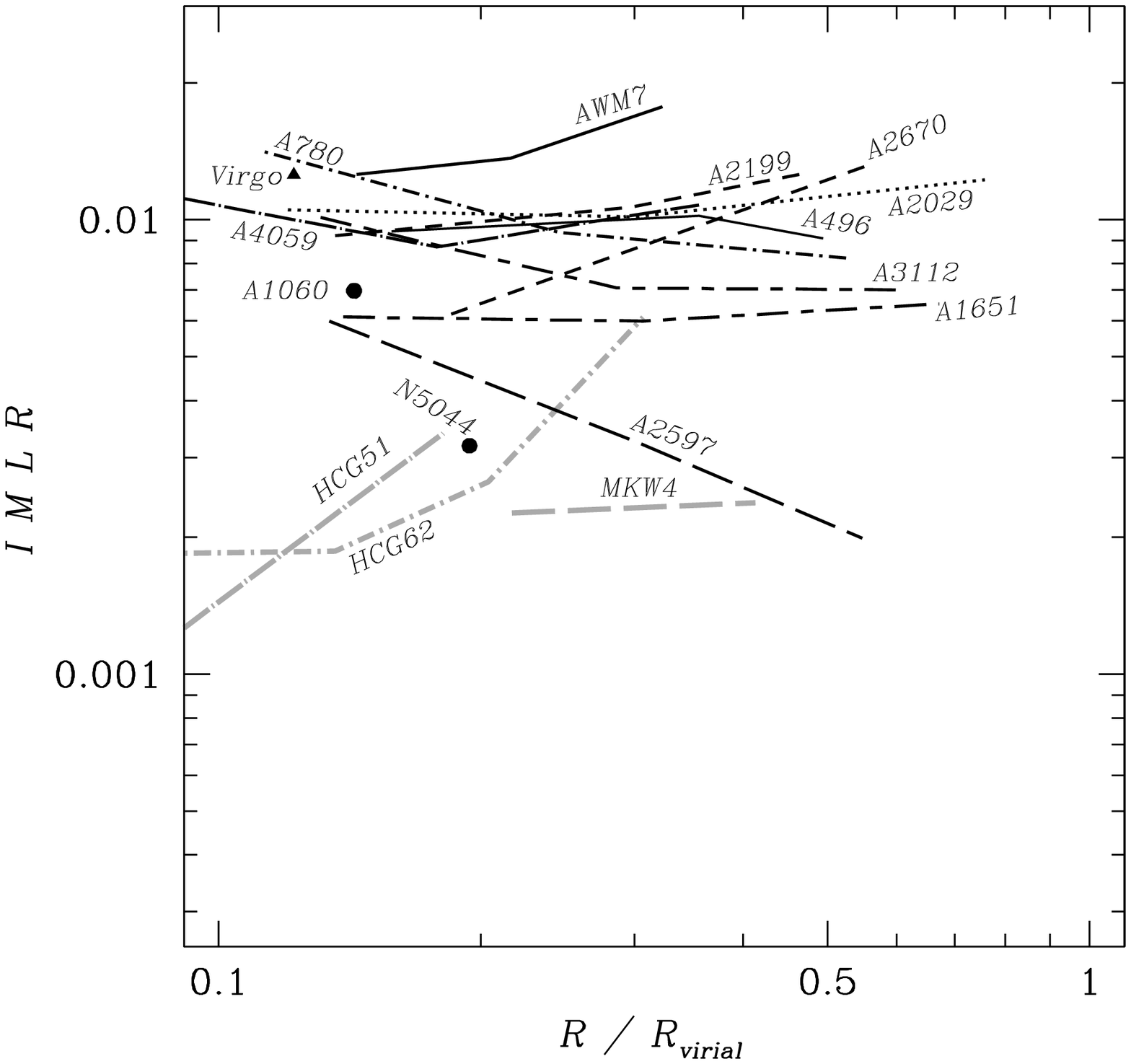} \hfill \hspace*{0.1cm}
\includegraphics[width=2.3in]{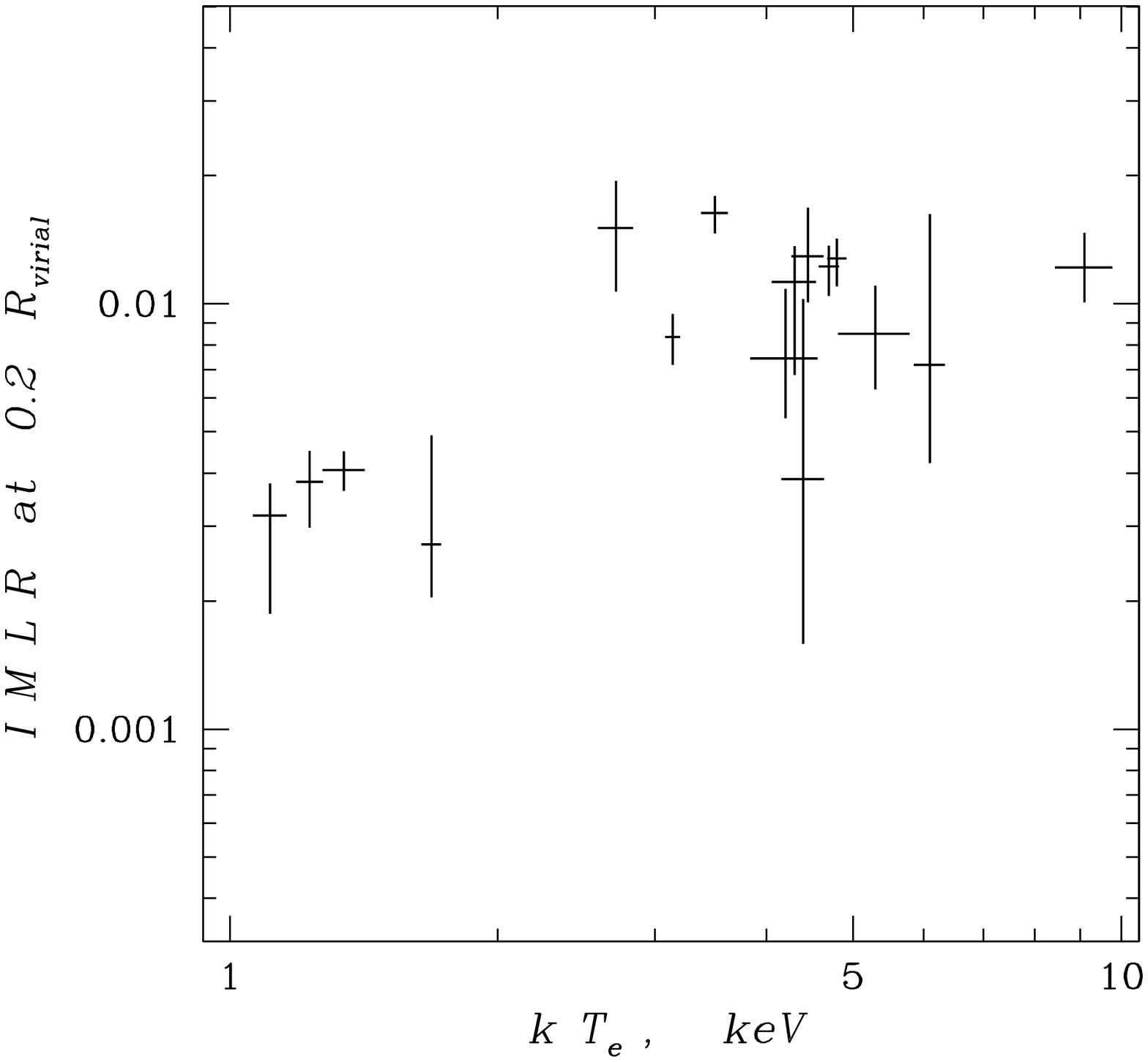} \hfill \hspace*{0.1cm}
\includegraphics[width=2.3in]{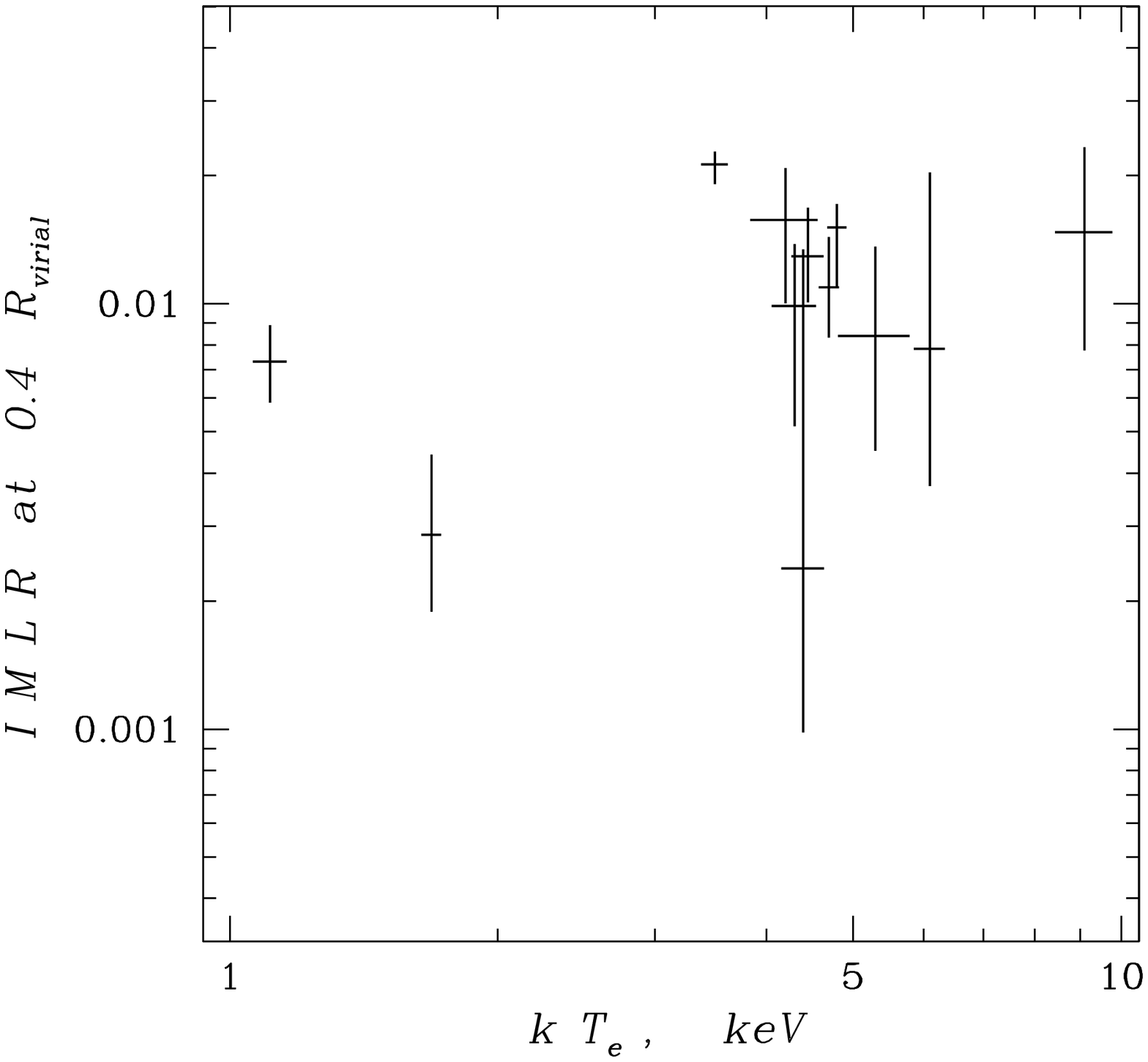}

\includegraphics[width=2.3in]{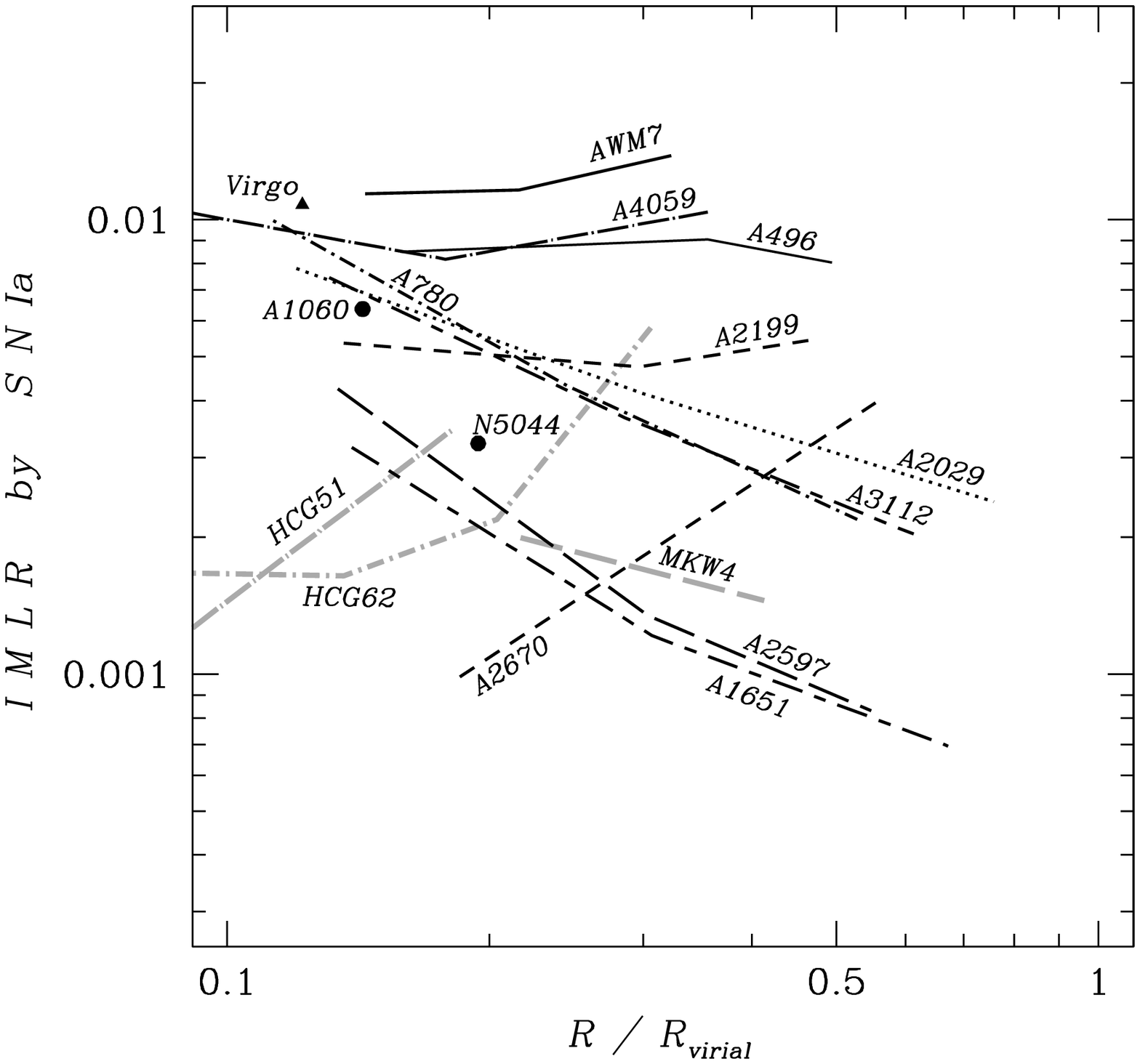} \hfill \hspace*{0.1cm}
\includegraphics[width=2.3in]{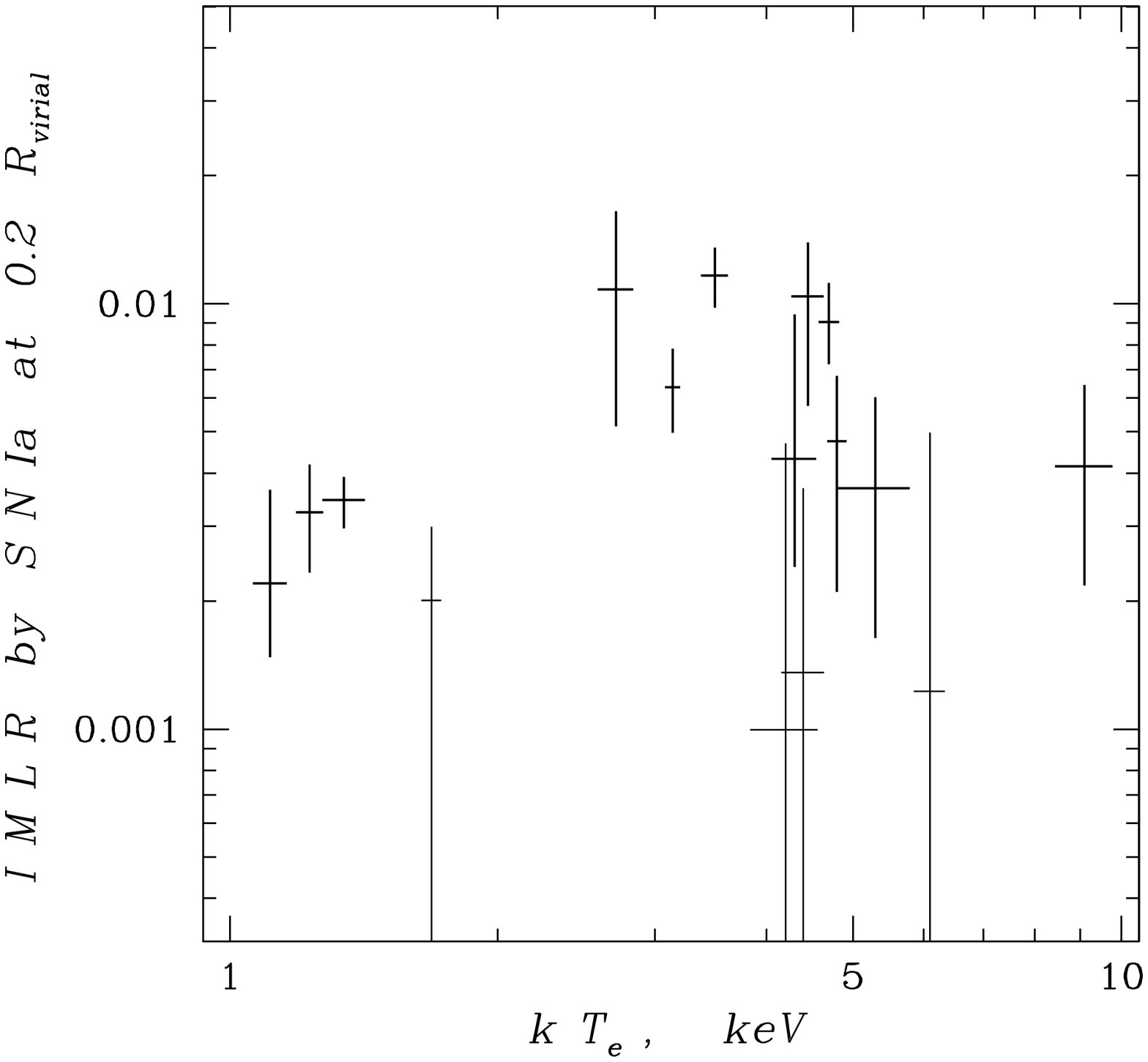} \hfill \hspace*{0.1cm}
\includegraphics[width=2.3in]{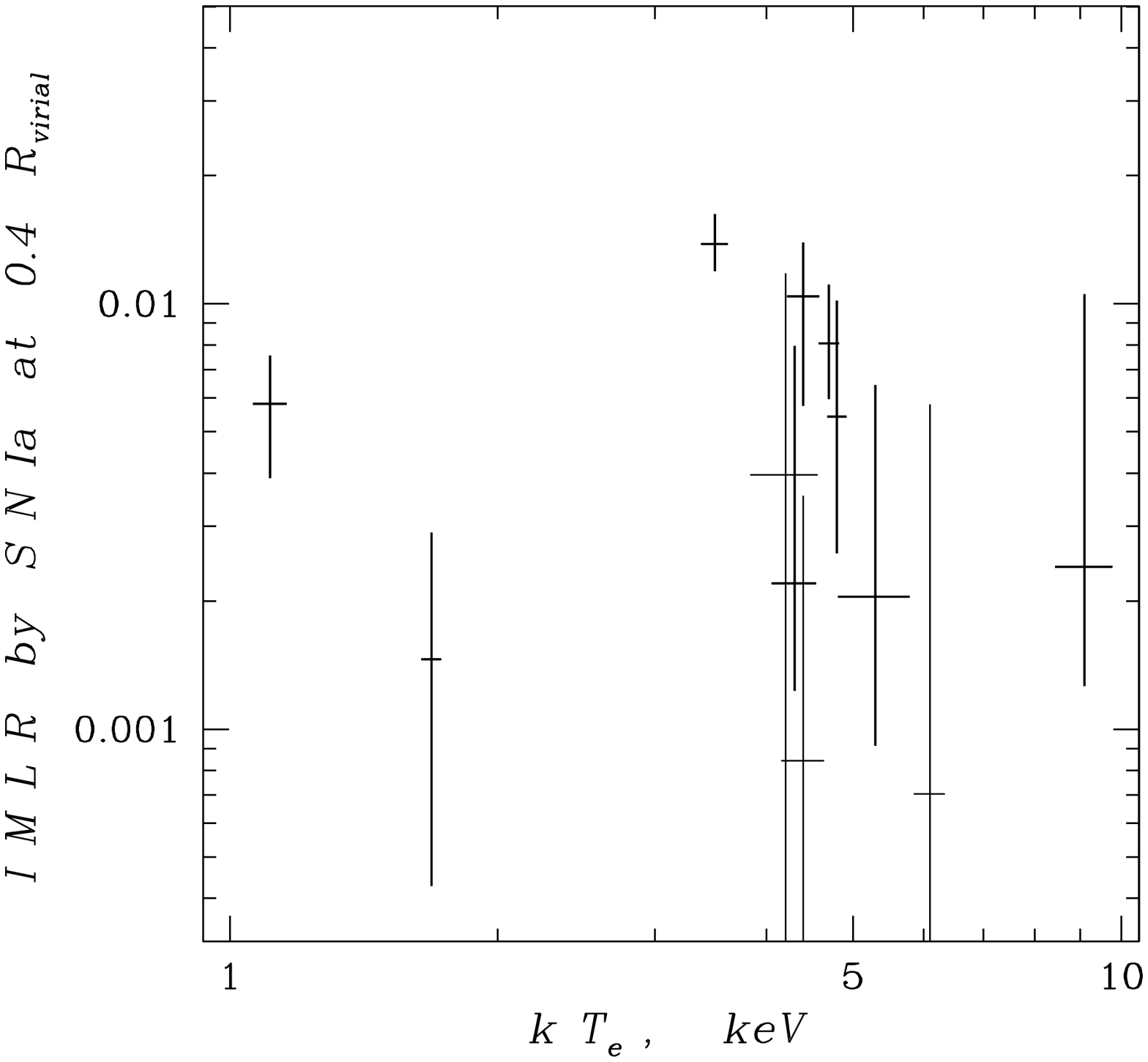}

\includegraphics[width=2.3in]{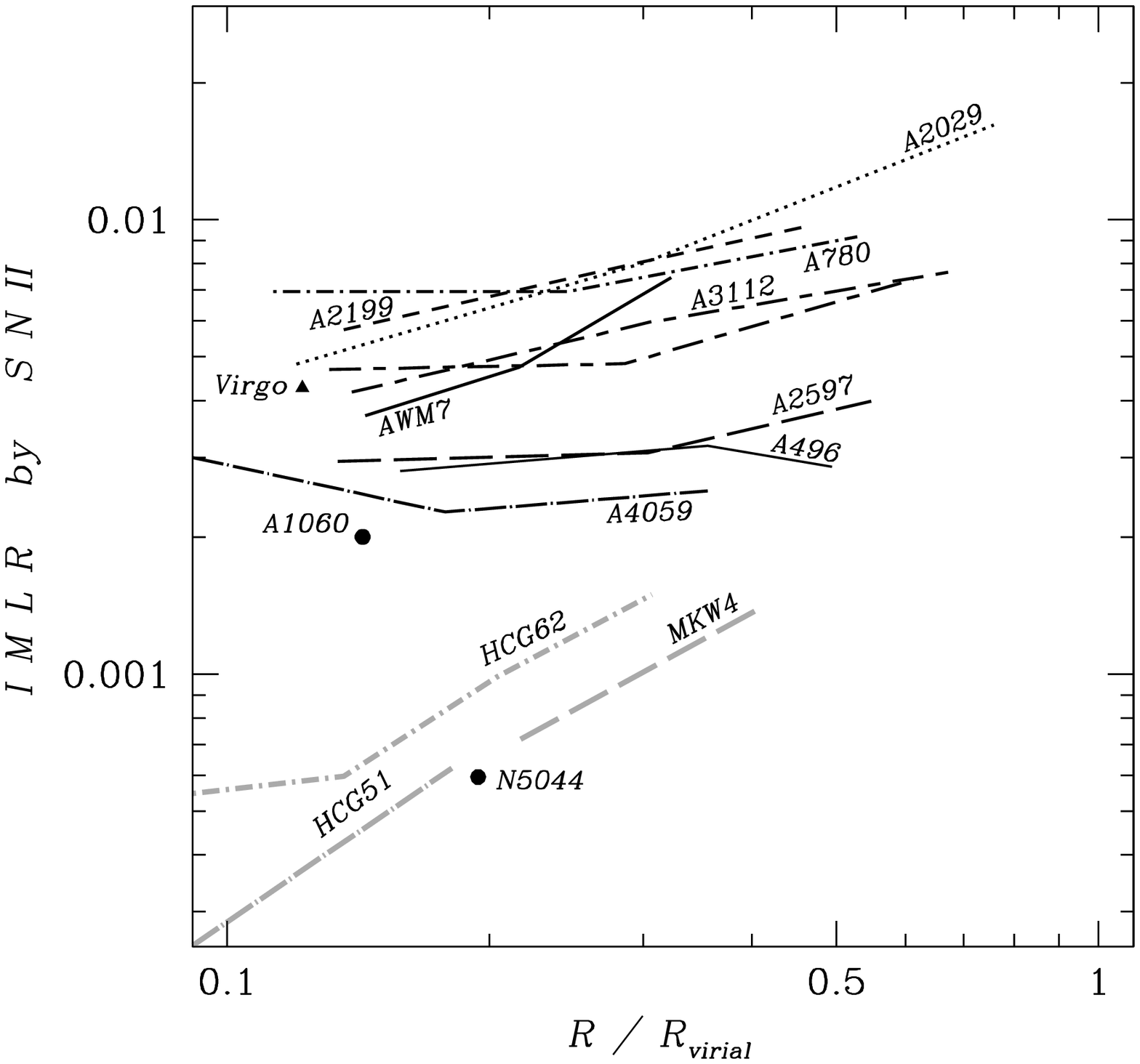} \hfill \hspace*{0.1cm}
\includegraphics[width=2.3in]{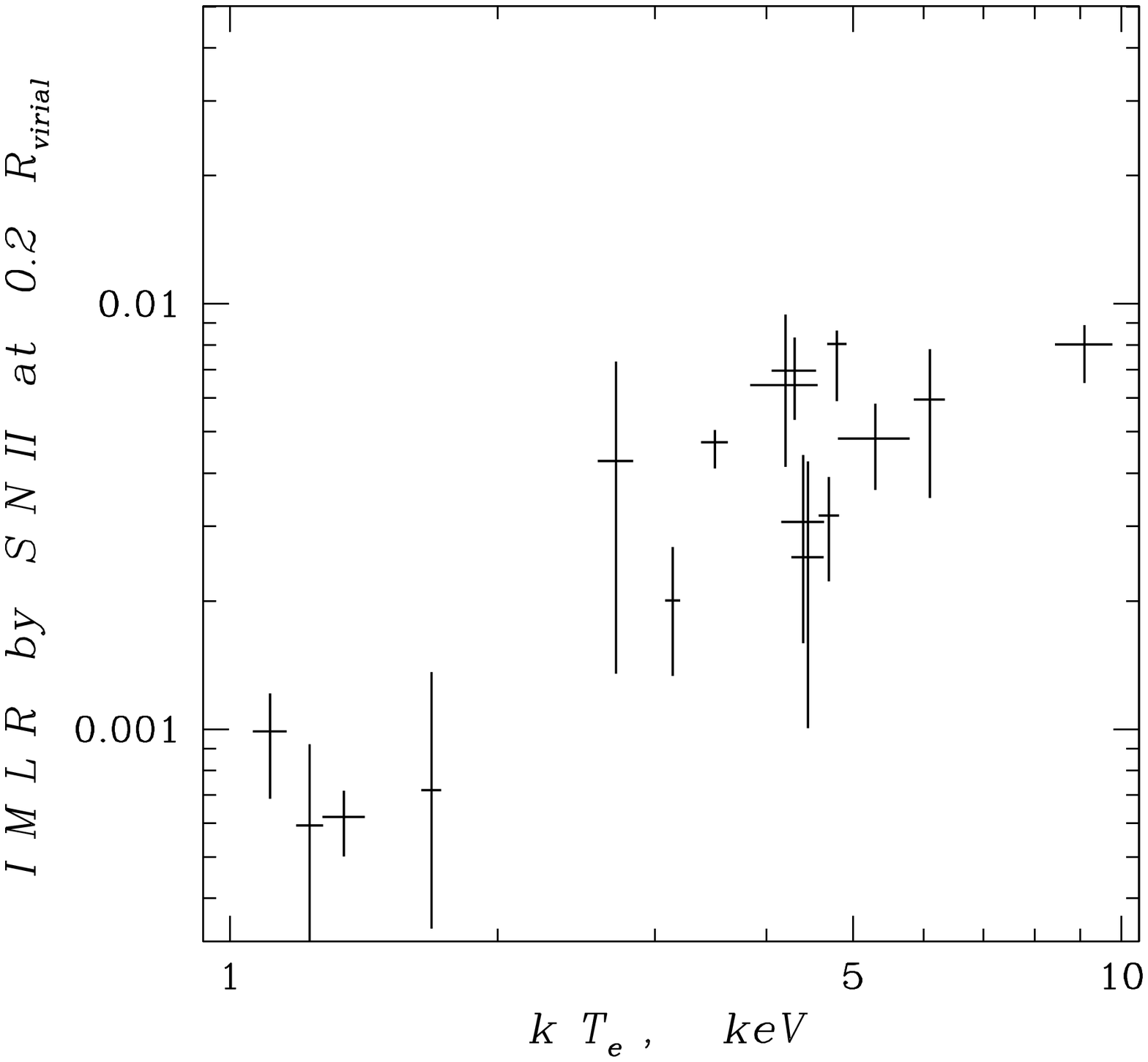} \hfill \hspace*{0.1cm}
\includegraphics[width=2.3in]{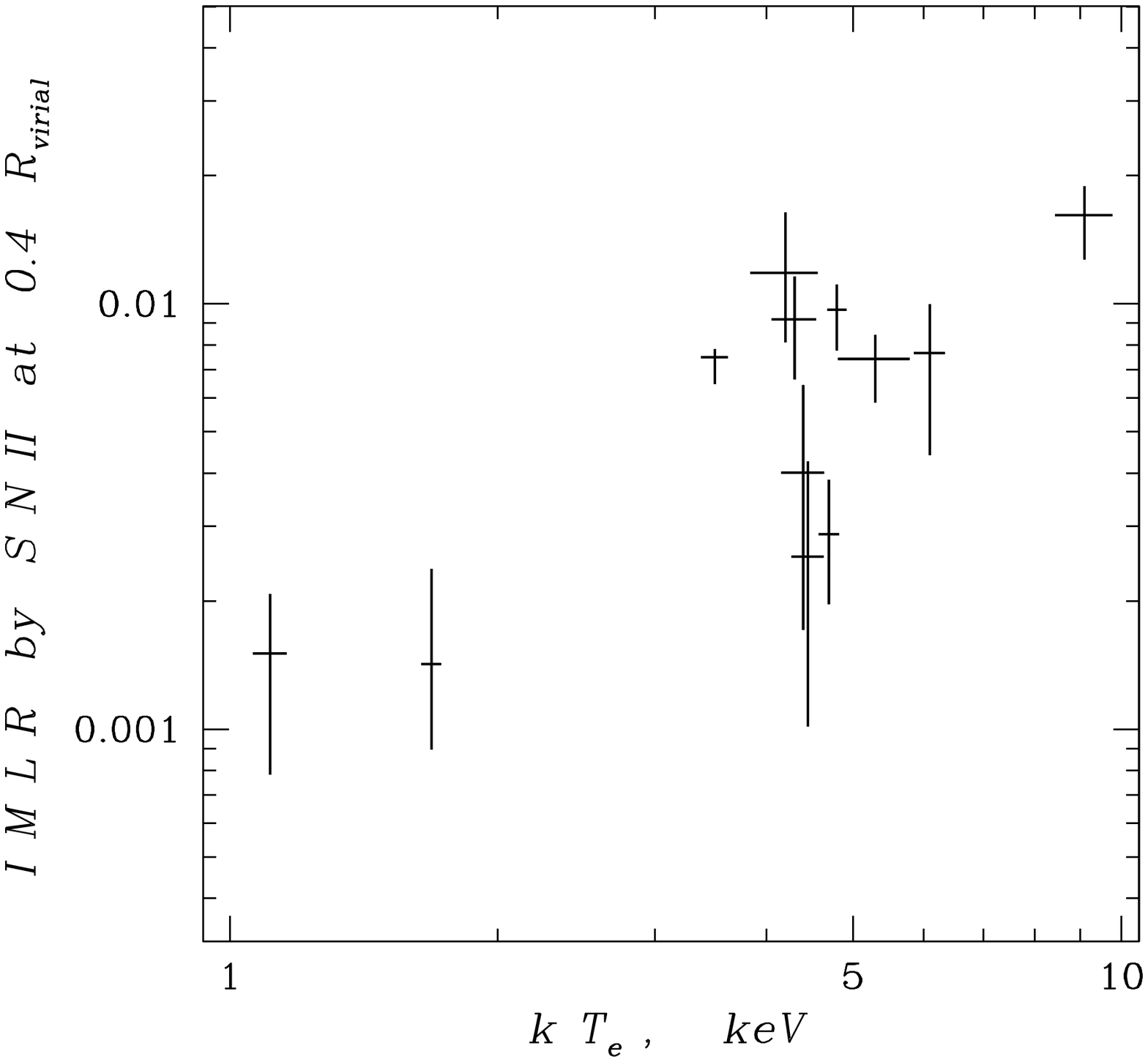}

\includegraphics[width=2.3in]{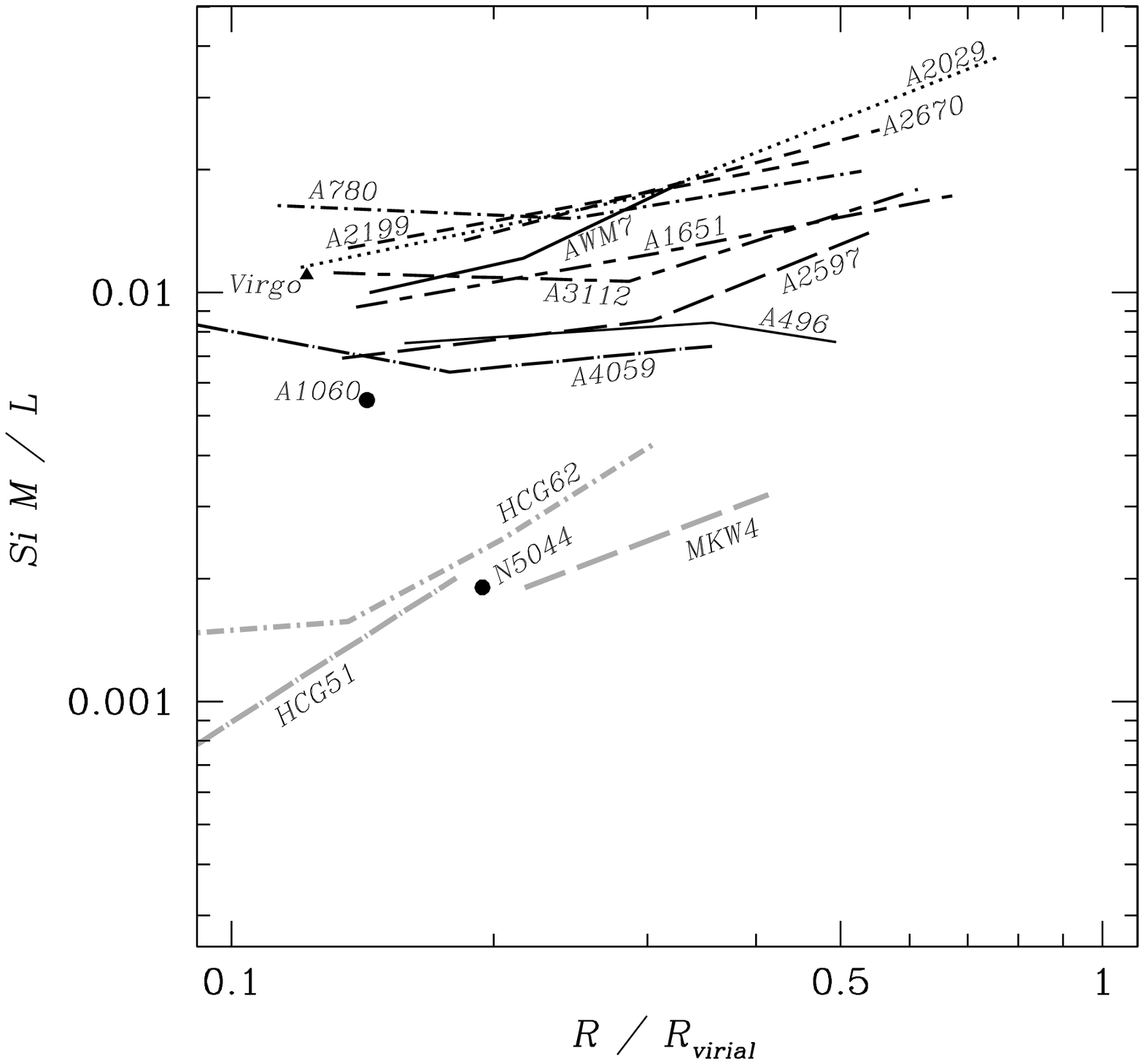} \hfill \hspace*{0.1cm}
\includegraphics[width=2.3in]{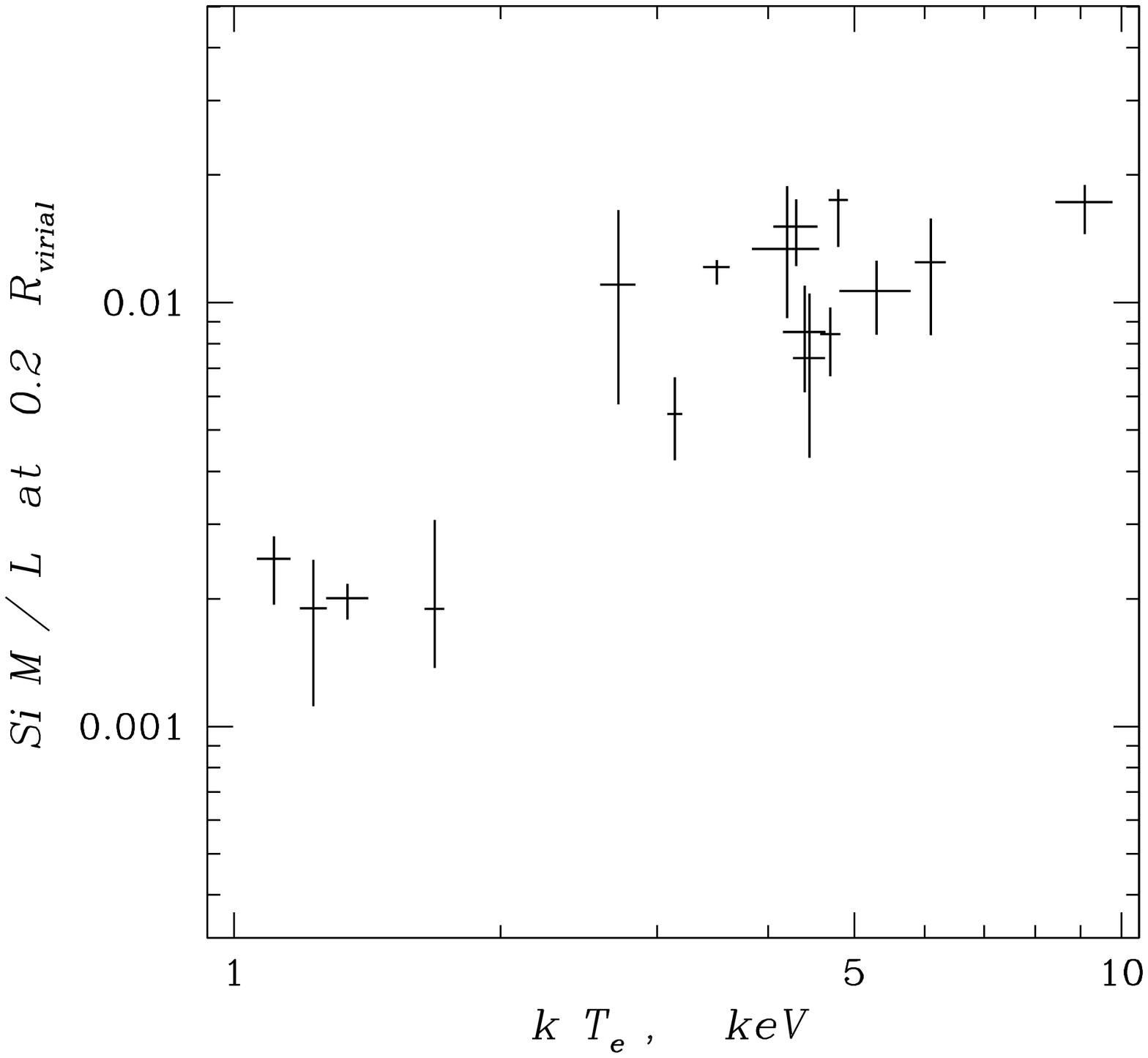} \hfill \hspace*{0.1cm}
\includegraphics[width=2.3in]{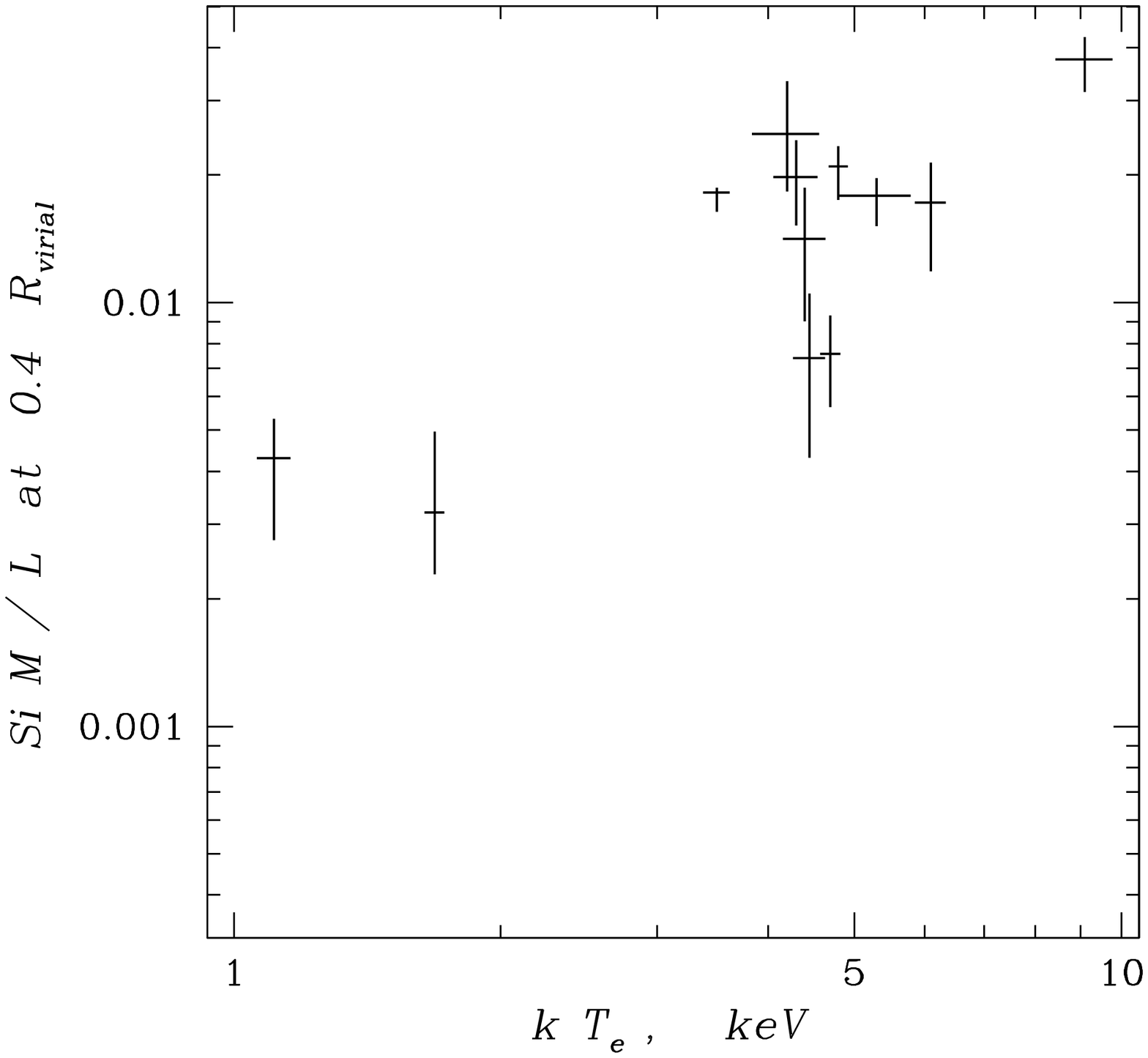}

\figcaption{M/L ratios for total Fe mass, Fe from SN Ia, Fe from SNII, and
Si vs cluster radius in units of virial radii (left column).  The middle and
right columns show the same M/L ratios at fixed radii of $0.2R_{virial}$ and
$0.4R_{virial}$. The central 200 kpc was omitted to exclude the effects of cD
galaxies, except for HCG51 and HCG62 which do not contain cD galaxies.
 \label{fig:imlr:nocd}}

\end{figure*}

\begin{figure*}

\includegraphics[width=3.5in]{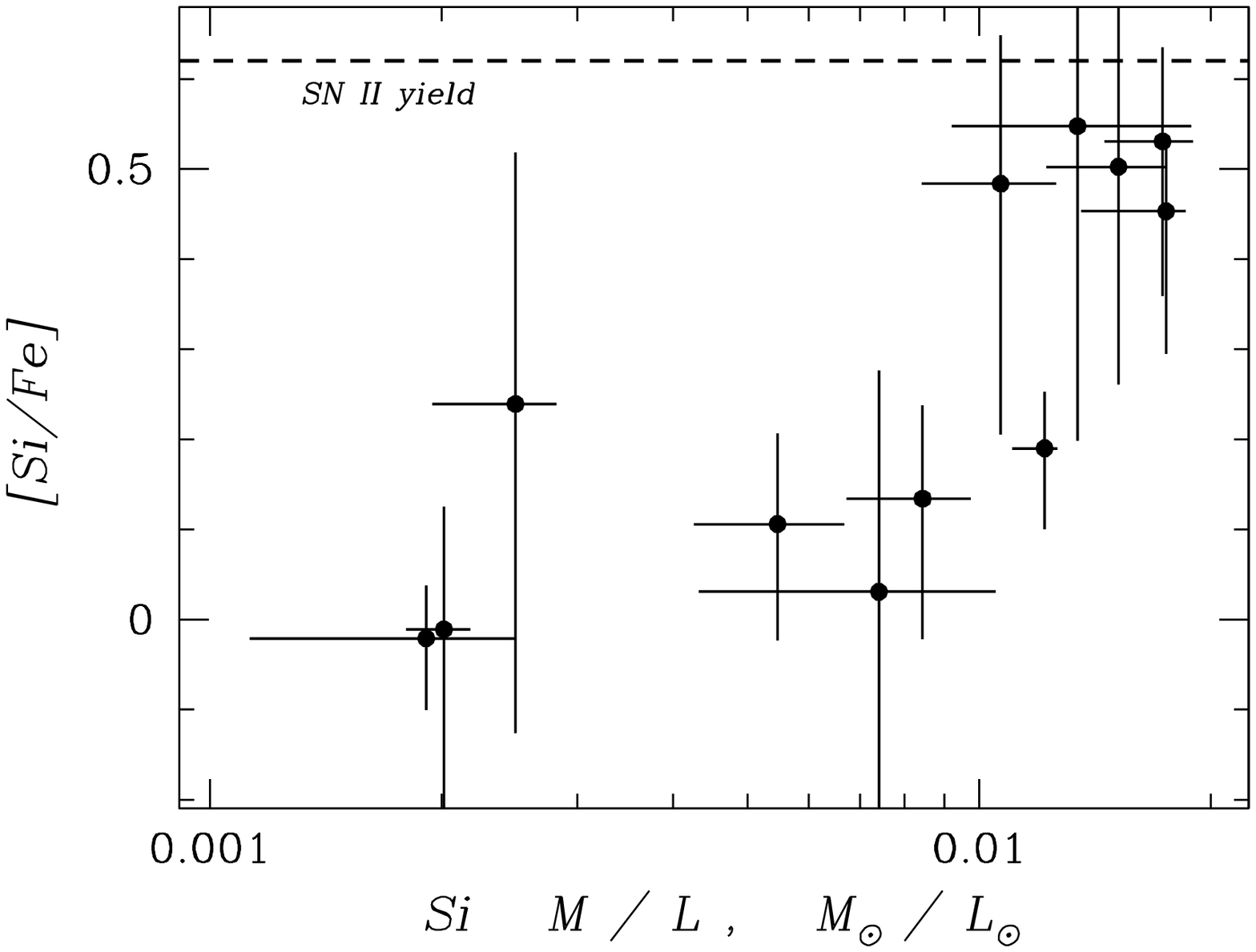} \hfill \includegraphics[width=3.5in]{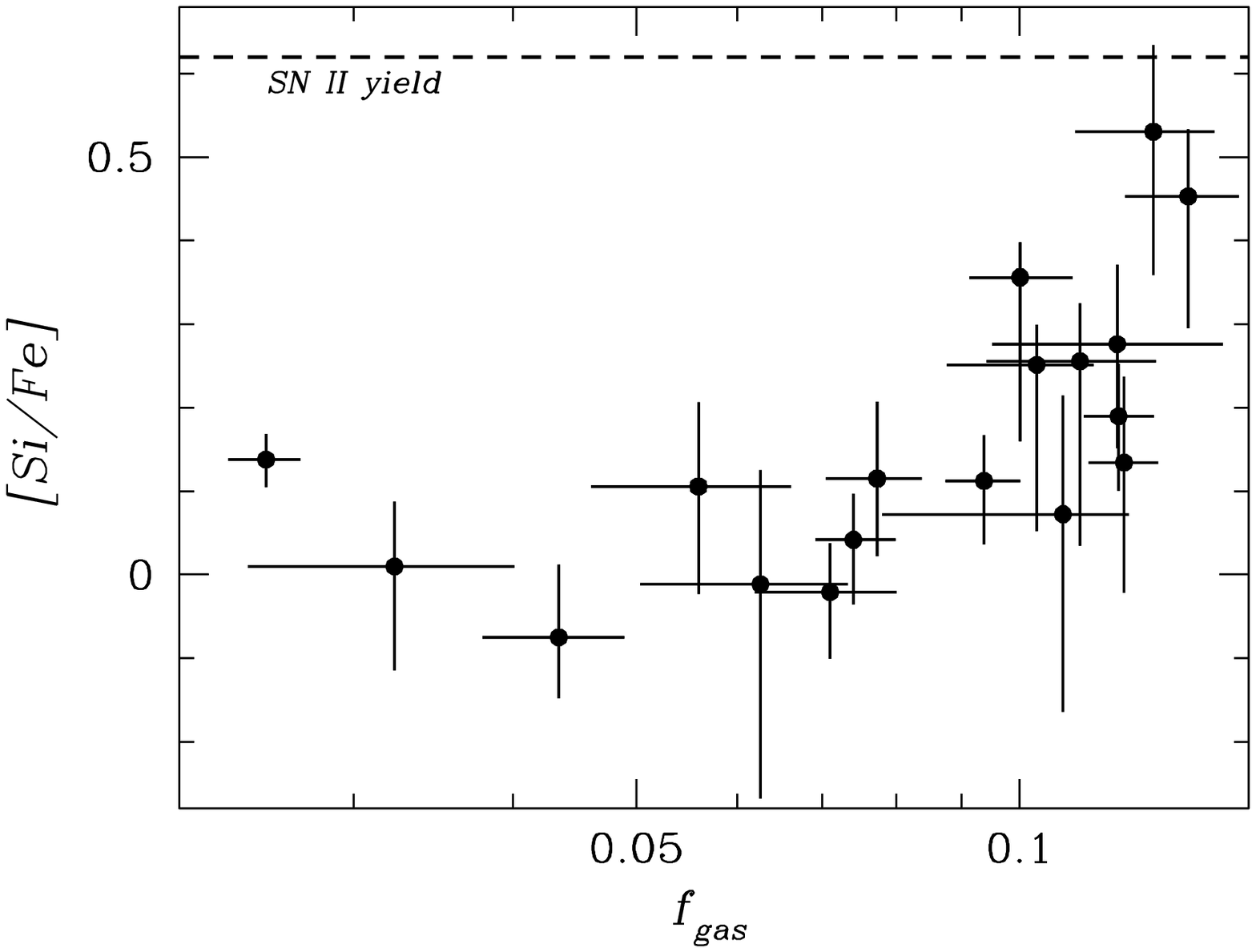} 

\figcaption{[Si/Fe] vs Si MLR (left panel) and [Si/Fe] vs. gas
mass fraction (right panel). All values are evaluated at 20\%
of the virial radius. Dashed lines indicate the adopted SN II yields.
\label{si2fe-imlr}}

\end{figure*}

In the following analysis we exclude the central 200 kpc in the computation
of the Fe mass and optical light.  This allows us to include the IMLR for
A780 and A1651 for which the luminosities of the cD galaxies are not
available. This exclusion only leaves one remaining point for Virgo, A1060,
and NGC5044.  In Fig.\ref{fig:imlr:nocd} we show the total IMLR, its
decomposition into type II and Ia SN, and the Si M/L.  This figure shows
that Fe synthesized in SN Ia is centrally concentrated in groups and
clusters, while Fe synthesized in SN II is dominant in the outer regions of
clusters.  The distribution of the Si M/L ratio is very similar to the Fe
M/L ratio from type II SNe and is, on average, flat or increases slightly
with radius. Clusters also have larger values of Si M/L than groups.  The
radial behavior of the Si M/L and Fe M/L from SN II profiles is consistent
with the scenario in which SN II ejecta is only weakly retained (or
captured) in groups.

We also show in Fig.\ref{fig:imlr:nocd} how the IMLR varies with cluster
temperature at radii of $0.2R_{virial}$ and $0.4R_{virial}$.  At a given
fraction of a virial radius, the IMLR from SN~II and the Si M/L increases
with gas temperature.  This trend implies a greater retention ability of Fe
and Si released by SN~II in hotter systems. The average value of the IMLR
from SN~II among the clusters is 0.005 at 0.2$R_{virial}$ and 0.009 at
0.4$R_{virial}$.

Fig.\ref{si2fe-imlr} shows how the Si/Fe abundance ratio varies with Si M/L
and gas mass fraction at a fixed radius of 0.2$R_{virial}$.  The large
scatter in gas mass fraction in our analysis occurs at a much smaller radius
than studied by Ettori \& Fabian (1998) and Vikhlinin, Forman \& Jones
(1999) who observed very little scatter in the cluster gas mass
fraction. These results are discussed in Finoguenov, Arnaud, David (2000)
where we find that the scatter in gas mass fraction decreases strongly with
radius, as the central entropy floor of a system becomes negligible compared
to the total entropy of the system at large radii. As follows from
Fig.\ref{si2fe-imlr}, systems that retained or accreted the most gas also
have the largest Si M/L ratios and Si/Fe ratio representative of SN II
ejecta.  Finoguenov \& Ponman (1999) came to a similar conclusion by
examining the radial behavior of the IMLR and gas mass fraction in
clusters. These arguments also provide strong constraints on the relative
importance of AGNs in heating the cluster gas (Wu, Fabian and Nulsen 1999).

The Si/Fe ratio in groups and clusters varies by 0.8dex, fully spanning the
range from pure SN Ia ejecta to pure SN II ejecta.  In some clusters,
contribution of SN Ia to element enrichment reaches 0.01 in terms of
IMLR. Using the modeling of the cosmic history of SN Ia explosion (Tab.1 in
Renzini et al. 1993) we conclude the role of SNe~Ia in the iron enrichment
process of the ICM is consistent with the present-day SN~Ia rate found in
optical searches (Capellaro \etal 1997) and the abundances determined from
X-ray observations of early-type galaxies (cf Finoguenov \& Jones 2000), but
requires higher SN Ia rates in the past.

\section{Summary}\label{sec:sum}

We have derived the distribution of heavy elements in a sample of groups and
clusters and compared these distributions to the 
galaxy distribution.  Our main results are:

\begin{itemize} 

\item For clusters with the best photon statistics, we find that the total
Fe abundance decreases significantly with radius, while the Si abundance is
either flat or decreases slightly.  This results in an increasing Si/Fe
ratio with radius and implies a radially increasing predominance of type II
SNe enrichment in clusters. 

\item SN~Ia products show a strong central concentration in most clusters,
suggesting an enrichment mechanism with a strong density dependence (e.g.,
tidal / ram pressure stripping of infalling gas rich galaxies). Moreover,
the corresponding IMLR for SN~Ia decreases with radius in most clusters.

% \item In an alternative scenario suggested by the simulations of Metzler \&
% Evrard (1994), a flattening of the gas profile relative to the galaxy
% distribution will naturally produce an abundance gradient if the process
% gas is injected after the difference in density profiles is established.
% Within this scenario, we expect that the central gas mass fractions would
% decrease after the initial epoch of SN II enrichment and produce the
% observed differences in the radial distributions of Fe and $\alpha$-chain
% elements in general agreement with the findings of Ponman \etal
% (1999). Given the radial behavior of the IMLR attributed to SN Ia, we
% suggest that both mechanisms may be important in the formation of the
% observed Fe abundance gradient.

\item Our results strongly favor a model in which SNe II explode mostly
before cluster collapse, whilst SNe Ia ejecta are removed from galaxies by
e.g. a density-dependent mechanism after cluster collapse, in general
agreement with the findings of Ponman \etal (1999). In simulations of the
formation of abundance gradients this has not yet been taken into account.

In a scenario suggested by the simulations of Metzler \& Evrard (1994),
flattening of the gas profile relative to the galaxy distribution will
naturally produce an abundance gradient if the processed gas is injected
{\it after} the difference in density profiles is established. Therefore
this model should be applied only to the Fe contributed later by SN Ia. If
this scenario is correct, the distribution of IMLR from SN Ia (as opposed to
the Fe abundance) should be flat with radius, whilst if density-dependent
mechanisms are at work, IMLR should decrease with radius. Given the radial
behavior of the IMLR attributed to SN Ia (Fig.11), we suggest that both
mechanisms may be important in the formation of the observed Fe abundance
gradient.

\item The dependence of [Si/Fe] vs [Fe/H] reveals a larger role of SN~Ia in
the enrichment of groups compared with that in clusters. Due to differences
in the elemental abundance patterns, it is not possible to produce a cluster
simply by coadding groups. The lack of metals attributed to SN~II in groups
suggests that SN~II products were only weakly captured (or retained) by the
shallower potential wells of groups due to the high entropy of the preheated
gas.

\item As an alternative to a top heavy IMF, we propose an explanation
for the observed supersolar Si/Fe ratios involving preferential loss
of SN II products from galaxies, arising from either very short
duration star-bursts or from enrichment by metal-poor galaxies.
Observations of subsolar S/Fe ratios argues in favor of the latter
suggestion.

%Gas with high Si/Fe ratios are
%preferentially found in clusters with large gas mass fractions, so it seems
%likely that this gas is released into the ICM during the early epochs of
%cluster formation.

% We have identified an additional component in the chemical enrichment
% of clusters that is associated with gas with an extremely high Si/Fe ratio.
% We suggest, based on the observed relation between [Si/Fe] vs [Fe/H] in the
% stars in our galaxy, that this component arises from galaxies with short
% periods of star formation.  The morphological type of galaxies associated
% with this enrichment mechanism remains to be resolved. Gas with high Si/Fe
% ratios are preferentially found in clusters with large gas mass fractions,
% so it seems likely that this gas is released into the ICM during the early
% epochs of cluster formation.

\end{itemize}

\section{Acknowledgments}

AF acknowledges support from Alexander von Humboldt Stiftung during
preparation of this work. The authors acknowledge the devoted work of the
ASCA operation and calibration teams, without which this paper would not be
possible. We thank Monique Arnaud for communicating to us the idea regarding
Si M/L ratio, Maxim Markevitch for his help in simulating the systematic
effects of ASCA XRT PSF, Alexey Vikhlinin for allowing gas mass comparison
with his ROSAT/PSPC results, and the referee for stimulating us to check the
effects of systematics in our modeling and for a number of useful
suggestions regarding theoretical interpretation of the data. Simulations
required for this work were performed using computer resources of ISDC, CfA
and AIP.

%\end{document}

%\clearpage

%\endrefs
%\clearpage
\appendix

\section{Drawbacks of application to CCD-quality spectra of
  MEKAL modeling of the iron L-shell complex as an abundance indicator.}

Discordant results have been reported for element abundance determinations
using X-ray emission of early-type galaxies and galaxy groups.  Supersolar
metallicities have been derived by Buote (1999, 2000a), fitting
two-temperature models to the integrated hot plasma emission from within the
central regions of galaxies and groups. Lower metallicities are found using
single-temperature models (e.g. Matsushita 1998), and also from spatially
resolved single-temperature fits combining ROSAT and ASCA data (Finoguenov
\etal 1999; Finoguenov and Ponman 1999; Finoguenov and Jones 2000).

Single-temperature models, when used to fit an intrinsically two-temperature
plasma, can be shown to underestimate element abundances (Buote 1999;
Finoguenov and Ponman 1999). However, real spectra from hot gas are unlikely
to be two-temperature, and fitting two-temperature models can also give
misleading results. Consider, for example, the following simple simulation:
using a MEKAL model with kT=1.35, Z=0.33, $N_H=0.7\times10^{21}$ (and other
parameters as for HCG51), we simulate an ASCA SIS spectrum and then fit a
two-temperature MEKAL model.  The final reduced $\chi^2$ is about 2/3
instead of 1, and the abundance is biased high (0.4--0.6 in different runs,
with varying exposure times). Now, when we have to fit the real spectra, the
MEKAL code does not exactly match the source emission, due to the complex
physics of iron L-shell emission. Recent high-energy resolution data has
been used to benchmark the MEKAL code (Phillips \etal 1999), revealing at
least 10\% systematics for the CCD energy resolution, while mismatches in
line positions will be a dominant problem at a resolution of few ev. Given
the presence of such systematics, overfitting the spectrum with a
two-temperature model may compensate for data-model mismatches, and give 
reduced $\chi^2$ values $\sim1$.

Taking a more realistic case, we use our derived radial distributions of
temperature and emission measure for HCG62, together with an iron abundance
dropping from 0.5 solar at the center down to 0.2 solar at
$r=$6\amin. Comparing the integrated spectrum from this region with that
from a two-temperature MEKAL model with parameters fixed at the values
published by Buote (2000a), we find that the fit does not deviate from the
simulated data by more than 10\%, which is well within the systematic
uncertainty of the MEKAL model. However, the Fe abundance in this model is
solar, deviating by factors 2--5 from the abundance input into these
simulations. The problem arises because Buote's results are derived from
spectral fits to integrated spectra from a region which is generally
dominated by a central cooling flow.

In a more recent paper, Buote (2000b) has performed a spatially resolved
spectral analysis of a number of the systems in our study using ROSAT PSPC
data, this time employing single-temperature model fits.  Again he finds
extremely high central abundances in many of his systems -- for example, his
central Fe abundance for HCG62 lies in the range 5-15 solar, in strong
conflict with our results. Buote (2000b) suggests that our low abundance
determination at the center of HCG62 is due to `over-regularization'. Since
it is certainly true that regularization would tend to suppress such a huge
central spike in abundance, if observed at low statistical significance, we
have checked the effect of completely turning off the regularization for Fe
in our analysis of HCG62. The result can be seen in Fig.\ref{fe-fig}. As
expected, the abundance is subject to larger fluctuations, however, our
basic results that the abundance is subsolar, and drops at large radii, are
unchanged.  In Finoguenov, Arnaud \& David (2000) we have analyzed Centaurus
cluster, which has a very steep abundance gradient profile and obtained
similar results to Ikebe \etal (1999). In addition we have simulated and
shown that an extremely peaked abundance profiles, e.g. 10 times solar as
reported in Buote (2000b), could not remain unseen by our analysis if it
were present.

To check the effects on our results of possible systematic effects of using
the MEKAL code, we included a 10\% systematic error into the fitting process
and reanalyzed the ASCA observation of HCG62. This analysis naturally
produces larger error bars in metallicity in the 0.1--0.4 Mpc region,
permitting values up to 0.6 solar at 0.1 Mpc at the 90\% confidence limit
(previously 0.4) and up to 0.3 at 0.2 Mpc (previously 0.2).  With the
revised ASCA error estimates, the mild discrepancy between our ASCA result
and the ROSAT Fe determinations reported in Finoguenov \& Ponman (1999) (and
shown in Fig.\ref{fe-fig}) is removed. Also, a Si abundance of 0.35 solar is
permitted at 0.4 Mpc (previously 0.25). We have also analyzed the 1998 ASCA
observations of HCG62 as an independent check. Both ASCA sets of
observations independently yield Fe abundance between 0.3 and 0.5 Mpc at the
$0.15\pm0.10$ level.

The conclusions that we draw from the above are that in the central cooling
regions of groups and clusters the temperature structure may be complex at
each radius, and also varies with radius. We believe that there is
considerable doubt in the credibility of abundance determinations derived
from integrated CCD-quality spectra of such regions using {\it any}
modeling. Results derived from the lower spectral resolution spectra
available from the ROSAT PSPC are even more suspect. Higher resolution X-ray
spectra will be required to derive reliable results for such regions. At
larger radii, where gas density is lower and cooling times long, it is
reasonable to represent the gas as a single temperature plasma at each
radius.  In this regard it is encouraging that temperature distributions
from ROSAT and ASCA analyses are generally in good agreement (Finoguenov \&
Ponman 1999, Finoguenov \& Jones 2000).  Here we believe that results from
our analysis should be correct to within systematic uncertainties resulting
from the use of the MEKAL model.

\end{document}